%% file: article.tex
\documentclass[onefignum,onetabnum]{siamart171218}

\usepackage{booktabs}
\usepackage{tikz}  
\usetikzlibrary{matrix,shapes,decorations.pathreplacing,external}
\usepackage{todonotes}
\input{shared}
\usepackage{makecell}

\usepackage{microtype}
\usepackage{seqsplit}
\newcommand{\code}[1]{\texttt{\seqsplit{#1}}}

\SetKwFor{For}{for}{in order do}{}
\SetKwFor{ForRev}{for}{in reverse order do}{}
\SetKwFor{ForPar}{for}{do}{}

\newcommand{\complexity}[2]{\tcp*{#1$(#2)$}}
\newcommand{\complexitytwo}[4]{\tcp*{#1$(#2)$\textrm{\textit{,}} #3$(#4)$}}
\newcommand{\complexitythree}[6]{\tcp*{#1$(#2)$\textrm{\textit{,}} #3$(#4)$\textrm{\textit{,}} #5$(#6)$}}
\newcommand{\comment}[1]{\tcp*{\textrm{\textit{#1}}}}

\newcommand{\triu}{\mathrm{triu}}

\newcommand{\htriu}{\mathrm{htriu}}

\newcommand{\hstril}{\mathrm{hstril}}
\newcommand{\PO}{\mathring{P}}
\newcommand{\po}{\mathring{p}}
\newcommand{\FO}{\mathring{F}}

\ifpdf
\hypersetup{
  pdftitle={A Portable and Versatile Limited-Memory BFGS Implementation in PETSc/TAO},
  pdfauthor={H. Suh, T. Isaac, A. Dener, T. Munson, H. Zhang, R. T. Mills}
}
\fi

\newsiamremark{remark}{Remark}
\newsiamremark{hypothesis}{Hypothesis}
\crefname{hypothesis}{Hypothesis}{Hypotheses}
\newsiamthm{claim}{Claim}
\headers{Portable and Versatile L-BFGS in PETSc/TAO}{H.\ Suh, T.\ Isaac, A.\ Dener, T.\ Munson, H.\ Zhang, R. T.\ Mills}

\begin{document}

\maketitle

\begin{abstract}
  The limited-memory BFGS (\LBFGS) Hessian update scheme 
  is the critical kernel in many quasi-Newton optimization algorithms.
  The most common approach to implementing \LBFGS uses $2m$ sequential rank-1 updates as part of solving a linear system when there are $m$ history steps.
  The performance of this approach suffers when the latency of synchronization is significant, and its poor temporal locality increases the memory traffic when vectors do not fit in cache.
  The compact dense representation of \LBFGS 
  results in an approach that has minimal synchronization latency and better temporal locality, but it requires an additional pass over the basis vectors and an additional basis that must be recomputed when the $B_0$ matrix changes
  as in variable-metric methods.
  In the Portable Extensible Toolkit for Scientific Computation and the Toolkit for Advanced Optimization (PETSc/TAO), we have implemented an intermediate dense formulation of BFGS that retains most of the good characteristics of both the recursive and compact dense approaches.
  We report single-node performance tests of these implementations on the U.S. Department of Energy's Polaris and Frontier machines, testing both GPU-based and CPU-based computations.
\end{abstract}



\section{Introduction}

The limited-memory BFGS (\LBFGS) Hessian update scheme \cite{Liu1989} is an
important tool in numerical optimization: it is the workhorse of quasi-Newton
unconstrained and bound-constrained smooth optimization. It is implemented in
PETSc/TAO (the Portable, Extensible Toolkit for Scientific Computation and the
Toolkit for Advanced Optimization \cite{petsc-web-page,petsc-user-ref}) as a
particular instance of a \emph{limited-memory variable metric} operator, which
can be used in quasi-Newton methods to solve nonlinear systems of equations and
unconstrained or bound-constrained optimization problems. The BFGS update to a
base Hessian $B_0$ or Hessian inverse $H_0 = B_0^{-1}$ can also appear in other
quasi-Newton optimization algorithms, such as quasi-Newton Krylov methods,
globalized by linesearch or trust-region methods.  All of these optimizers are
available in PETSc/TAO.


\section{Background}\label{sec:background}

\LBFGS is sometimes taken to mean a complete optimization algorithm for the unconstrained minimization of a smooth function $f(x)$.
%
In this work we take \LBFGS to mean only the approximation scheme $H_k \approx
(\nabla^2 f(x_k))^{-1}$ for the inverse Hessian based on a sequence $\{x_k\}$ of
solution vectors and the corresponding sequence $\{g_k\}$ of gradients. In this
scheme, $H_0 \in \mathbb{R}^{N\times N}$ is taken to be a base symmetric
positive definite (SPD) matrix (often a scaled identity but with
other common choices as well; see \cref{sec:bfgs-use-cases}). Given two
successive approximate solutions and their gradients, $(x_{k-1}, g_{k-1})$ and
$(x_k, g_k)$, the full BFGS update is defined by $H_0^{\text{full}} \defeq H_0$ and
\begin{equation}\label{eq:bfgs-full}
H_k^{\text{full}} \defeq (I - \Phi_{k-1})^T H_{k-1}^{\text{full}} (I - \Phi_{k-1}) + s_{k-1} d_{k-1}^{-1} s_{k-1}^T,
\end{equation}
where
\begin{align}
s_{k-1} &\defeq x_k - x_{k-1}, & y_{k-1} &\defeq g_k - g_{k-1}, \\
d_{k-1} &\defeq s_{k-1}^T y_{k-1}, & \Phi_{k-1} &\defeq y_{k-1} d_{k-1}^{-1} s_{k-1}^T.
\end{align}
For large problems it is not efficient to form $H_k^{\text{full}}$ explicitly, as this requires $O(N^2)$ work and storage.  It is more common to choose a fixed history size $m \ll N$ and implement an approximation $H_k^m$ where only the last $m$ pairs of $s$ and $y$ vectors are kept.  Formally, the \LBFGS approximate inverse Hessian $H_k^m$ is
defined by
\begin{equation}\label{eq:lbfgs-defn}
H_k^m \defeq
\begin{cases}
H_0, & m = 0 \text{ or } k=0, \\
(I - \Phi_{k-1})^T H_{k-1}^{m - 1} (I - \Phi_{k-1}) + s_{k-1} d_{k-1}^{-1} s_{k-1}^T & \text{otherwise}.
\end{cases}
\end{equation}
Letting $B_k^m \defeq (H_k^m)^{-1}$ be the corresponding approximate Hessian, one can  show that
\begin{equation}\label{eq:lbfgs-mult-defn}
B_k^m = B_0 + \sum_{i = \max\{0,k-m\}}^{k-1} y_i d_i^{-1} y_i^T - p_{m-(k-i),i} f_{m-(k-i),i}^{-1} p_{m-(k-i),i}^T,
\end{equation}
%
where
\begin{align}\label{eq:pfdef}
p_{m,k} &\defeq B_k^m s_k, & f_{m,k} &\defeq s_k^T B_k^m s_k.
\end{align}
In this work we compare the efficiency of different approaches of implementing \cref{eq:lbfgs-defn}.  We assume that an implementation {\tt B} of \LBFGS
must provide the following operations required of all quasi-Newton Hessian approximations:

\begin{enumerate}
    \item {\tt B.update}$(x_k, g_k)$:  Update {\tt B} from representing $B_{k-1}^m$/$H_{k-1}^m$ to $B_{k}^m$/$H_{k}^m$.
    \item {\tt B.solve}$(u, v)$:  Compute $v = H_k^m u$.
    \item {\tt B.mult}$(u, v)$: Compute $v = B_k^m u$.
\end{enumerate}

Different optimization algorithms use these core operations in different ways. An important one is unconstrained quasi-Newton optimization globalized by a line search.

\begin{algorithm2e}\label{alg:nls}%
\caption{Quasi-Newton optimization with line search}
\Until(\textbf{convergence}){%
    $g \gets \nabla f(x)$\;

    {\tt B.update}$(x, g)$\;

    {\tt B.solve}$(g, p)$\;

    $x \gets x - \alpha p$, where $\alpha$ is determined by line search on $f(x - \alpha p)$.
}%
\end{algorithm2e}

In \cref{alg:nls}, {\tt B.mult} is not used, but a more complex algorithm like a bound-constrained quasi-Newton method globalized by a trust region may require {\tt B.mult} and may use {\tt B.solve}$(v, p)$ with a generic right-hand side vector $v$ unrelated to the gradient. 


\section{\LBFGS use cases in PETSc/TAO} \label{sec:bfgs-use-cases}

PETSc/TAO is used to solve problems of vastly different sizes and supports MPI parallelism and GPU acceleration \cite{Mills2021}.
While the extensible software pattern of the library allows for an algorithm like \LBFGS to have multiple implementations for different architectures, the maintenance and tuning of a large number of implementations become a burden for the library developers, and users may not know how to choose the best implementation for their problem.
It is important, then, to provide a default implementation of \LBFGS that is portable (the same algorithm achieves close-to-optimal performance on different architectures) and versatile (different problems achieve close-to-optimal performance on the same architecture).  With that in mind, we describe what kinds of performance, portability, and versatility matter when we assess \LBFGS algorithms.


\paragraph{The time to update and solve is the most relevant performance metric}
An \LBFGS approximation can be used by any optimization algorithm that needs an approximate Hessian, but \cref{alg:nls} is the most common choice,
so when comparing the performance of different implementations of \LBFGS, we focus on the sum of the time of lines 3 and 4, which we refer to together as the \emph{quasi-Newton update}.

\paragraph{The location and movement of vectors are the most important determinants of performance}

All our algorithms are built from vector-vector and matrix-vector operations with low arithmetic intensity, so memory bandwidth and the latency of synchronization are critical.
In PETSc/TAO there are three important cases for where the $\mathbb{R}^N$ vectors reside, corresponding to different implementations of the \petscmanlink{Vec}{Vec} interface:

\begin{enumerate}
    \item
        \emph{In the host memory of a single process (\petscmanlink{Vec}{VECSEQ}).}
        No synchronization between memory domains is required, so here latency is negligible,
        and the primary concerns for \LBFGS are the sizes and bandwidths of the cache hierarchy and how well algorithms exploit locality.
    
    \item
        \emph{In the device memory of a single GPU (\petscmanlink{Vec}{VECSEQCUDA}, \petscmanlink{Vec}{VECSEQHIP}).} 
        In addition to bandwidth and cache concerns, reduction operations across multiple threadblocks within GPU memory incur a synchronization latency that elementwise vector updates do not: this manifests as lower bandwidths for inner products at small vector sizes in streaming benchmarks such as \cite[Figure 1]{Anzt2020} measuring performance on the NVIDIA A100.
         
        Potentially more significant, however, are latencies that are caused by the programming model.
        PETSc's function \petscmanlinkfn{Vec}{VecDot} writes the value of an inner product to a scalar on the host: this introduces latency from the transfer of the value from device to host memory and from the pipeline stall caused by the host thread blocking before it launches subsequent kernels to the stream of computation.
        This source of latency can be avoided in other programming models that return a reference to the value in device memory, such as pytorch's {\tt \href{https://pytorch.org/docs/stable/generated/torch.dot.html}{tensor.dot()}}.
        At the time of writing, a similar {\tt ManagedMemory} interface is in development for PETSc/TAO \cite{Faibussowitsch2023}, but that feature is not yet represented in the performance results in this work.
    \item
        \emph{Distributed across multiple memory domains, whether host or device memory (\petscmanlink{Vec}{VECMPI}, \petscmanlink{Vec}{VECMPICUDA}, \petscmanlink{Vec}{VECMPIHIP}).}
        Here the interconnect between memory domains is a significant source of latency in reductions and broadcasts regardless of the programming model.
\end{enumerate}

\paragraph{The base inverse Hessian $H_0$ may be cheap or expensive, constant or variable}
While it is common to choose $H_0 = I$ in the absence of better information, in physics-based simulations $H_0$ may be a relatively expensive preconditioner, such as a multigrid preconditioner for elasticity \cite{Brown2013,Brown2022}.
In such cases the application of $H_0$ takes the majority of the time in {\tt B.solve()}, and extraneous applications of $H_0$ should be avoided.

Although the analysis of quasi-Newton methods is often simplified by the base matrix $H_0$ being constant, in practice \emph{variable metric} versions of quasi-Newton methods are used, where $H_0$ is actually $H_{0,k}$, which is adapted from $H_{0,k-1}$ based on new data.
The main implication for \LBFGS implementations is that products with $H_{0,k}$ cannot be reused after updates and must be recomputed.
If the variability is only scalar variability (if, for instance, $H_{0,k} = (s_{k-1}^T y_{k-1})(y_{k-1}^T y_{k-1})^{-1} I$ as suggested in \cite{NocedalWright2006}), then such quantities can be reused with the appropriate rescaling, but reuse is not possible for more general variable metric methods.
Work comparing optimization algorithms from PETSc/TAO has demonstrated the superiority of diagonal variable metric methods over constant and scalar variable metric methods for a wide range of problems \cite{Dener2019}.

\section{Storage of vectors}
\label{sec:storage-vectors}

As described in \cref{sec:background}, \LBFGS uses sliding intervals of past vectors. To describe the intervals used in these algorithms, we define
\begin{align}
 m_k &\defeq \min\{m,k\}, \\
 \iota(m,k) &\defeq \{k-m_k, \dots, k-1\}, \\
 \omega(m,k) &\defeq \{0, \dots, m_k - 1\},
\end{align}
and we use subscripts to indicate index sets, as in $x_{\iota(m,k)} \defeq \{x_i\}_{i\in\iota(m,k)}$.  We will continue to use $N$ as the size of the vectors.  When vectors are spread across host processes using MPI, we use $n$ for length of the largest subvector assigned to any one process ($n$ is the more useful value for describing the time complexity of the algorithms we will present).

An implementation of $H_k^m$ must store the latest vectors $s_{\iota(m,k)}$ and $y_{\iota(m,k)}$, and some algorithms use other auxiliary vectors with matching numbering.
For these vectors it is common to use arrays of vectors kept in \emph{history order}: for example, an array $\tilde{s}$ of $m$ vectors with $s_{\iota(m,k)}$ stored in $\tilde{s}_{\omega(m,k)}$.
When updating from $H_{k-1}^m$ to $H_k^m$ it is important to maintain history order by relabeling the vectors and not by copying so that the time complexity is $O(n)$, as shown in \cref{alg:lbfgs-history-order-update}, and not $O(mn)$.  (In \cref{alg:lbfgs-history-order-update} and subsequent algorithms, we annotate nontrivial operations with the corresponding BLAS-like routine and input sizes.)
\begin{algorithm2e}[h]\label{alg:lbfgs-history-order-update}
\caption{Store vector $s_k$ in $\tilde{s}$, history-order storage with history size $m$}
\If{$k \geq m$}{%
$\tilde{s}^{\text{temp}} = \tilde{s}_0$ \comment{reference, no copy}

\lFor*{$j \in \{0, \dots, m-2\}$}{%
$\tilde{s}_j = \tilde{s}_{j+1}$%
}
\comment{"}

$\tilde{s}_{m-1} = \tilde{s}^{\text{temp}}$ \comment{"}
}%

$\tilde{s}_{\min\{m-1,k\}} \gets s_k$\complexity{copy}{n}
\end{algorithm2e}

In \cref{sec:algorithms} we will present algorithms that format \LBFGS
operations in terms of Level 2 BLAS and Level 3 BLAS instead of Level 1 BLAS
(or equivalent cuBLAS or hipBLAS routines).
Arrays of vectors are incompatible with these dense routines, so we must use matrices of column vectors for the history
vectors: for example, $S = [S_0|\cdots|S_{m-1}]\in \mathbb{R}^{N \times m}$ to
store the $s$ vectors.  It would be convenient if the column vectors of $S$ were
in history order ($s_{\iota(m,k)}$ stored in $S_{\omega(m,k)}$), but maintaining this invariant would have $O(mn)$ time complexity.
Instead, we store all history vectors in cyclic order, with $s_k$ copied into $S_{k \bmod m}$, so that updating $S$ has the same complexity as \cref{alg:lbfgs-history-order-update}.

An alternative to cyclic ordering would be to have extra columns in $S$.  If
$S$ had $m + r$ columns, then $s_{\iota(m,k)}$ could be stored in a sliding
interval of columns, where the active columns would have to be copied back to
the start of the matrix every $r + 1$ steps.  This would give updating $S$ an amortized
complexity of $O((1 + \tfrac{m-1}{r+1})n)$ per step.  The drawback of
this approach is the extra spatial complexity, which is why we did not implement
it.

In PETSc/TAO, the column-vector matrices are instances of dense implementations of the \petscmanlink{Mat}{Mat} interface%
\footnote{%
\petscmanlink{Mat}{MATSEQDENSE},
\petscmanlink{Mat}{MATSEQDENSECUDA},
\petscmanlink{Mat}{MATSEQDENSEHIP},
\petscmanlink{Mat}{MATMPIDENSE},
\petscmanlink{Mat}{MATMPIDENSECUDA},
\petscmanlink{Mat}{MATMPIDENSEHIP}.
}
that use the same type of memory as the vectors; and when there are multiple MPI ranks, they are row-partitioned to match the vectors.

\section{Storage of inner-product matrices}
\label{sec:storage-inner-products}

In addition to the history vectors, the algorithms in \cref{sec:algorithms} use the inner products between different sets of history vectors.  Several, for example, use the values $r_{i,j} \defeq s_i^T y_j$ for $i,j\in \iota(m,k)$.  As with the basis vectors, we must store these values in a dense matrix $R \in \mathbb{R}^{m \times m}$ to take advantage of the appropriate linear algebra routines.  There are several design choices in how $R$ is stored that impact performance.

\paragraph{Host vs.\ device memory} $R$ resides in the same type of memory as the
$\mathbb{R}^N$ vectors and column-vector matrices like $S$ described in
\cref{sec:storage-vectors}.  This means that products such as $V \gets S R$ can
be computed without transferring $R$ between the host and the device.  It
also means that computing a matrix of inner products such as $R \gets S^T Y$ does
not synchronize the host and the device. 

\paragraph{Multiple MPI ranks: duplicate vs.\ single storage} In PETSc/TAO,
\petscmanlinkfn{Vec}{VecDot} computes a dot product using an {\tt allreduce} operation
so that the result is duplicated on every MPI rank, but
\petscmanlinkfn{Mat}{MatTransposeMatMult} computes a matrix of inner products
using a {\tt reduce} operation and stores each inner product in just one
location.\footnote{In all cases, GPU-aware MPI is used whenever possible
\cite{Zhang2022}.}  \LBFGS algorithms could be implemented with matrices like $R$ being duplicated on each MPI rank, or with $R$ being stored on a single rank: the choice affects the communication complexity of products with column-vector matrices.
\begin{itemize}
    \item If $R$ is duplicated on each process, $R \gets S^T Y$ has the communication complexity of an {\tt allreduce} of size $m^2$, and $V \gets SR$ is communication-free.
    \item If $R$ is stored once, $R \gets S^T Y$ has the communication complexity of a {\tt reduce} of size $m^2$ and $V \gets SR$ the complexity of a {\tt bcast} of size $m^2$.
\end{itemize}
In PETSc/TAO, we choose to store $R$ once on the first MPI rank because this
allows us to use more of the existing \petscmanlink{Mat}{Mat} routines. We
recognize, however, that in a self-consistent MPI implementation 
$T({\tt reduce}) + T({\tt bcast}) \geq T({\tt allreduce})$ \cite{LarssonTraff2010}, and for small
message sizes many MPI implementations will have $T({\tt reduce}) + T({\tt bcast}) \approx 2 T({\tt allreduce})$.

\begin{figure}
\centering
\input{figure/ordering.tex}
\caption{%
Cyclic ordering.  Left: cyclic ordering of the history vectors $s_{\iota(m,k)}$ in matrix $S$.  Right: the history-order upper-triangular
storage of $r_{i,j} = s_i^T y_j$ inner-products stored in the $R$ matrix, annotated with the steps used to apply $\htriu_k(R)^{-1}$ in 
\cref{alg:trsv}.%
}\label{fig:cyclic-ordering}
\end{figure}
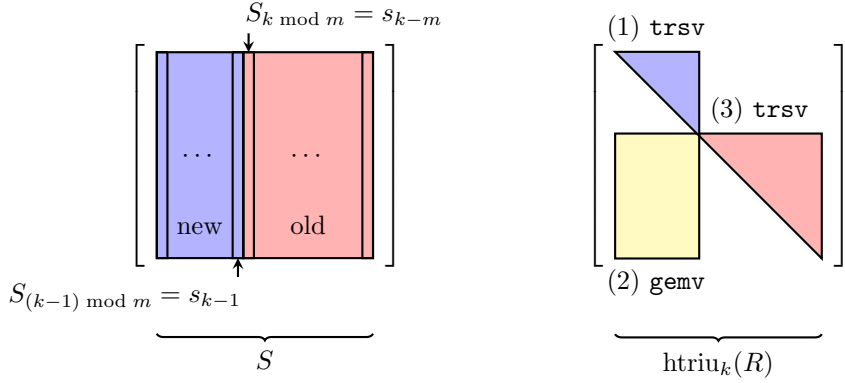

\paragraph{Cyclic ordering} Like matrices of column vectors,  matrices
of inner products in cyclic order ($r_{i,j}$ are stored in $R_{i\bmod m, j\bmod
m}$ for $i,j\in\iota(m,k)$).   Because of the matching indexing, a product such as 
$v_i \gets \sum_{j\in \iota(m,k)} r_{i,j} s_j$ can be accomplished in a single
{\tt gemm} operation.  The same is not true of products that are triangular with
respect to history order, like
$v_i \gets \sum_{j\in\iota(m,k),\ j\geq i}r_{i,j} s_j.$  We introduce
$\htriu_k(R)\in\mathbb{R}^{m \times
m}$ to refer to the entries in $R$ that are upper-triangular with
respect to history order (see also \cref{fig:cyclic-ordering}): given $i,j\in\iota(m,k)$,
\begin{equation}
\htriu_k(R)_{i\bmod m,j\bmod m} \defeq \begin{cases}
R_{i\bmod m,j\bmod m}, & i \geq j, \\
0 & \text{otherwise}.
\end{cases}
\end{equation}
Matrix solves with $\htriu_k(R)$ are accomplished by using three BLAS-like routines, as demonstrated
in {\tt trsv\_cyclic} (\cref{alg:trsv}).

\begin{algorithm2e}\label{alg:trsv}
    \caption{{\tt trsv\_cyclic}}
    \KwData{$R\in\mathbb{R}^{m\times m}, u\in\mathbb{R}^m$}
    \KwResult{$u \gets \htriu_k(R)^{-1} u$}

    $i \gets k\bmod m$\;

    $\iota_{\text{new}} \gets \{0, \dots, i - 1\}$\;
    $\iota_{\text{old}} \gets \{i, \dots, m - 1\}$\;
    
    $u_{\iota_{\text{new}}} \gets \triu(R_{\iota_{\text{new}},\iota_{\text{new}}})^{-1} u_{\iota_{\text{new}}}$\complexity{trsv}{\hat k}
    \If{$k \geq m$}{%
        $u_{\iota_{\text{old}}} \gets u_{\iota_{\text{old}}} - R_{\iota_{\text{old}},\iota_{\text{new}}} u_{\iota_{\text{new}}}$\complexity{gemv}{m - \hat k, \hat k}
        $u_{\iota_{\text{old}}} \gets \triu(R_{\iota_{\text{old}},\iota_{\text{old}}})^{-1} u_{\iota_{\text{old}}}$\complexity{trsv}{m - \hat k}
    }
\end{algorithm2e}

\Cref{alg:trsv} requires the same work as a single {\tt trsv} or GPU-based equivalent, but when the data resides on a GPU \cref{alg:trsv} is implemented using three sequential kernel launches instead of one. We call this an \emph{inplace} strategy.
Because triangular operations appear frequently in \LBFGS algorithms, we also considered storing $R$ and related matrices in history order, which we call the \emph{reorder} strategy, to make triangular operations as fast as possible.
In our performance measurements, this was never beneficial: the speedup of triangular operations was outweighed by the cost of maintaining history order.
Consequently, all the data presented in this paper were collected with the \emph{inplace} strategy.


\section{\LBFGS algorithms}
\label{sec:algorithms}

The dense storage of column-vector and inner-product matrices described in
\cref{sec:storage-vectors,sec:storage-inner-products} can be used by multiple
different algorithms to implement \LBFGS.  We start by describing the common features of all of them.

\begin{table}[]
    \caption{Common data structures across all \LBFGS implementations}
    \begin{tabular}{>{\centering}p{0.07\textwidth} p{0.86\textwidth}} \toprule
        $m$ & history size, constant \\
        $k$ & number of completed updates, initially $0$ \\
        $B_0$ & SPD $\mathbb{R}^{N\times N}$ matrix \\
        $x^{\text{prev}}$ & $\mathbb{R}^N$ vector, the latest solution value, initially $x_0$ \\
        $g^{\text{prev}}$ & $\mathbb{R}^N$ vector, the latest gradient, initially $g_0$ \\
        $S$ & $\mathbb{R}^{N \times m}$ cyclic-order column-vector matrix $[S_0 | \cdots | S_{m-1}]$ that stores $s_{\iota(m,k)}$ \\
        $Y$ & $\mathbb{R}^{N \times m}$ cyclic-order column-vector matrix $[Y_0 | \cdots | Y_{m-1}]$ that stores $y_{\iota(m,k)}$ \\
        $d$ & $\mathbb{R}^m$ cyclic-order vector of inner products, $d_i = S_i^T Y_i$ \\
        \bottomrule
    \end{tabular}
    \label{tab:common-data}
\end{table}

\Cref{tab:common-data} shows that all implementations
store the base Hessian $B_0$ (which can be any SPD PETSc
\petscmanlink{Mat}{Mat}, including a \petscmanlink{Mat}{MATSHELL} that wraps
user-supplied code). If $B_0 = \gamma I$ or $B_0$ is diagonal, then $H_0$ will be a direct inverse; otherwise $H_0$ is the application of a preconditioner (\petscmanlinkfn{PC}{PCApply}) or an iterative method (\petscmanlinkfn{KSP}{KSPSolve}).  Each implementation stores the last $x$ and $g$
vectors used to update {\tt B} and the $s_{\iota(m,k)}$ and
$y_{\iota(m,k)}$ vectors.  \Cref{alg:update} shows that all
implementations use a curvature condition $s_k^T y_k \geq \epsilon_{\text{tol}} \|y_k\|^2$ to ensure a valid update, so the $s_k^T y_k$ values are stored in a vector $d$ for reuse.
\Cref{alg:update} also shows how we use $\dots = {\tt B.}(\dots)$ to indicate which variables in the algorithm are references to data structures of the implementation and how we annotate steps with parallel communication with MPI-like routines and sizes.

\begin{algorithm2e}\label{alg:update}%
\caption{{\tt B.update}$(x, g)$}

  $m, k, B_0, x^{\text{prev}}, g^{\text{prev}}, S, Y, d = {\tt B.}(\dots)$\;

  \lIf{$B_0$ is variable metric}{$B_0{\tt.update}(x,g)$}

  $s \gets x - x^{\text{prev}}$ \complexity{axpy}{n}
  
  $y \gets g - g^{\text{prev}}$ \complexity{axpy}{n}

  $\begin{bmatrix} \delta \\ \eta \end{bmatrix} \gets \begin{bmatrix} s^T \\ y^T \end{bmatrix} y$ \complexitytwo{$2$\,dot}{n}{allreduce}{2}

  \If{$\delta > \epsilon_{\text{tol}} \eta$}{%
    
    $i \gets k \bmod m$\;

    $S_i \gets s$ \complexity{copy}{n}
    
    $Y_i \gets y$ \complexity{copy}{n}
    
    $d_i \gets \delta $\;
    
    $k \gets k + 1$\;
    
    ${\tt B.update\_solve}(x, g)$
    
    \lIf{{\tt B.mult} is used}{${\tt B.update\_mult}(x, g)$}

  }%
  $x^{\text{prev}} \gets x$ \complexity{copy}{n}
  
  $g^{\text{prev}} \gets g$ \complexity{copy}{n}
\end{algorithm2e}

We note that in this scheme the data structures used by {\tt B.mult} are not eagerly updated to enhance the performance of algorithms like \cref{alg:nls} that do not use it.  The cost of {\tt B.update}, exclusive of {\tt B.update\_solve} and {\tt B.update\_mult}, is
\begin{multline}\label{eq:B-update}
T({\tt B.update}) = 2T({\tt axpy}(n)) + 2T({\tt dot}(n)) + 4T({\tt copy}(n)) + T({\tt allreduce}(2)) \\ + O(1).
\end{multline}

\subsection{The recursive approach}\label{sec:algorithms-recursive}

We first describe the best performance that can be achieved with recursive algorithms that use only Level 1 BLAS.

\subsubsection{Recursive {\tt B.solve}}

An approach to {\tt B.solve} that directly translates \cref{eq:lbfgs-defn} into a recursive algorithm has two main advantages: no additional data structures are needed, so {\tt B.update\_solve} is a no-op, and  the algorithm does not change if $B_0$ varies from iteration to iteration.

%

\begin{algorithm2e}\label{alg:rbfgs-solve}%
\caption{{\tt Recursive::B.solve$(v, z)$}}

$m, k, B_0, S, Y, d = {\tt B}.(\dots)$\;

$a \in \mathbb{R}^{m_k}$ \comment{temporary work vector}

$z \gets v$\complexity{copy}{n}

\ForRev{$j \in \iota(m,k)$}{\label{line:rbfgsloop1start}
    $i \gets j \bmod m$\;
    
    $a_i \gets (S_i^T z) / d_i$ \complexitytwo{dot}{n}{allreduce}{1}
    
    $z \gets z - a_i y_i$ \complexity{axpy}{n}
}

$z \gets B_0^{-1} z$ \complexity{$B_0$.solve}{}

\For{$j \in \iota(m,k)$}{\label{line:rbfgsloop2start}
    $i \gets j \bmod m$\;
    
    $\beta \gets (Y_i^T z) / d_i$ \complexitytwo{dot}{n}{allreduce}{1}
    
    $z \gets z + (a_i - \beta) S_i$ \complexity{axpy}{n}
}
\end{algorithm2e}

\Cref{alg:rbfgs-solve} shows that the main disadvantage of this approach is the for-loops that cannot be unrolled because of  data dependencies, which results in poor temporal locality of memory access and reduction operations that cannot be coalesced.  The cost of {\tt Recursive::B.solve} is
\begin{multline}\label{eq:rbfgs-solve-complexity}
T({\tt Recursive::B.solve}) =
T(B_0{\tt .solve}) 
\\+ T({\tt copy}(n))
+ 2m_k\{T({\tt axpy}(n)) + T({\tt dot}(n)) + T({\tt allreduce(1)})\}
+ O(m_k).
\end{multline}

\subsubsection{Recursive {\tt B.mult}}
\Cref{eq:lbfgs-mult-defn} shows that $B_k^m$ can be expressed non-recursively as
a rank-$2m_k$ update of $B_0$: this form is good for performance because
the updates can be performed in parallel.  In particular, this means that $2m_k$
separate reductions can be combined into a single larger reduction, as described
in \cref{alg:rbfgs-mult}, where the column-vector matrix $P$ stores the $p$
vectors \cref{eq:pfdef} in cyclic order.\footnote{\Cref{alg:rbfgs-mult} shows that we use {\tt diagm}
and {\tt diags} to annotate the cost for elementwise multiplication and
division, respectively, even though these are not BLAS routines.}

\begin{algorithm2e}\label{alg:rbfgs-mult}%
\caption{{\tt Recursive::B.mult($v, z$)}}
$m, k, B_0, S, Y, d, P, f = {\tt B}.(\dots)$\;

$a, b \in \mathbb{R}^{m_k}$ \comment{temporary work vectors}

$\omega \gets \omega(m, k)$\comment{active history indices}

$z \gets B_0 v$ \complexity{$B_0$.mult}{}

$\begin{bmatrix} a \\ b \end{bmatrix}\gets \begin{bmatrix} Y_{\omega}^T \\ P_{\omega}^T\end{bmatrix} v$
\complexitytwo{$2 m_k$\,dot}{n}{allreduce}{2 m_k}



$\begin{bmatrix} a \\ b \end{bmatrix}\gets \begin{bmatrix} \mathrm{diag}(d_{\omega}) & 0 \\ 0 & \mathrm{diag}(f_{\omega})\end{bmatrix}^{-1} \begin{bmatrix} a \\ b \end{bmatrix}$
\complexity{$2$\,diags}{m_k}

$z \gets z +  \begin{bmatrix} Y_{\omega} & -P_{\omega}\end{bmatrix}
        \begin{bmatrix} a \\ b \end{bmatrix}$
\complexity{$2 m_k$\,axpy}{n}

\end{algorithm2e}

The cost of this algorithm is 
\begin{multline}
T({\tt Recursive::B.mult}) =
T(B_0{\tt .mult}) 
\\+ 2m_k\{T({\tt axpy}(n)) + T({\tt dot}(n))\} + T({\tt allreduce}(2m_k))
+ O(m_k).
\end{multline}
The main drawback is that the vectors $\{p_{m-(k-i),i}\}_{i\in\iota(m,k)}$ are defined recursively and must be computed in {\tt B.update\_mult}.  These vectors also have limited reuse: after updating from $B_{k-1}^m$ to $B_k^m$, the $p$ vectors cannot be reused if $k > m$ (because dropping the oldest $p$ vector from the recursion changes all subsequent vectors), and they cannot be reused if $H_0$ has changed.
To handle these two cases that invalidate the $p$ vectors separately, a recursive \LBFGS implementation can store (\cref{tab:rbfgs-mult-data}) the $p$ vectors as well as the auxiliary vectors
\begin{equation}
\po_{\iota(m,k)} \defeq \{B_0 s_i\}_{i\in\iota(m,k)},
\end{equation}
which need to be updated only if $B_0$ has changed, and which do not need to be computed at all if $B_0$ is a scaled identity matrix.

%
%

\begin{table}[]
    \centering
    \caption{Data structures for {\tt B.mult} for the recursive approach}
    \begin{tabular}{>{\centering}p{0.07\textwidth} p{0.86\textwidth}} \toprule
        $\PO$ & $\mathbb{R}^{N \times m}$ matrix to store $B_0 S$ \\
        $P$ & $\mathbb{R}^{N \times m}$ cyclic-order column-vector matrix $[P_0 | \cdots | P_{m-1}]$ that stores $\{p_{m-(k-i),i}\}_{i\in\iota(m,k)}$ \cref{eq:pfdef} \\
        $f$ & $\mathbb{R}^{m}$ cyclic-order vector of inner products, $f_i = P_i^T S_i$ \\
        \bottomrule
    \end{tabular}
    \label{tab:rbfgs-mult-data}
\end{table}

We give {\tt Recursive::update\_mult} as \cref{alg:rbfgs-update-mult} in the appendix.  We can summarize its performance when $k \geq m$ as
\begin{multline}\label{eq:brfgs-update-mult-complexity}
T({\tt Recursive::B.update\_mult}) =
m_{\text{update}}T(B_0{\tt .mult}) 
\\+ mT({\tt axpy}(n)) + m^2T({\tt dot}(n)) + (m^2-m)T({\tt axpy}(n)) + (2m-1)T({\tt allreduce}(O(m)))
\\+ O(m^2),
\end{multline}
where
\begin{equation}\label{eq:m-update}
m_{\text{update}} = \begin{cases}
0, & B_0 = B_{0,k} = \gamma_k I, \\
1, & B_0 \text{ is constant}, \\
m, & \text{otherwise}.
\end{cases}
\end{equation}
Regardless of the value of $m_{\text{update}}$, {\tt Recursive::B.update\_mult} has $O(m^2 n)$ time complexity just
in its BLAS routines, as opposed to $O(mn)$ for {\tt
Recursive::B.solve}. Thus  it is important to
avoid eagerly calling this routine in {\tt B.update}.

%
%
%
%
%
%
%
%

\subsection{The compact dense approach}\label{sec:algorithms-compact-dense}

Byrd, Nocedal, and Schnabel \cite{Byrd1994}  presented a compact dense representation of $H_k^m$ and developed an approach to using it in \cref{alg:nls} in a way that is work-equivalent to using the recursive approach from the preceding section.  Here we present this approach, summarizing their analysis for the case $B_0 = \gamma I$ and analyzing the case when $B_0$ is a general variable-metric matrix.

We define a new set of vectors $q_\iota(m,k) \defeq \{H_0 y_i\}_{i\in\iota(m,k)}$ and let $Q\in\mathbb{R}^{N\times m}$ be a cyclic-order column-vector matrix for storing them. We define the inner-product matrices as
\begin{align}
    R &\defeq \htriu_k(S^T Y), & D &\defeq \diag(S^TY), & Z &\defeq Q^T Y = Y^T H_0 Y.
\end{align}
With these matrices, the compact dense representation of $H_k^m$ is
\begin{equation}\label{eq:compact-dense}
H_k^m = H_0 + \begin{bmatrix}S & Q\end{bmatrix}
\begin{bmatrix}
    R^{-T}(Z + D)R^{-1} & -R^{-T} \\ -R^{-1} & 0
\end{bmatrix}
\begin{bmatrix}S^T \\ Q^T\end{bmatrix}.
\end{equation}

\subsubsection{Compact dense {\tt B.solve}}

The inner matrix in the product in \cref{eq:compact-dense} 
can be applied to a vector in a handful of BLAS operations (including {\tt
trsv\_cyclic} (\cref{alg:trsv}) and its transpose) using $O(m^2)$ work.  The
majority of the work in ${\tt B.solve}$ is in the products with $S$, $Q$,
and their transposes,
\begin{multline}
T({\tt CompactDense::B.solve}) = T(B_0{\tt .solve}())\\
+ 4T({\tt gemv}(n, m_k)) + T({\tt reduce}(2 m_k)) + T({\tt bcast}(2 m_k)) + O(m_k^2),
\end{multline}
where reductions and broadcasts can be coalesced because they occur in parallel.

In ${\tt B.update\_solve}$ the $Q$, $R$, and $Z$ matrices must be updated. Like
the $\PO$ matrix in the recursive approach, updating $Q$ can be avoided if $H_0
= \gamma I$.  $R$ can be updated with a single ${\tt gemv}(n, m_k-1)$ between
$y_k$ and the older columns of $S$ (the diagonal value $s_k^T y_k$ was computed
in ${\tt B.update}$).  Because $Z$ is symmetric, it can also be updated
with a single ${\tt gemv}(n, m_k)$ between $y_k$ and the active columns of $Q$.

If we sum the BLAS work just described in updating and solving using the compact dense approach (using $\sim\!2mn$
flops for ${\tt gemv}(n,m)$ and $\sim\!2n$ flops for ${\tt dot}(n)$), we get
$\sim\!(12m_k -2)n$ flops, more than the $\sim\!8 m_k n$ flops for {\tt B.solve} in the recursive approach.
Byrd et al.\ solved this problem by recognizing that in the quasi-Newton update in \cref{alg:nls} the same gradient vector
$g_k$ is an argument to both {\tt B.update} and {\tt B.solve}.  If the $S^T g_k$ and $Q^T g_k$ products are cached during {\tt B.update}$(x_k, g_k)$ (\cref{tab:cdbfgs-solve-data}), then they do not need to be recomputed in {\tt B.solve}$(g_k, p)$, as shown in \cref{alg:cdbfgs-solve}.

\begin{table}[]
    \caption{Data structures for {\tt B.solve} for the compact dense approach}
    \begin{tabular}{>{\centering}p{0.07\textwidth} p{0.86\textwidth}} \toprule
      $R$ & $\mathbb{R}^{m\times m}$ matrix to store $\htriu_k(S^T Y)$ \\
      $Q$ & $\mathbb{R}^{N\times m}$ matrix to store $H_0 Y$ \\
      $Z$ & $\mathbb{R}^{m\times m}$ matrix to store $Y^T H_0 Y$ \\
      $a^{\text{cache}}$ & $\mathbb{R}^{m}$ vector to cache the product $S^T g^{\text{prev}}$ \\
      $b^{\text{cache}}$ & $\mathbb{R}^{m}$ vector to cache the product $Q^T g^{\text{prev}}$ \\
      \bottomrule
    \end{tabular}
    \label{tab:cdbfgs-solve-data}
\end{table}

\begin{algorithm2e}[h]\label{alg:cdbfgs-solve}%
\caption{{\tt CompactDense::B.solve($v, z$)}}

    $m, k, B_0, S, Y, d, Q, R, Z, g^{\text{prev}}, a^{\text{cache}}, b^{\text{cache}} = {\tt B}.(\dots)$\;
    
    $a, b, c, e \in \mathbb{R}^{m_k}$\comment{temporary work vectors}
    
    $\omega \gets \omega(m, k)$\comment{active history indices}

    $(\alpha, \tilde Q) =$ \lIf{$B_0 = \beta I$}{$(\beta^{-1}, Y)$ \textbf{else} $(1, Q)$\hfill{\tt //} $\alpha \tilde Q = Q$}
    
    $z \gets B_0^{-1} v$ \complexity{$B_0$.solve}{}

    \If{$v \neq g^{\text{prev}}$}{%
      
      $\begin{bmatrix}a \\ b\end{bmatrix} \gets \begin{bmatrix}S_{\omega}^T \\ \alpha \tilde Q_{\omega}^T\end{bmatrix} v$ \complexitytwo{$2$\,gemv}{n, m_k}{reduce}{2 m_k}

    }%
    \lElse{
        $(a, b) \gets (a^{\text{cache}}_{\omega}, b^{\text{cache}}_\omega)$\hfill{\tt // $2$\,copy$(m_k)$}
    }

    $a \gets -\htriu_k(R_{\omega,\omega})^{-1} a$
    \complexitytwo{trsv\_cyclic}{m_k}{scal}{m_k}

    $\begin{bmatrix}c \\ e \end{bmatrix} \gets 
    \begin{bmatrix}
      Z_{\omega,\omega} + \mathrm{diag}(d_{\omega}) & I \\ 
      I & 0
    \end{bmatrix}
    \begin{bmatrix} a \\  b\end{bmatrix}$ \complexitythree{gemv}{m_k,m_k}{diagm}{m_k}{axpy}{m_k}

    $c \gets -\htriu_k(R_{\omega,\omega})^{-T} c$ \complexitytwo{trsv\_cyclic}{m_k}{scal}{m_k}
    
    $z \gets z + \begin{bmatrix}S_{\omega} & \alpha \tilde Q_{\omega}\end{bmatrix}  \begin{bmatrix}c \\ e \end{bmatrix}$ \complexitytwo{$2$\,gemv}{n, m_k}{bcast}{2 m_k}
\end{algorithm2e}

When we are given $g_{k+1}$ in the next update, $S^T y_k = S^T g_{k+1} - S^T g_k$ can be computed when $S^T g_{k+1}$ is computed with only $O(m)$ additional work.  The details of how this is accomplished are in \cref{alg:cdbfgs-update-solve} in the appendix. Here we summarize the performance of {\tt B.update\_solve} in isolation when $k \geq m$ as
\begin{multline}
T({\tt CompactDense::update\_solve}) = m_{\text{update}} T(B_0{\tt .solve}) \\
+ T({\tt gemm}(m_{\text{update}},m_{\text{update}}, n))
+ 2T({\tt gemv}(n, m))
+ T({\tt dot}(n)) \\
+ T({\tt reduce}(m_{\text{update}}^2))
+ 2T({\tt reduce}(m))
+ T({\tt reduce}(1))
+ O(m^2),
\end{multline}
where $m_{\text{update}}$ is the same as \cref{eq:m-update}.
When the $S^T g_k$ and $Q^T g_k$ products are cached, the complexity of the quasi-Newton update using \cref{alg:cdbfgs-update-solve,alg:cdbfgs-solve} becomes, for $k \geq m$,
\begin{multline}\label{eq:cdbfgs-update-and-solve-complexity}
T({\tt CompactDense::update\_solve}(x_k,g_k) + {\tt CompactDense::solve}(g_k,p)) = \\
(1 + m_{\text{update}}) T(B_0{\tt. solve}) \\
+ T({\tt gemm}(m_{\text{update}},m_{\text{update}}, n))
+ 4T({\tt gemv}(n, m)
+ T({\tt dot}(n)) \\
+ T({\tt reduce}(m_{\text{update}}^2))
+ T({\tt reduce}(2m))
+ T({\tt bcast}(2 m))
+ T({\tt reduce}(1))
+ O(m^2).
\end{multline}
When we compare this to $T({\tt Recursive::B.solve})$ \cref{eq:rbfgs-solve-complexity}, we see the majority of the BLAS operations are now in Levels 2 and 3 BLAS, and we also see that the number of parallel reductions and broadcasts has been reduced from $O(m)$ to $O(1)$, which should translate to superior machine performance.  Whether the compact dense approach is superior, however, depends on $B_0$ and $m_{\text{update}}$.

If $B_0 = \gamma I$, then $m_{\text{update}} = 0$, and the BLAS work in \cref{eq:cdbfgs-update-and-solve-complexity} is $\sim\!(8m + 2)n$ flops.  Byrd et al.\ further optimized this algorithm by reusing the cached inner products to check the curvature condition in \cref{alg:update} and the stopping conditions for the algorithm, so that
the BLAS work in the compact dense approach can be further reduced to the optimal $\sim 8mn$ flops in this case.

If $B_0$ is constant but not diagonal, then $m_{\text{update}} = 1$, and the BLAS work is $\sim\!(8m + 4)n$, which is still nearly optimal.  
However, the number of applications of $B_0${\tt .solve} is now 2 instead of 1, so the compact dense approach will be worse if $B_0${\tt .solve} is a relatively expensive operation.

If $B_0$ is variable-metric, then $m_{\text{update}} = m$, and the $Q$ and $Z$ matrices must be recomputed at each iteration, leading to
$m$ applications of $B_0${\tt .solve} and $\sim (2 m^2 + 8m + 2)n$ flops of BLAS work.  For large $m$ the compact dense approach could require so much extra work that its superior machine performance does not make it faster than the recursive approach.

\subsubsection{Compact dense {\tt B.mult}}  The Sherman--Morrison--Woodbury formula can be used on \cref{eq:compact-dense}
to compute a compact dense representation of $B_k^m$.  Using $\PO = B_0 S$ as previously defined and letting $\FO \defeq \PO^T S = S^T B_0 S$,
we have
\begin{equation}\label{eq:compact-dense-B0}
    B_k^m = B_0 -
    \begin{bmatrix} Y & \PO \end{bmatrix}
    \begin{bmatrix} -D & L^T \\ L & \FO \end{bmatrix}^{-1}
    \begin{bmatrix} Y^T \\ \PO^T \end{bmatrix},
\end{equation}
where $L$ is the history-order strictly lower-triangular portion of $S^T Y$,
\[
L = \hstril_k(S^T Y) \defeq S^T Y - \htriu_k(S^T Y).
\]
The Cholesky factorization $JJ^T \defeq \FO + L D L^T$ is guaranteed to exist \cite[Theorem 2.4]{Byrd1994}, which allows us to write
\cref{eq:compact-dense-B0} as
\begin{equation}\label{eq:compact-dense-B0-factored}
    B_k^m = B_0 -
    \begin{bmatrix} Y & \PO \end{bmatrix}
    \begin{bmatrix} I & D^{-1}L^T \\ 0 & I \end{bmatrix}
    \begin{bmatrix} I & 0 \\ 0 & J^{-T}J^{-1}\end{bmatrix}
    \begin{bmatrix} I & 0 \\ -L & I \end{bmatrix}
    \begin{bmatrix} D^{-1} & 0 \\ 0 & I \end{bmatrix}
    \begin{bmatrix} -Y^T \\ \PO^T \end{bmatrix}.
\end{equation}
If these matrices are stored (\cref{tab:cdbfgs-mult-data}) and updated in {\tt B.update\_mult}, then \cref{eq:compact-dense-B0-factored} shows how to efficiently implement this representation of $B_k^m$ in {\tt B.mult} (\cref{alg:cdbfgs-mult}).

\begin{table}[]
    \caption{Data structures for {\tt B.mult} for the compact dense approach}
    \begin{tabular}{>{\centering}p{0.07\textwidth} p{0.86\textwidth}} \toprule
      $\PO$ & $\mathbb{R}^{N \times m}$ matrix to store $B_0 S$ \\
      $\FO$ & $\mathbb{R}^{m \times m}$ matrix to store $\PO^T S$ \\
      $L$ & $\mathbb{R}^{m \times m}$ matrix to store $\hstril_k(S^T Y)$ \\
      $J$ & $\mathbb{R}^{m \times m}$ matrix to store the Cholesky factor $JJ^T = \FO + L\diag(d)^{-1} L^T$ \\
      \bottomrule
    \end{tabular}
    \label{tab:cdbfgs-mult-data}
\end{table}

\begin{algorithm2e}[h]\label{alg:cdbfgs-mult}%
\caption{{\tt CompactDense::B.mult($v, z$)}}

    $m, k, B_0, S, Y, d, \PO, L, J = {\tt B}.(\dots)$\;

    $a, b \in \mathbb{R}^{m_k}$ \comment{temporary work vectors}
    
    $\omega \gets \omega(m, k)$\comment{active history indices}
    
    $(\alpha, \tilde P) =$ \lIf{$B_0 = \beta I$}{$(\beta, S)$ \textbf{else} $(1, \PO)$\hfill{\tt //} $\alpha \tilde P = \PO$}
    
    $z \gets B_0 v$ \complexity{$B_0$.mult}{}
    
    $\begin{bmatrix}a \\ b\end{bmatrix} \gets \begin{bmatrix}-Y_{\omega}^T \\ \alpha \tilde P_{\omega}^T\end{bmatrix} v$ \complexitytwo{$2$\,gemv}{n,m_k}{reduce}{2 m_k}

    $a \gets \diag(d_\omega)^{-1} a$\complexity{diags}{m_k}
    
    $b \gets J_{\omega,\omega}^{-T}J_{\omega,\omega}^{-1}(b - L_{\omega,\omega} a)$\complexitytwo{gemv}{m_k, m_k}{$2$\,trsv}{m_k}
    
    
    $a \gets a + \diag(d_\omega)^{-1} L_{\omega,\omega}^T b$\complexitytwo{gemv}{m_k,m_k}{diags}{m_k}
    
    $z \gets z - \begin{bmatrix}Y_{\omega} & \alpha \tilde P_{\omega}\end{bmatrix}  \begin{bmatrix}a \\ b \end{bmatrix}$
    \complexitytwo{$2$\,gemv}{n,m_k}{bcast}{2m_k}
\end{algorithm2e}

The time complexity of \cref{alg:cdbfgs-mult} is
\begin{multline}
T({\tt CompactDense::B.mult}) =
T(B_0{\tt .mult})  \\
+ 4T({\tt gemv}(n,m_k)) + T({\tt reduce}(2m_k)) + T({\tt bcast}(2m_k))
+ O(m_k^2).
\end{multline}

\Cref{eq:compact-dense-B0} is a rank-$2m$ update; but unlike the $p$ vectors in
the rank-$2m$ update defining $B_k^m$ \cref{eq:lbfgs-mult-defn}, the $\po$
vectors are not recursively defined.  The cost of updating $\PO$ was discussed
in \cref{sec:algorithms-recursive}. The cost of updating $\FO$ is dependent on the
variability of $B_0$ in the same way as the $Z$ matrix in
\cref{tab:cdbfgs-solve-data}, and only one row of $L$ needs to be updated with
$m_k - 1$ new inner products per iteration.  Once these matrices are computed,
the $J$ matrix can be computed in $O(m_k^3)$ work.  Therefore the cost of {\tt
B.update\_mult}, given in the appendix as \cref{alg:cdbfgs-update-mult}, is for $k \geq
m$,
\begin{multline}
T({\tt CompactDense::update\_mult}) =
m_{\text{update}} T(B_0{\tt .mult}) \\
+ T({\tt gemv}(n, m - 1)) + T({\tt reduce}(m - 1)) \\
+ T({\tt gemm}(\max\{1,m_{\text{update}}\},m,n)) + T({\tt reduce}(\max\{1,m_{\text{update}}\}m))
+ O(m^3).
\end{multline}

If $B_0 = \gamma I$ or if $B_0$ is constant, then this is superior to the complexity of {\tt Recursive::B.update\_mult} \cref{eq:brfgs-update-mult-complexity} because it requires only $O(mn)$ BLAS work as opposed to $O(m^2 n)$.
If $B_0$ is a general variable-metric matrix, however, the compact dense approach also requires $O(m^2 n)$ BLAS work.

\subsection{Intermediate dense approach}

If we multiply out the blocks in the compact dense representation \cref{eq:compact-dense}, we see
\begin{equation}\label{eq:blocks-expanded}
\begin{aligned}
    H_k^m &= H_0 + S R^{-T} (Y^T H_0 Y + D) R^{-1} S^T - SR^{-T}Q^T - Q R^{-1} S^T \\
    &= (I - SR^{-T} Y^T)H_0(I - Y R^{-1} S^T) + SR^{-T} D R^{-1} S^T,
\end{aligned}
\end{equation}
which looks like the sum of an oblique projection of $H_0$ plus a correction, similar to the original definition in \cref{eq:bfgs-full}.
In this form $H_0$ appears only once, and there is no need for the $Q = H_0 Y$
matrix of column vectors nor the $Z = Y^T H_0 Y$ inner product matrix.  An implementation of \LBFGS that does not require those
data structures avoids the drawbacks of the compact dense approach discussed in \cref{sec:algorithms-compact-dense}.

\subsubsection{Intermediate dense {\tt B.solve}}
The last line in \cref{eq:blocks-expanded} can be reorganized into a form we call the \emph{intermediate dense form},
\begin{equation}\label{eq:intermediate-dense}
H_k^m = 
\begin{bmatrix}I & S R^{-T} \end{bmatrix}
\begin{bmatrix}I & 0 \\ -Y^T & I\end{bmatrix}
\begin{bmatrix}
    H_0 & 0 \\ 0 & D
\end{bmatrix}
\begin{bmatrix}I & -Y \\ 0 & I\end{bmatrix}
\begin{bmatrix}I \\ R^{-1} S^T \end{bmatrix},
\end{equation}
which shows the steps involved in computing $z \gets H_k^m v$.   This approach only requires
that the $R = \htriu_k(S^T Y)$ matrix be kept up to date (\cref{tab:idbfgs-solve-data}); but in order to optimize for the quasi-Newton update,
the implementation should cache the $S^T g_k$ product in {\tt B.update\_mult} for reuse in {\tt B.mult}, using the same caching strategy as the
compact dense approach in \cref{sec:algorithms-compact-dense}. The result is \cref{alg:idbfgs-solve}.

\begin{table}[]
    \centering
    \caption{Data structures for {\tt B.solve} for the intermediate dense approach}
    \begin{tabular}{>{\centering}p{0.07\textwidth} p{0.86\textwidth}} \toprule
      $R$ & $\mathbb{R}^{m\times m}$ matrix to store $\htriu_k(S^T Y)$ \\
      $a^{\text{cache}}$ & $\mathbb{R}^{m}$ vector to cache the product $S^T g^{\text{prev}}$ \\
      \bottomrule
    \end{tabular}
    \label{tab:idbfgs-solve-data}
\end{table}

\begin{algorithm2e}[h]\label{alg:idbfgs-solve}%
\caption{{\tt IntermediateDense::B.solve$(v, z)$}}

    $m, k, B_0, S, Y, d, R, g^{\text{prev}}, a^{\text{cache}} = {\tt B}.(\dots)$\;

    $a \in \mathbb{R}^{m_k}$ \comment{temporary work vector}
    
    $\omega \gets \omega(m, k)$\comment{active history indices}

    \If{$v \neq g^{\text{prev}}$}{%
      
      $a \gets S_{\omega}^T v$ \complexitytwo{gemv}{n,m_k}{reduce}{m_k}

    }%
    \lElse{
      $a \gets  a_{\omega}^{\text{cache}}$ \hfill{\tt // copy$(m_k)$\hspace{-0.35em}}
    }

    $a \gets \htriu_k(R_{\omega,\omega})^{-1} a$ \complexity{trsv\_cyclic}{m_k}
    
    $z \gets v - Y_{\omega} a$ \complexitytwo{gemv}{n,m_k}{bcast}{m_k}
    
    $z \gets B_0^{-1} z$ \complexity{$B_0$.solve}{}
    
    $a \gets \mathrm{diag}(d_{\omega}) a$ \complexity{diagm}{m_k}
    
    $a \gets a - Y_{\omega}^T z$ \complexitytwo{gemv}{n, m_k}{reduce}{m_k}
    
    $a \gets \htriu_k(R_{\omega,\omega})^{-T} a$ \complexity{trsv\_cyclic}{m_k}
    
    $z \gets z + S_{\omega} a$ \complexitytwo{gemv}{n, m_k}{bcast}{m_k}
\end{algorithm2e}

The complexity of \cref{alg:idbfgs-solve} applied to a generic vector $v$ is
\begin{multline}
    T({\tt IntermediateDense::B.solve}) = T(B_0{\tt .solve}) \\
    + 4T({\tt gemv}(n, m_k)) + 2T({\tt reduce}(m_k)) + 2 T({\tt bcast}(m_k))
    + O(m_k^2).
\end{multline}

The complexity of {\tt B.update\_solve} (\cref{alg:idbfgs-update-solve} in the appendix) is
\begin{multline*}
    T({\tt IntermediateDense::B.update\_solve}) = \\
    T({\tt gemv}(n,m_k)) + T({\tt dot}(n)) \\
    + T({\tt reduce(n, m_k)}) + T({\tt reduce}(1)) + O(m_k).
\end{multline*}

When the $S^T g_k$ product is cached, the cost of the quasi-Newton update is
\begin{multline*}
    T({\tt IntermediateDense::B.solve}(x_k, g_k) \\ + {\tt IntermediateDense::B.update\_solve}(g_k, p)) =
    T(B_0{\tt .solve}) \\
    + 4T({\tt gemv}(n, m_k)) + T({\tt dot}(n)) \\
    + 2T({\tt reduce}(m_k)) + 2 T({\tt bcast}(m_k)) + T({\tt reduce}(1))
    + O(m_k^2).
\end{multline*}

Compared with the cost of $T({\tt Recursive::B.solve})$
\cref{eq:rbfgs-solve-complexity}, this uses only $\sim\!2n$ additional flops, and the majority of the BLAS work is in Level 2 BLAS, like
the compact dense approach.  The reductions and broadcasts are not as coalesced as in the compact dense approach \cref{eq:cdbfgs-update-and-solve-complexity}, because the intermediate formulation is essentially two sequential rank-$m$ updates, as opposed to one rank-$2m$ update, but the performance is not affected by an expensive or variable $B_0$.

\subsubsection{Non-existence of an intermediate dense {\tt B.mult}}
There cannot be an analog to the intermediate dense form
\cref{eq:intermediate-dense} for $B_k^m$ that avoids the drawbacks of the
compact dense approach to implementing {\tt B.mult}.  By an analog we mean an
algorithm to compute $z \gets B_k^m x$ exactly that uses only $O(1)$ applications of
$B_0${\tt .mult} and $B_0${\tt .solve} for every choice of $B_0$, $S$, and $Y$.

Consider the case where $S = Y = U$ is a set of orthonormal vectors.
Then \cref{eq:compact-dense-B0} simplifies to
\[
\begin{aligned}
B_k^m &= B_0 - \begin{bmatrix} U & B_0 U \end{bmatrix}
\begin{bmatrix} -I & 0 \\ 0 & U^T B_0 U \end{bmatrix}^{-1}
\begin{bmatrix} U^T \\ U^T B_0 \end{bmatrix} \\
&=
B_0 + U U^T - B_0 U (U^T B_0 U)^{-1} U^T B_0.
\end{aligned}
\]
If we then consider the product $U^T H_0 B_k^m H_0 U$, we see
\[
\begin{aligned}
U^T H_0 B_k^m H_0 U
&=
U^T H_0 \left[B_0 + U U^T - B_0 U (U^T B_0 U)^{-1} U^T B_0\right] H_0 U \\
&= 
U^T H_0 U + (U^T H_0 U)^2 - (U^T B_0 U)^{-1},
\end{aligned}
\]
which when rearranged yields
\[
(U^T B_0 U)^{-1} = U^T H_0 U + (U^T H_0 U)^2 - U^T H_0 B_k^m H_0 U.
\]
The existence of an algorithm with the desired properties would thus imply the
existence of an algorithm that exactly computes $z \gets (U^T B_0 U)^{-1}x$
using $O(1)$ applications of $B_0${\tt .mult} and $B_0${\tt .solve} for every
SPD $B_0$ and every set of orthonormal vectors $U$. The $O(1)$ limit allows
neither the formation of $(U^T B_0 U)$ for factorization nor for the solution
via an iterative method that could take up to $m$ iterations, so this is
generally impossible.

Because of this negative result, \code{CompactDense::B.mult} (\cref{alg:cdbfgs-mult}) is the only
dense algorithm for {\tt B.mult} that we consider.

\section{Quasi-Newton update performance comparison}\label{sec:performance}

\subsection{Performance model}\label{sec:performance-model}
\begin{table}[]
    \caption{Quasi-Newton update performance ({\tt B.update}$(x_k, g_k)$ + {\tt B.solve}$(g_k, p)$).  In the {\tt reduce} and {\tt bcast} columns, an entry ``$a$~$(b)$'' denotes $a$ collective operations that coalesce into $b$ synchronizations (one latency each); $b$ is omitted when $a=b$.}
    \centering
    \begin{tabular}{lccccccc} \toprule
                                       & $B_0^{-1}$ & \multicolumn{3}{c}{BLAS memops}          & \multicolumn{3}{c}{MPI synchronizations} \\
         \cmidrule(lr){3-5}                         
         \cmidrule(lr){6-8}            &            
                                                    & Lvl.\ 1      & Lvl.\ 2     & Lvl.\ 3     & {\tt allr.} & {\tt reduce} & {\tt bcast} \\
         \midrule                                    
         {\tt B.update}                & 0          & $\sim\!18n$  & 0           & 0           & 1           & 0            & 0           \\
         \cmidrule(lr){1-8}                             
         {\tt Recursive}               & 1          & $\sim\!10mn$ & 0           & 0           & $2m$        & 0            & 0           \\
         \cmidrule(lr){1-8}              
         {\tt CompactDense}                 &            &              &             &             &             &              &             \\
         \ $B_{0,k} = \gamma_k I$      & 1          & $\sim\!4n$   & $\sim\!4mn$ &             & 0           & 4 (1)        & 2 (1)       \\
         \ $B_0$ constant              & 2          & $\sim\!4n$   & $\sim\!4mn$ & 0           & 0           & 4 (1)        & 2 (1)       \\
         \ $B_0 = B_{0,k}$ var.        & $1 + m$    & $\sim\!2n$   & $\sim\!4mn$ & $\sim\!2mn$ & 0           & 4 (1)        & 2 (1)       \\
         \cmidrule(lr){1-8}      
         {\tt Intermediate}            & 1          & $\sim\!2n$   & $\sim\!4mn$ & 0           & 0           & 3 (2)        & 2           \\
         \bottomrule
    \end{tabular}
    \label{tab:performance-summary}
\end{table}

Throughout we analyze the limited-memory BFGS (\LBFGS) method at an iteration $k \geq m$, so that all the history vectors $(s_{k-m}, y_{k-m}),\allowbreak \dots,\allowbreak (s_{k-1}, y_{k-1})$ are present.
The \LBFGS inverse Hessian $H_k$ is a rank-$2m$ update of $H_0$, so there exists a matrix $U_k \in\mathbb{R}^{n \times 2m}$ such that
$H_k = H_0 + U_k U_k^T.$  If we did not count the setup time to compute $U_k$, this representation would always be the most efficient for computing $p \gets H_kg$, because the update term
$U_k U_k^Tg$ is computed in only two dense matrix-vector products and requires reading only one additional matrix.  With this representation in mind, we define the \emph{effective bandwidth}
of an \LBFGS implementation as
\begin{equation}\label{eq:model}
\mathsf{B}_e \defeq \frac{wn(2m + 2)}{T_{\text{update}} + T_{\text{solve}}}.
\end{equation}

%
%
%

\subsection{Performance measurements}
\label{sec:performance-measurements}

\paragraph{Compute environments}

In this paper we present CPU- and GPU-based results on OLCF's Frontier and ALCF's Polaris. 
Frontier is an HPE Cray EX machine. Each node is equipped with one 64-core AMD EPYC 7A53 (Trento) CPU and 4 AMD Instinct MI250X cards, each with 2 Graphics Compute Dies (GCDs) for a total of 8 GCDs per node. The CPUs and GPUs are connected via Infinity Fabric.
Polaris is an HPE Apollo 6500 machine, equipped with 32 core AMD EPYC 7543P Milan, with 4 NVIDIA A100 GPU cards, per node. The CPUs and GPUs are connected via PCIe-4.0, and the GPUs are connected via NVLink.
For Polaris, both single- and four-GPU regimes were tested. However, for Frontier, despite each node offering up to eight GCDs (usable as eight separate GPUs), this paper only investigates code efficiency on a single GCD.
\Cref{table:env} describes the modules and environments used on each machine. 



\begin{table}
    \centering
    \caption{Environments for each machine}\label{table:env}
    \begin{tabular}{ll}
        \toprule
        machine & {modules and environment} \\
        \midrule
        Frontier & PrgEnv-gnu/8.6.0, rocm/6.3.1, craype-x86-trento,\\
                 & cray-mpich/8.1.31, craype-accel-amd-gfx90a \\
        Polaris & PrgEnv-nvidia/8.6.0, nvidia/24.11,\\
                & craype-x86-milan\\
        \bottomrule
    \end{tabular}
\end{table}

\paragraph{Solve and update step performance}

We measure the timings of update and solve when the right-hand side is $g_k$, which we call $T_{\text{step}}$.  The goal is to compare the implementation to the performance model in \cref{eq:model}.
The program that generated these timings is included in the PETSc test suite.\footnote{\tt ksp\_ksp\_utils\_lmvm\_tests-solve\_performance\_0.}

The value $T_{\text{step}}$ is the average iteration time of a loop that repeatedly calls {\tt Update($x_k, g_k$)} followed by {\tt Solve($g_k$)} (but without a linesearch or even any object evaluation).
The loop only includes steps for which $k \geq m$ so that the full history size is used,
because the initial steps for $k < m$ will have different performance characteristics, making it harder to compare the measurements with the performance model.

For performance testing, we used two types of $H_0$: a \emph{constant diagonal} matrix, $\alpha I$, and a \emph{diagonal} matrix. For scaling $H_0$, we used two types: \emph{no scaling} and \emph{diagonal}.
Diagonal scaling uses a full-memory restricted Broyden update formula to construct a diagonal matrix for $H_0$ initialization \cite{gilbert1989some}.
In all, we conducted tests on two possible combinations: constant diagonal matrix without scaling and diagonal matrix with diagonal scaling.
The former represents the simplest strategy for $H_0$, while the latter represents the most complex case, with which we illustrate
shortfalls of the compact dense method and superiority of our intermediate dense strategy.

\paragraph{Mult and update step performance} We perform the same test for {\tt Mult}. 
We use  the same setup as {\tt Solve}, including the PETSc test suite, loop structure, and types of $H_0$ matrices. 

We note that {\tt Mult} timings vary markedly between the Recursive and compact dense methods.
We therefore show two representative cases: 4 MPI ranks on CPUs and 4 MPI ranks on GPUs.

\subsection{CPU}\label{sec:cpu}
In this subsection we present performance measurements for the CPU case.
We have constructed time/efficiency plots that compare $T_{\text{step}}$ for different L-BFGS algorithms at
different problem sizes $n$ and different history sizes $m$ with different strategies for $H_0$.
Efficiency is measured by using the effective bandwidth $B_e$. Solid lines connect different measurements with the same
problem size $n$ at history sizes $m\in\{5,10,20,50\}$; the problem size $n$ is annotated, and the history size $m$ is indicated with the marker size.

We note that all CPU performance measurements were conducted on ALCF's Polaris.
We present four plots measuring {\tt Solve()} and 
one plot measuring {\tt Mult()} done on the CPU.
We have MPI rank counts of 1 and 4; and, for each given MPI rank, we present both
constant diagonal $H_0$ without scaling and diagonal $H_0$ with diagonal scaling.
We choose 4 MPI ranks because L3 cache can become saturated with the problem size at this rank count.
For {\tt Mult()}, we present one plot with 4 MPI ranks, with constant diagonal $H_0$ without scaling.

\begin{figure}[h]
\centering%
\input{figure/polaris-solve-host-bfgs-condiag-user.tex}
\caption{BFGS {\tt Solve} comparison, constant diagonal $H_0$ without scaling. $B_e$ is the effective bandwidth, as defined in \cref{eq:model}. Polaris (EPYC 7543P), 1 MPI CPU process}\label{fig:polaris-host-bfgs-cd-u}%
\end{figure}
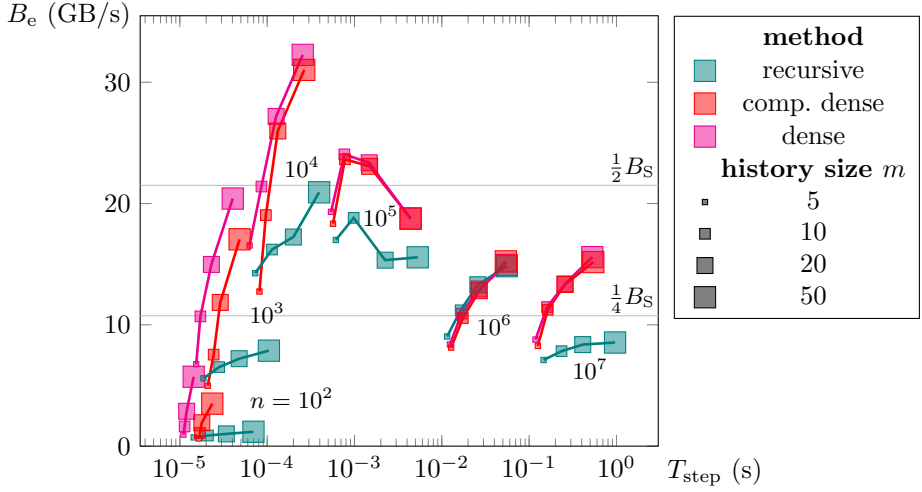

In \Cref{fig:polaris-host-bfgs-cd-u}, one can observe that
at $n=10^2$ the problem size is so small that the latency dominates.
The efficiency of the recursive algorithm is nearly constant, while both dense algorithms exhibit nearly constant runtime.  

At $n=10^4$, the latency is no longer dominant, and all the problem data fits into the 32 MiB L3 cache for each of the history sizes tested.
At this problem size the 256 KiB L2 cache is only $\approx 3\times$ the size of a single vector. We speculate that the intermediate dense algorithm performs better than the recursive algorithm for larger values of $m$ because it makes better use of this cache.

At $n=10^6$, the 32 MiB L3 cache is only $\approx 4\times$ bigger than a single vector.
For the recursive algorithm, if its rank-1 updates keep the vectors being modified in cache, it is predicted that
the recursive algorithm will have the same effective bandwidth as the intermediate dense algorithm, which matches our measurements.


At $n=10^7$ a single vector does not fit in the L3 cache, so we predict that the recursive algorithm will behave poorly compared with the dense algorithm, as the rank-1 update no longer keeps the vectors being modified in cache, which also matches our measurements.
We measured the bandwidth to be approximately $43$\,GB/s for a single core.

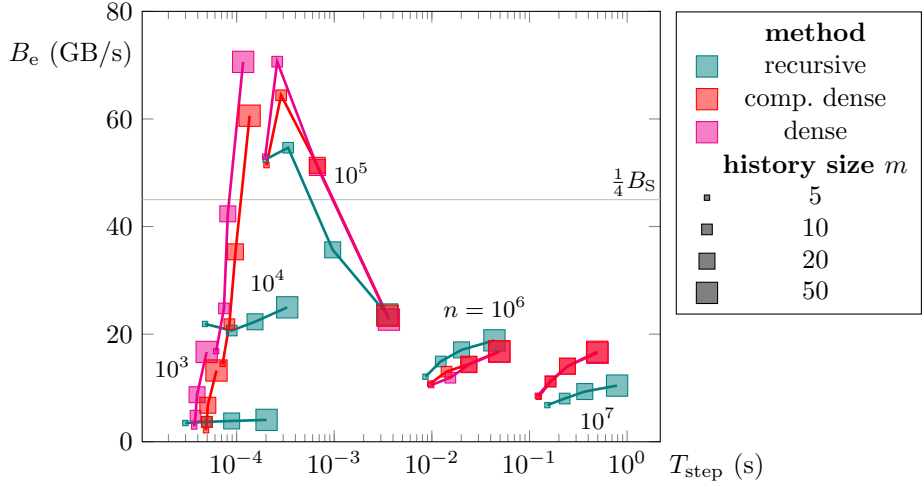
\begin{figure}%
\centering%
\input{figure/polaris-solve-host-bfgs-mpifour-condiag-user.tex}
\caption{BFGS {\tt Solve} comparison, constant diagonal $H_0$ without scaling. Polaris (EPYC 7543P), 4 MPI CPU processes}\label{fig:polaris-host-mpifour-bfgs-cd-u}%
\end{figure}

One can observe similar behavior for 4 MPI ranks. In \Cref{fig:polaris-host-mpifour-bfgs-cd-u,fig:polaris-host-mpifour-bfgs-dd}, 
we note that, for $n=10^6$, the recursive algorithm very slightly outperforms both dense algorithms. We observed that when $n=10^6$, {\tt Update()} was slightly more expensive for dense algorithms, 
possibly due to extra memory movement to assemble appropriate internal objects, while such extra cost can be avoided for recursive algorithms. 
Regardless, we still observe that the dense algorithm outperforms the recursive algorithm for the largest problem size of $n=10^7$.

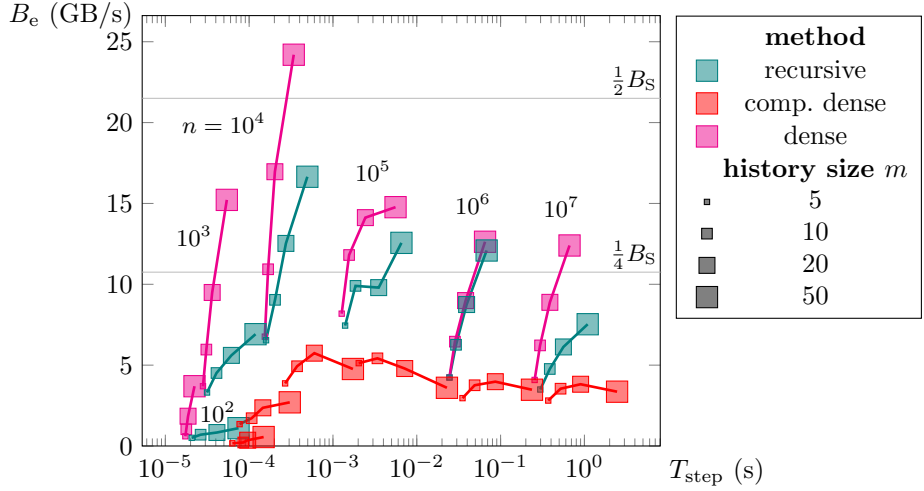
\begin{figure}%
\centering%
\input{figure/polaris-solve-host-bfgs-diag-diag.tex}
\caption{BFGS {\tt Solve} comparison, diagonal $H_0$ with diagonal scaling. Polaris (EPYC 7543P), 1 MPI CPU process}\label{fig:polaris-host-bfgs-dd}%
\end{figure}

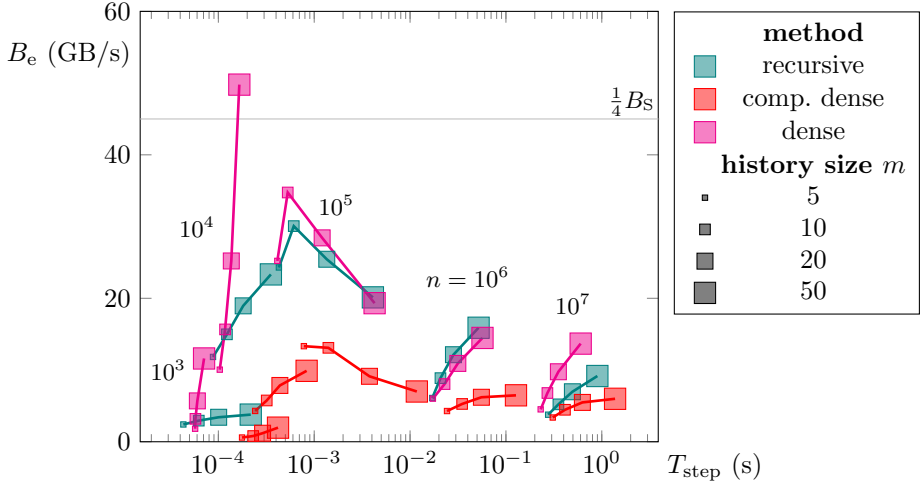
\begin{figure}%
\centering%
\input{figure/polaris-solve-host-bfgs-mpifour-diag-diag.tex}
\caption{BFGS {\tt Solve} comparison, diagonal $H_0$ with diagonal scaling. Polaris (EPYC 7543P), 4 MPI CPU processes}\label{fig:polaris-host-mpifour-bfgs-dd}%
\end{figure}

When $H_0$ is not constant, the compact‑dense formulation performs substantially worse. 
For a diagonal $H_0$ with diagonal scaling, \Cref{fig:polaris-host-bfgs-dd,fig:polaris-host-mpifour-bfgs-dd} show that the compact dense algorithm is the slowest of the three across all problem and history sizes, for both 1 and 4 MPI processes. 
This behavior is expected: as summarized in \Cref{tab:performance-summary}, a variable $H_0$ forces the compact dense method to perform $m$ additional evaluations of $H_0$ and 
$\sim (2m^2+8m+2)n$ flops of BLAS work.
By contrast, the intermediate‑dense formulation performs much better than the compact‑dense one and typically outperforms the recursive algorithm. 
The exceptions occur with 4 MPI processes for $n=10^5$ with $m=50$ and for $n=10^6$, where the recursive method is slightly faster. 
We hypothesize that storing the $R$ matrix on the first MPI rank ({\tt reduce} and {\tt bcast}), rather than duplicating it on each process ({\tt allreduce}), introduces overhead that slows the intermediate dense method in these cases. 
Even so, for the largest problem size, $n=10^7$, the intermediate dense formulation remains faster than the recursive algorithm.

\begin{figure}%
\centering%
\input{figure/polaris-mult-host-mpifour-dbfgs-bfgs.tex}
\caption{BFGS {\tt Mult} comparison, diagonal $H_0$ without scaling,  Polaris (EPYC 7543P), 4 MPI CPU processes}\label{fig:polaris-mult-host-mpifour-dbfgs-bfgs}
\end{figure}
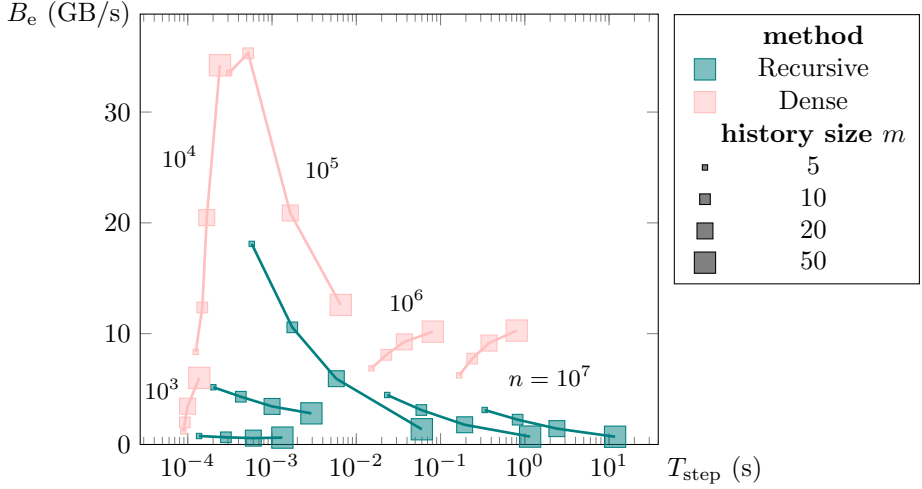

We also discuss {\tt Mult()} timings.  We present {\tt Mult()} timings with diagonal $H_0$ without scaling in \Cref{fig:polaris-mult-host-mpifour-dbfgs-bfgs}, with 4 CPU MPI ranks.
We note that for all cases, dense versions significantly outperform the recursive version, as expected. 
We note that the speedup is much more pronounced for higher history size $m$, with
almost an order of magnitude speedup for $m=50$ for $n > 10^3$. This speedup  is 
expected because this problem type has $O(mn)$ BLAS work, compared with $O(m^2n)$ BLAS work for the recursive method.

\subsection{GPU}
In this subsection we present performance measurements for the GPU;
notation and plotting conventions are identical to \Cref{sec:cpu},
except for using circle markers instead of square markers.

We present four plots measuring {\tt Solve()} timings and one plot measuring {\tt Mult()}
timings on the GPUs of ALCF's Polaris machine.
As in the CPU experiments, for each GPU MPI rank, we consider two setups:  
constant diagonal $H_0$ without scaling and diagonal $H_0$ with diagonal scaling.
In addition, we present one plot measuring {\tt Solve()} conducted on OLCF's Frontier, with constant diagonal $H_0$ without scaling.

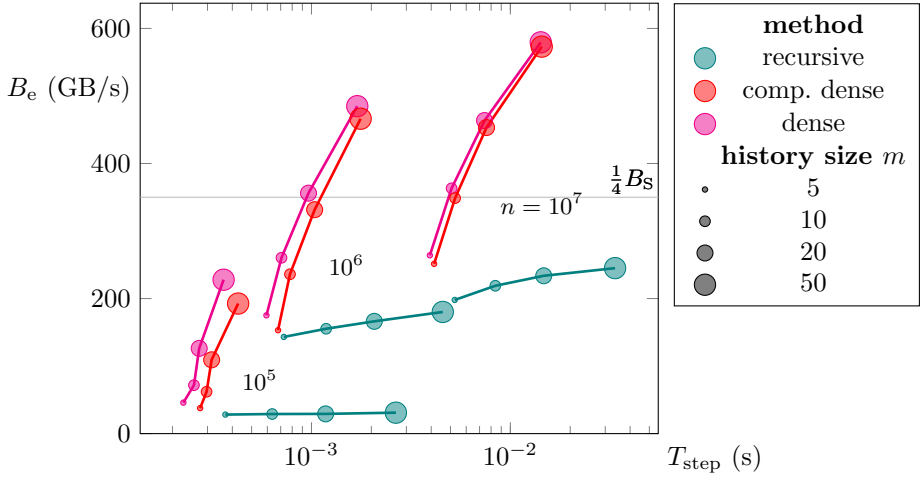
\begin{figure}%
\centering%
\input{figure/polaris-solve-device-bfgs-condiag-user.tex}
\caption{BFGS {\tt Solve} comparison, constant diagonal $H_0$ without scaling. Polaris A100, 1 GPU}\label{fig:polaris-device-bfgs-cd-u}%
\end{figure}

\begin{figure}%
\centering%
\input{figure/frontier-solve-device-dbfgs-bfgs-cd-user.tex}
\caption{BFGS {\tt Solve} comparison, constant diagonal $H_0$ without scaling. Frontier MI250X, 1 GPU}\label{fig:frontier-device-dbfgs-bfgs}
\end{figure}
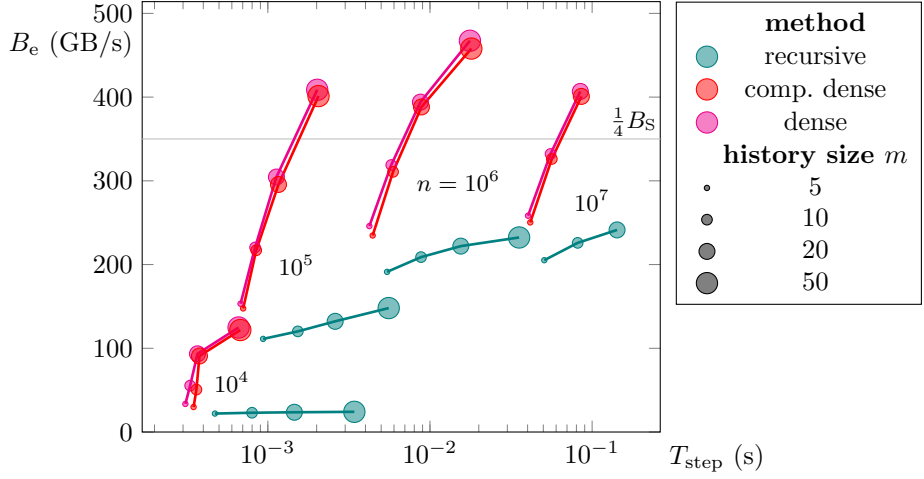

\begin{figure}%
\centering%
\input{figure/polaris-solve-device-bfgs-diag-diag.tex}
\caption{BFGS {\tt Solve} comparison, diagonal $H_0$ with diagonal scaling. Polaris A100, 1 GPU}\label{fig:polaris-device-bfgs-dd}%
\end{figure}
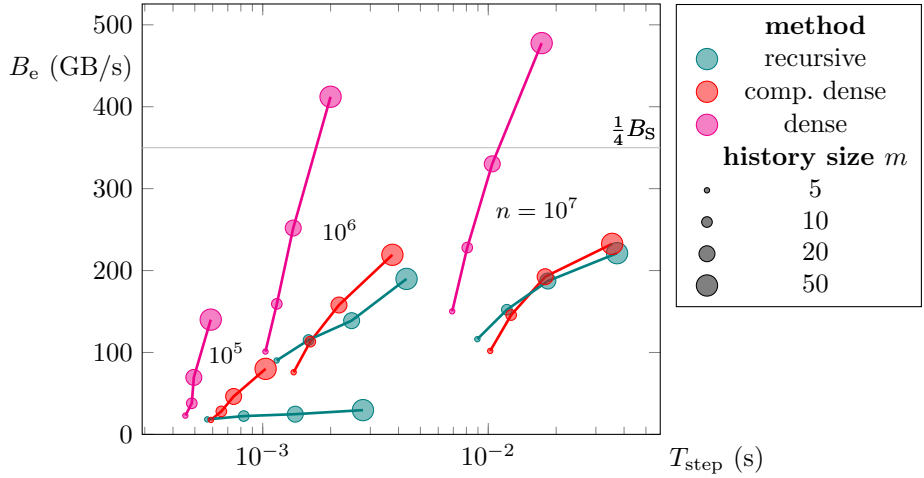

\begin{figure}%
\centering%
\input{figure/polaris-solve-device-bfgs-mpifour-condiag-user.tex}
\caption{BFGS {\tt Solve} comparison, constant diagonal $H_0$ without scaling. Polaris A100, 4 GPUs}\label{fig:polaris-device-mpifour-bfgs-cd-u}%
\end{figure}
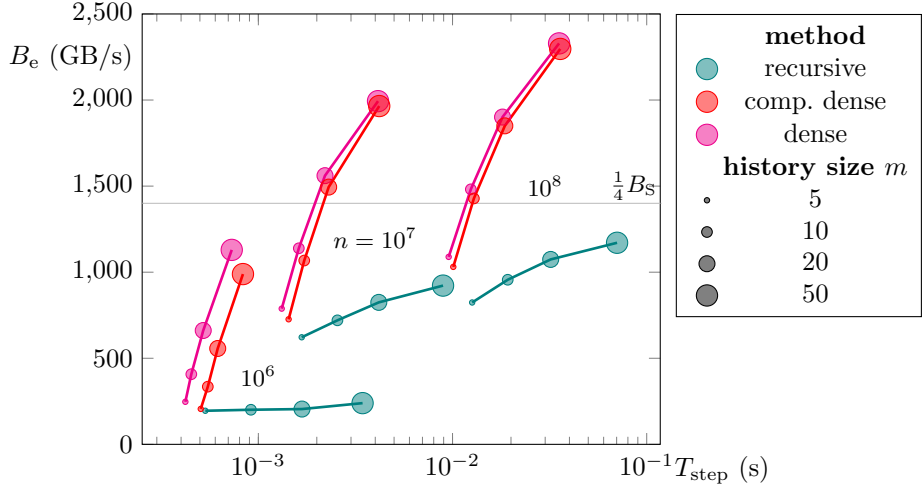


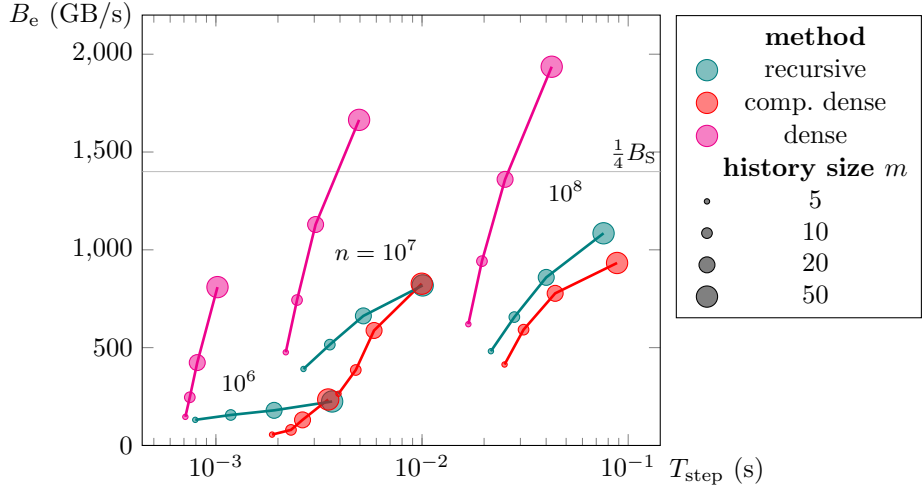
\begin{figure}%
\centering%
\input{figure/polaris-solve-device-bfgs-mpifour-diag-diag.tex}
\caption{BFGS {\tt Solve} comparison, diagonal $H_0$ with diagonal scaling. Polaris A100, 4 GPUs}\label{fig:polaris-device-mpifour-bfgs-dd}%
\end{figure}

In \Cref{fig:polaris-device-bfgs-cd-u} one can observe that
for all problem sizes, dense algorithms vastly outperform the recursive algorithm, as expected. 
Since $H_0$ is constant with no scaling, the compact dense and intermediate dense algorithms' performances are similar, although intermediate dense is slightly faster at smaller problem sizes.
The recursive algorithm exhibits an almost constant effective bandwidth across $m$, for each problem size, consistent with its low arithmetic intensity: runtime is bounded by memory bandwidth and synchronization latency rather than compute.
We find that for constant diagonal $H_0$ without scaling, the highest speedup factor of 7.35 is achieved, with $n=10^5$ with $m=50$, and the speedup factor for the biggest problem size is 2.36 for $n=10^7$ with $m=50$.

Similar to the CPU-based results, we observe the compact dense approach exhibiting much worse performance when $H_0$ is not constant.
This is not surprising, because different architecture does not change the fundamental additional algorithmic overhead of having to recompute over all the basis vectors.
The compact dense algorithm, which performed much worse than the recursive algorithm on the CPU, performs on-par with the recursive method on GPUs, even outperforming it for the small problem size of $n=10^5$.
We suspect that, despite its higher computational burden, the compact dense algorithm with variable $H_0$ incurs lower synchronization latency and places less pressure on memory bandwidth, which may explain its slightly better performance relative to the recursive algorithm.
We find that for variable $H_0$ with diagonal scaling, the highest speedup factor of 4.73 is achieved, with $n=10^5$ with history size $m=50$, and the speedup factor for largest problem size is 2.16 for $n=10^7$ with $m=50$.

One can observe similar behavior for 4 GPUs. In \Cref{fig:polaris-device-mpifour-bfgs-cd-u,fig:polaris-device-mpifour-bfgs-dd},
we note that the intermediate dense algorithm greatly outperforms the recursive algorithm.
For constant diagonal $H_0$ without scaling, intermediate dense and compact dense algorithms show similar results, as also observed in CPU-based results and results from one GPU card. 
But for diagonal $H_0$ with diagonal scaling, while the intermediate dense algorithm still greatly outperforms the recursive algorithm, the compact dense algorithm performs slightly worse than the recursive one.

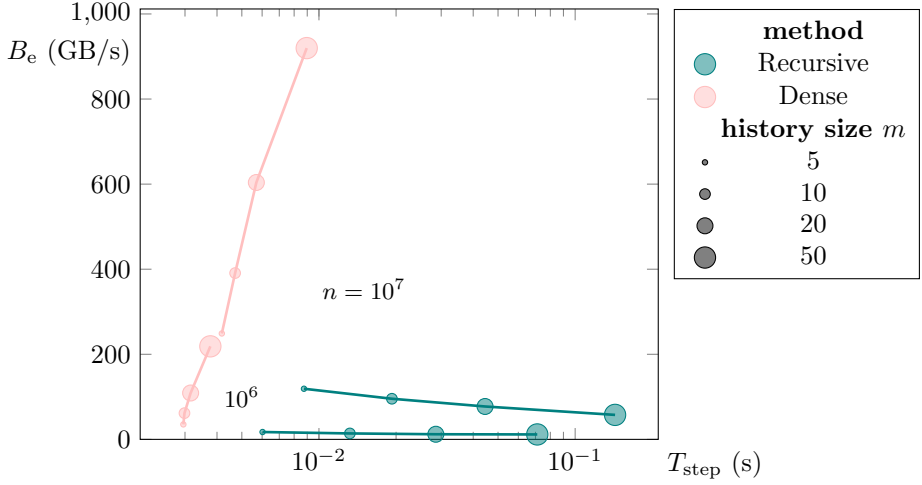
\begin{figure}%
\centering%
\input{figure/polaris-mult-device-mpifour-dbfgs-bfgs.tex}
\caption{BFGS {\tt Mult} comparison, diagonal $H_0$ without scaling. Polaris A100, 4 GPUs}\label{fig:polaris-mult-device-mpifour-dbfgs-bfgs}
\end{figure}

For the {\tt Mult()} timings in \Cref{fig:polaris-mult-device-mpifour-dbfgs-bfgs}, 
we observe that the dense method outperforms the recursive method by roughly an order of magnitude as the history size $m$ increases.
As noted in \Cref{sec:cpu}, this gap is expected because the recursive method requires $O(m^2n)$ BLAS work for update, compared with $O(mn)$ work for the dense method.


\section{Performance on a convolutional neural network training benchmark}
In this section we evaluate the performance of the 
\textit{dense BFGS} solver against the baseline \textit{recursive BFGS} implementation on a supervised learning problem: training a convolutional neural network on the MNIST image classification benchmark. 

We integrated PETSc/TAO's BFGS solvers with PyTorch, enabling their use for training a neural network (NN).
The MNIST training set contains $60{,}000$ grayscale images, each of size $28\times28$ pixels, labeled across ten classes.
The objective function minimizes the cross-entropy loss between the model predictions and the ground-truth labels. We use a simple CNN with two convolutional layers followed by two fully connected layers. It contains $1{,}199{,}882$ trainable parameters, resulting in a BFGS Hessian matrix of size $1{,}199{,}882\times1{,}199{,}882$.

Training was performed with a mini-batch size of $1024$, and the maximum number of BFGS iterations per mini-batch was set to $20$.
\Cref{tab:method_hist_speedup} reports the solver times for different history sizes.
The solver time includes all operations performed by TAO except for the loss function evaluation (forward propagation) and the gradient computation (backpropagation).

As shown in \Cref{tab:method_hist_speedup}, the dense BFGS solver consistently outperforms the recursive implementation across all history sizes, achieving up to a $4.7\%$ speedup.

\begin{table}[h!]
\centering
\setlength{\tabcolsep}{5pt}
\caption{Comparison of recursive and dense BFGS under different history sizes on the MNIST benchmark.
Values are mean$\pm$std of per epoch time cost (in seconds) over three repeated runs.
Improvement = ($\text{Mean}_{\text{recursive}} - \text{Mean}_{\text{dense}})/ \text{Mean}_{\text{recursive}}$.}
\label{tab:method_hist_speedup}
\begin{tabular}{@{}cccc@{}}
\toprule
\makecell{\textbf{History Size}} &
\makecell{\textbf{Recursive BFGS}} &
\makecell{\textbf{Dense BFGS}} &
\makecell{\textbf{Improvement}} \\
\midrule
5  & 8.882$\pm$0.11  & \textbf{8.804$\pm$0.10}  & \textbf{0.9\%} \\
10 & 9.111$\pm$0.09  & \textbf{8.825$\pm$0.09}  & \textbf{3.1\%} \\
20 & 9.272$\pm$0.09  & \textbf{8.838$\pm$0.07}  & \textbf{4.7\%} \\
\bottomrule
\end{tabular}
\end{table}

\section{Generalization to other quasi-Newton methods}
\label{sec:generalization}

\subsection{Davidon--Fletcher--Powell}
As indicated in \Cref{eq:compact-dense-B0-factored}, \code{CompactDense::B.mult} requires a Cholesky factorization.
Although  we found Cholesky variant to be faster than recursive version,
as noted in \Cref{fig:polaris-mult-host-mpifour-dbfgs-bfgs,fig:polaris-mult-device-mpifour-dbfgs-bfgs},
for certain $H_0$ scaling strategies, one may encounter zero pivot during Cholesky decomposition.
\footnote{{\tt -mat\_lmvm\_J0\_mat\_type diagonal -mat\_lmvm\_scale\_type diagonal -mat\_lmvm\_alpha 0.25 -mat\_lmvm\_theta 1.0} for the test used in the paper}

When Cholesky decomposition becomes numerically unstable,
yet the computational efficiency of the dense method remains essential, the
Davidon--Fletcher--Powell (DFP) method provides a viable alternative.
It is well known that the DFP method is dual to BFGS method \cite{erway2013shifted}.
Therefore, from BFGS, by exchanging $s_k$ with $y_k$, $B_k$ with $H_k$, and $L_k$ with $R_k^T$, one can obtain the DFP method. \cite[Appendix D]{Erway2017}
With this, one can substitute \code{CompactDense::B.mult} with \code{DenseDFP::B.solve} and avoid numerically unstable Cholesky decomposition.
Such a hybrid approach is implemented in \petscmanlink{Mat}{MATLMVMDQN} matrix type.
However, this practice comes at the cost of a slight duality gap.

\subsection{Broyden family}

PETSc/TAO provides several types of limited-memory variable-metric (LMVM) matrices.
In developing our dense BFGS formulations, we identified reusable design patterns common to the \petscmanlink{KSP}{MATLMVM} family,
enabling the techniques used for dense BFGS to be applied to other \petscmanlink{KSP}{MATLMVM} variants as well. 
Consequently, the following matrices in PETSc/TAO now support dense formulations aimed at performance portability: Broyden, Bad Broyden, Symmetric Broyden, Symmetric Bad Broyden, and SR1.~\footnote{%
\petscmanlink{Mat}{MATLMVMBROYDEN},
\petscmanlink{Mat}{MATLMVMBADBROYDEN},
\petscmanlink{Mat}{MATLMVMSYMBROYDEN},
\petscmanlink{Mat}{MATLMVMSYMBADBROYDEN},
\petscmanlink{Mat}{MATLMVMSR1}.
}
\footnote{{\tt -mat\_lmvm\_mult\_algorithm <recursive, compact\_dense, dense>} options flag can be used}. For brevity, we do not report benchmarks for these variants.

\section{Discussion}
\label{sec:discussion}

\paragraph{Quasi-Newton Krylov trust region}

This work has focused on \LBFGS globalized by a line search, which requires only one application of $H_k$
per iteration.  When using \LBFGS within a trust-region optimization method, an approximation to the forward Hessian $B(x_k)$ is also needed for the Steihaug-CG bounded Krylov subspace iteration.
Both the recursive and dense formulations of \LBFGS require additional setup computations to construct the data structures needed to efficiently apply $H_k^{-1}$ to a vector.
One approach that would avoid these costs would be to perform Steihaug-CG with a limited memory L-DFP approximation of $B_k$ preconditioned by the \LBFGS $H_k$ approximation.
This approach would require minimal setup costs.
It is not implemented in PETSc/TAO but could be the subject of further investigation.

\section{Conclusion}\label{sec:conclusion}

We have introduced a new \LBFGS representation---\allowbreak the \textit{intermediate dense} formulation---and
implemented it in PETSc/TAO alongside the compact dense formulation of \cite{Byrd1994}.
The \textit{intermediate dense} method preserves the main performance benefits of dense formulations
(improved locality and reduced synchronization relative to the recursive two-loop recursion)
while avoiding a key limitation of compact dense methods in variable-metric settings: 
extra passes over the history and repeated applications of a changing base inverse Hessian $H_0$. 
We have evaluated three PETSc/TAO implementations (recursive, compact dense, and intermediate dense) on CPUs and GPUs on the ALCF Polaris and OLCF Frontier.
Results show that dense methods consistently outperform the recursive approach at large problem sizes, 
that compact dense is most effective when $H_0$ is constant or inexpensive,
and that the intermediate dense approach provides the most robust performance across architectures and base-matrix choices. 
On the MNIST neural network training benchmark, we have demonstrated end-to-end gains from the dense implementation, achieving up to a $4.7\%$ solver-time speedup over recursive BFGS.


\section*{Acknowledgments}

This material is based upon work supported by the U.S. Department of Energy (DOE) under Contract No. DE-AC02-06CH11357.  This material is based upon work partially supported by the U.S. DOE, Office of Science, Office of Advanced Scientific Computing Research by the Competitive Portfolios for Advanced Scientific Computing Research
Program, by the Scientific Discovery through Advanced Computing (SciDAC) Program through the FASTMath Institute, and by the Exascale Computing Project (ECP).

This research used supporting resources at the Argonne and the Oak Ridge Leadership Computing Facilities. The Argonne Leadership Computing Facility at Argonne National Laboratory is supported by the Office of Science of the U.S. DOE under Contract No. DE-AC02-06CH11357. The Oak Ridge Leadership Computing Facility at the Oak Ridge National Laboratory is supported by the Office of Science of the U.S. DOE under Contract No. DE-AC05-00OR22725.

\bibliographystyle{siamplain}
\bibliography{references.bib}

\appendix

\section{Recursive algorithms}

In this appendix we present auxiliary algorithms for the recursive algorithm.

\begin{algorithm2e}[H]\label{alg:update-P0}
\caption{{\tt update\_P0}$(m, k, B_0, S, \PO)$}
    \KwData{$m, k$: history size and current iterate}
    \KwData{$S\in \mathbb{R}^{n\times m}$ column-vector matrix, $S_{i\bmod m} = s_i, i\in\iota(m,k)$}
    \KwData{$\PO\in \mathbb{R}^{n\times m}$ column-vector matrix, $\PO_{i\bmod m} = \po_i, i\in\iota(m,k-1)$}
    \KwResult{$\alpha \in \mathbb{R}, \tilde P \in \mathbb{R}^{n \times m}$ such that
    $\alpha \tilde P_{i\bmod m} = \po_i, i\in\iota(m,k)$}

    \If{$B_0 = \gamma I$}{%
        \Return $(\gamma, S)$ \comment{return $S$ by reference, no copy}
    }
    \If {$B_0$ has changed}{%
        \ForPar{$j\in\iota(m,k-1)$}{%
            $i\gets j \bmod m$\;
            
            $\PO_i \gets B_0 S_i$\complexity{$B_0$.mult}{}
        }
    }
    $i\gets (k-1)\bmod m$\;
    $\PO_i \gets B_0 S_i$\complexity{$B_0$.mult}{}
    
    \Return $(1, \PO)$ \comment{return $\PO$ by reference, no copy}
\end{algorithm2e}

\begin{algorithm2e}[H]\label{alg:rbfgs-update-mult}
\caption{{\tt Recursive::B.update\_mult}$(x, g)$}
  $m, k, B_0, S, Y, d, P, \PO, f = {\tt B}.(\dots)$\;

  $(\alpha, \tilde P) \gets {\tt update\_P0}(m, k, B_0, S, \PO)$\comment{\cref{alg:update-P0}}

  \If{$B_0$ has changed or $k > m$}{
    $\iota \gets \iota(m,k)$\comment{all $p$ vectors must be recomputed}
  }
  \Else{
    $\iota \gets \{k-1\}$\comment{only $p_{m-1,k-1}$ must be computed}
  }
  
  \For{$j \in \iota$}{
        $i \gets j \bmod m$\;

        $\tilde m \gets m - (k-j)$\;
        
        $a, b \in \mathbb{R}^{\tilde m}$ \comment{temporary work vectors}
        
        $\omega \gets \{ \ell\bmod m: \ell \in \iota(\tilde m,j)\}$\;
        
        $P_i \gets \alpha \tilde P_i$\complexity{axpy}{n}

        $\begin{bmatrix} a \\ b \end{bmatrix}\gets \begin{bmatrix} Y_{\omega} &  P_{\omega}\end{bmatrix}^T S_i$
        \complexitytwo{$2\tilde m$\,dot}{n}{allreduce}{2\tilde m}
        
        $\begin{bmatrix} a \\ b \end{bmatrix}\gets \begin{bmatrix} \mathrm{diag}(d_{\omega}) & 0 \\ 0 & \mathrm{diag}(f_{\omega})\end{bmatrix}^{-1} \begin{bmatrix} a \\ b \end{bmatrix}$
        \complexity{$2$\,diags}{\tilde m}
        
        $P_i \gets P_i +  \begin{bmatrix} Y_{\omega} &  P_{\omega}\end{bmatrix}
        \begin{bmatrix} a \\ b \end{bmatrix}$
        \complexity{$2\tilde m$\,axpy}{n}
        
       $f_i \gets S_i^T P_i$ \complexitytwo{dot}{n}{allreduce}{1}
  }
\end{algorithm2e}

\section{Compact dense algorithms}

In this appendix we present auxiliary algorithms for the compact dense algorithm.

\begin{algorithm2e}[H]\label{alg:update-Atb}%
\caption{{\tt update\_Atb}$(m, k, \beta, A, b_{\text{old}}, b_{\text{new}}, c, e)$}
\KwData{$(m, k)$: history size and current iterate}
\KwData{$\beta \in \mathbb{R}$}
\KwData{$A$: $\mathbb{R}^{N \times m}$ cyclic-order column-vector matrix}
\KwData{$b_{\text{old}}, b_{\text{new}}$: $\mathbb{R}^n$ vectors}
\KwData{$c$: $\mathbb{R}^m$ vector where $c_{j\bmod m} = \beta A_{j\bmod m}^T b_{\text{old}}$ for $j\in\iota(m,k-1)$}
\KwResult{$c$: $\mathbb{R}^m$ vector where $c_{j\bmod m} = \beta A_{j\bmod m}^T b_{\text{new}}$ for $j\in\iota(m,k)$}
\KwResult{$e$: $\mathbb{R}^m$ vector where $c_{j\bmod m} = \beta A_{j\bmod m}^T (b_{\text{new}} - b_{\text{old}})$ for $j\in\iota(m,k)$}

$i \gets (k-1) \bmod m$\;

$\omega \gets \omega(m,k)$\;

\comment{$A_{i}^T b_{\text{old}}$ has not been computed yet, compute it now}

$c_i \gets \beta A_i^T b_{\text{old}}$ \complexitytwo{dot}{n}{reduce}{1}

$e_{\omega} = - c_{\omega}$ \complexity{axpy}{m_k}

$c_{\omega} \gets \beta A_{\omega}^T b_{\text{new}}$ \complexitytwo{gemv}{n,m_k}{reduce}{m_k}

$e_{\omega} \gets e_{\omega} + c_{\omega}$ \complexity{axpy}{m_k}

\end{algorithm2e}

\begin{algorithm2e}[H]\label{alg:cdbfgs-update-solve}%
\caption{{\tt CompactDense::B.update\_solve}$(x, g)$}
    $m, k, B_0, S, Y, d, Q, R, Z, g^{\text{prev}}, a^{\text{cache}}, b^{\text{cache}} = {\tt B}.(\dots)$\;

    $(\alpha, \tilde Q) \gets {\tt update\_Q}(m, k, B_0, Y, Q)$\comment{equivalent to \cref{alg:update-P0}, $\alpha \tilde Q = Q$}

    $i \gets (k-1) \bmod m$ \comment{newest history index}
    
    $\omega \gets \{ \ell\bmod m: \ell \in \iota(m,k)\}$\comment{active history indices}
    
    $\omega_{-} \gets \{ \ell\bmod m: \ell \in \iota(m-1,k-1)\}$\comment{active indices except the newest}
    
    {\tt update\_Atb}$(m, k, 1, S, g^{\text{prev}}, g, a^{\text{cache}}, R_{:,i})$
    \comment{$R$ and $a^{\text{cache}}$ are up to date}

    \If{$B_0$ did not change}{%
        {\tt update\_Atb}$(m, k, \alpha, \tilde Q, g^{\text{prev}}, g, b^{\text{cache}}, Z_{:,i})$\;
    }%
    
    \ElseIf{$B_0$ updated by a scalar factor $H_{0,k} \gets \gamma H_{0,k-1}$}{%
    
      $Z_{\omega_{-},\omega_{-}} \gets \gamma Z_{\omega_{-},\omega_{-}}$ \complexity{scal}{(m_k - 1)^2}
      
      $b^{\text{cache}}_{\omega_{-}} \gets \gamma b^{\text{cache}}_{\omega_{-}}$ \complexity{scal}{m_k - 1}
      
      {\tt update\_Atb}$(m, k, \beta, \tilde Q, g^{\text{prev}}, g, b^{\text{cache}}, Z_{:,i})$\;

    }%
    \Else{%
        
      $Q_{\omega} \gets B_0 Y_{\omega}$ \complexity{$m_k B_0$.mult}{}

      $Z_{\omega,\omega} \gets Q_{\omega}^T Y_{\omega}$ \complexitytwo{gemm}{m_k,m_k, n}{reduce}{m_k^2}
      
      $b^{\text{cache}}_{\omega} \gets Q_{\omega}^T g^{\text{prev}}$ \complexitytwo{gemv}{n, m_k}{reduce}{m_k}
        
    }%

    $Z_{i,\omega_{-}} \gets Z_{\omega_{-}, i}$\complexity{copy}{\tilde m - 1}
\end{algorithm2e}

\begin{algorithm2e}[h]\label{alg:cdbfgs-update-mult}%
\caption{{\tt CompactDense::B.update\_mult}$(x, g)$}

    $m, k, B_0, S, Y, d, \PO, L, \FO, J = {\tt B}.(\dots)\;$

    $M\in\mathbb{R}^{m_k \times m_k}$ \comment{temporary work matrix}
    
    $i \gets (k-1) \bmod m$\;
    
    $\omega \gets \omega(m, k)$\comment{active history indices}
    
    $\omega_{-} \gets \{ \ell\bmod m: \ell \in \iota(m-1,k-1)\}$\comment{active indices except the newest}
    
    $(\alpha, \tilde P) \gets {\tt update\_P0}(m, k, B_0, S, \PO)$\comment{\cref{alg:update-P0}}

    $L_{i,\omega_{-}} \gets S_i^T Y_{\omega_{-}}$\complexitytwo{gemv}{n, m_k - 1}{reduce}{m_k - 1}
    
    $L_{\omega_{-},i} \gets 0$\comment{zeroing this column ensures $L=\hstril_k(S^T Y)$}

    \If{$B_0$ did not change}{%
        $F_{\omega,i} \gets \alpha \tilde P_{\omega}^T S_i$\complexitytwo{gemv}{n, m_k}{reduce}{m_k}
        
        $F_{i,\omega_{-}} \gets F_{\omega_{-}, i}$\complexity{copy}{m_k - 1}
    }
    \ElseIf{$B_0$ updated by a scalar factor $B_{0,k} \gets \gamma B_{0,k-1}$}{
        $F_{\omega_{-},\omega_{-}} \gets \gamma F_{\omega_{-}, \omega_{-}}$\complexity{scal}{(m_k - 1)^2}
        
        $F_{\omega,i} \gets \alpha \tilde P_{\omega}^T S_i$\complexitytwo{gemv}{n, m_k}{reduce}{m_k}
        
        $F_{i,\omega_{-}} \gets F_{\omega_{-}, i}$\complexity{copy}{m_k - 1}
    }
    \Else{
        $F_{\omega,\omega} \gets \alpha \tilde P_{\omega}^T S_{\omega}$\complexitytwo{gemm}{m_k, m_k, m_k}{reduce}{m_k^2}
    }

    $M \gets L_{\omega,\omega} \diag(d_\omega)^{-1} L_{\omega,\omega}^T$\complexitytwo{$m_k$\,diags}{m_k}{gemm}{m_k,m_k,m_k}
    
    $J_{\omega,\omega} \gets {\tt Cholesky}(M + \FO_{\omega,\omega})$ \complexitytwo{axpy}{m_k^2}{potrf}{m_k}
\end{algorithm2e}

\section{Intermediate dense algorithms}

In this appendix section, we present auxiliary algorithms for intermediate dense algorithm.

\begin{algorithm2e}[h]\label{alg:idbfgs-update-solve}%
\caption{{\tt IntermediateDense::B.update\_solve($g$)}}

    $m, k, S, Y, R, d, g^{\text{prev}}, a^{\text{cache}} = {\tt B}.(\dots)\;$
    
    $i \gets k \bmod m$\;

    $l \gets \max\{m-1,k-1\}$\;
    
    \tcp{store $S^T g$ as part of computing $r = S^T y = S^T (g - g^{\text{prev}})$}

    {\tt update\_Atb}$(m, k, 1, S, g^{\text{prev}}, g, a^{\text{cache}}, R_{:,i})$\;
    
    \tcp{\textrm{\textit{the upper triangle of $R$ with respect to history index ($R_{(i \bmod m, j \bmod m)}, j \geq i$) now contains current values of $S^T Y$}}}
\end{algorithm2e}

%
%
\end{document}

%% file: figure/ordering.tex
\begin{tikzpicture}[
        >=latex,line width=1pt,
        Brace/.style={decorate,decoration={brace,raise=-7pt}}]

    \matrix[
        matrix of nodes,
        text height=3.35ex,
        text depth=0.25ex,
        text width=2ex,
        align=center,
        left delimiter={[},
        right delimiter={]},
        column sep=1pt,
        row sep=5pt,
        nodes in empty cells,
    ] at (0,0) (M){ 
    &   &   &   &       \\
    &   &   &   &       \\
    &   &   &   &       \\
    };
    \draw[thick,fill=blue!30,draw] (M-1-1.north west) rectangle (M-3-2.south east);
    \draw[thick,fill=red!30,draw](M-1-3.north west) rectangle (M-3-5.south east);
    \draw (M-2-4) node {$\cdots$\strut};
    \draw (M-2-1)++(8pt,0pt) node {$\cdots$\strut};
    \draw (M-3-4) node {\strut old};
    \draw (M-3-1)++(8pt,0pt) node {\strut new};
    
    \path (M-3-3.south west)++(4pt,0pt) node [coordinate] (colcorner1) {};
    \draw[thick] (M-1-3.north west) rectangle (colcorner1);
    
    \path (M-3-5.south east)++(-4pt,0pt) node [coordinate] (colcorner3) {};
    \draw[thick] (M-1-5.north east) rectangle (colcorner3);
    
    \path (M-3-2.south east)++(-4pt,0pt) node [coordinate] (colcorner2) {};
    \draw[thick] (M-1-2.north east) rectangle (colcorner2);
    
    \path (M-3-1.south west)++(4pt,0pt) node [coordinate] (colcorner4) {};
    \draw[thick] (M-1-1.north west) rectangle (colcorner4);
    
    \path (M-1-3.north west)++(2pt,0pt) node [coordinate] (coltop1) {};
    \path (coltop1) node [above=0.25cm,coordinate] (coltop1label) {};
    \path (coltop1label) node [anchor=south west,xshift=-0.5em,yshift=-0.1cm] {$S_{k\bmod m} = s_{k - m}$\strut};
    \draw [-stealth,thick] (coltop1label) -- (coltop1);
    
    \path (M-3-2.south east)++(-2pt,0pt) node [coordinate] (colbot2) {};
    \path (colbot2) node [below=0.25cm,coordinate] (colbot2label) {};
    \path (colbot2label) node [anchor=north east,xshift=0.5em,yshift=0.1cm] {$S_{(k-1)\bmod m} = s_{k - 1}$\strut};
    \draw [-stealth,thick] (colbot2label) -- (colbot2);

    \draw [thick,decoration={brace,mirror,raise=1cm},decorate] 
    (M-3-1.south west) -- (M-3-5.south east)
    node [pos=0.5,anchor=north,yshift=-1.1cm] {$S$}; 

%
%

    \matrix[
        matrix of nodes,
        text height=2ex,
        text depth=0.0ex,
        text width=2ex,
        align=center,
        left delimiter={[},
        right delimiter={]},
        column sep=0.1pt,
        row sep=0.1pt,
        nodes in empty cells,
    ] at (6,0) (M3){ 
    &   &   &   &       \\
    &   &   &   &       \\
    &   &   &   &       \\
    &   &   &   &       \\
    &   &   &   &       \\
    };
    \draw[thick,fill=red!30,draw] (M3-3-3.north west)
    -- (M3-3-5.north east)
    -- (M3-5-5.south east)
    -- cycle;
    \draw[thick,fill=yellow!30,draw](M3-3-1.north west)
    -- (M3-3-2.north east)
    -- (M3-5-2.south east)    
    -- (M3-5-1.south west)
    -- cycle;
    \draw[thick,fill=blue!30,draw](M3-1-1.north west)
    -- (M3-1-2.north east)
    -- (M3-2-2.south east)
    -- cycle;
    \draw (M3-3-4.north) node[above,anchor=south] {(3) {\tt trsv}};
    \path (M3-5-2.south west) node [below,anchor=north] {(2) {\tt gemv}};    
    \path (M3-1-2.north west) node[above,anchor=south] {(1) {\tt trsv}};

    \draw [thick,decoration={brace,mirror,raise=1cm},decorate] 
    (M3-5-1.south west) -- (M3-5-5.south east)
    node [pos=0.5,anchor=north,yshift=-1.1cm] {$\htriu_k(R)$}; 
    
\end{tikzpicture}

%% file: figure/polaris-solve-host-bfgs-condiag-user.tex
\begin{tikzpicture}
    \begin{semilogxaxis}[
        ymin=0,
        xlabel={\strut$T_{\text{step}}$ (s)},
        ylabel={$B_{\text{e}}$ (GB/s)},
        legend pos=outer north east,
        extra y ticks={10.75,21.5},
        extra y tick labels={{},{}},
        extra y tick style={grid=major,major tick length=0pt},
        ylabel style={anchor=east,rotate=-90,at={(0,1)}},
        xlabel style={anchor=north west,at={(1,0)}},
        ]

        \addlegendimage{empty legend}
        \addlegendentry{\textbf{method}}
        \addlegendimage{histseries,rbfgs,host,hist50}
        \addlegendentry{recursive}
        
        \addlegendimage{histseries,cd,host,hist50}
        \addlegendentry{comp.\ dense}


        \addlegendimage{histseries,ddense,host,hist50}
        \addlegendentry{dense}

        \addlegendimage{empty legend}
        \addlegendentry{\textbf{history size $m$}}
        
        \addlegendimage{histseries,host,hist5}
        \addlegendentry{5}
        
        \addlegendimage{histseries,host,hist10}
        \addlegendentry{10}
        
        \addlegendimage{histseries,host,hist20}
        \addlegendentry{20}
        
        \addlegendimage{histseries,host,hist50}
        \addlegendentry{50}
        
            \addplot[varseries,cd] table [ x=\xvar, y=\yvar, restrict expr to domain={\thisrow{num_variables}}{100:100}, unbounded coords=discard ] {\polarisXhostXbfgsXcdXcondiagXuserXdim};
            \addplot[varseries,cd] table [ x=\xvar, y=\yvar, restrict expr to domain={\thisrow{num_variables}}{1000:1000}, unbounded coords=discard ] {\polarisXhostXbfgsXcdXcondiagXuserXdim};
            \addplot[varseries,cd] table [ x=\xvar, y=\yvar, restrict expr to domain={\thisrow{num_variables}}{10000:10000}, unbounded coords=discard ] {\polarisXhostXbfgsXcdXcondiagXuserXdim};
            \addplot[varseries,cd] table [ x=\xvar, y=\yvar, restrict expr to domain={\thisrow{num_variables}}{100000:100000}, unbounded coords=discard ] {\polarisXhostXbfgsXcdXcondiagXuserXdim};
            \addplot[varseries,cd] table [ x=\xvar, y=\yvar, restrict expr to domain={\thisrow{num_variables}}{1000000:1000000}, unbounded coords=discard ] {\polarisXhostXbfgsXcdXcondiagXuserXdim};
            \addplot[varseries,cd] table [ x=\xvar, y=\yvar, restrict expr to domain={\thisrow{num_variables}}{10000000:10000000}, unbounded coords=discard ] {\polarisXhostXbfgsXcdXcondiagXuserXdim};

            \addplot[varseries,rbfgs] table [ x=\xvar, y=\yvar, restrict expr to domain={\thisrow{num_variables}}{100:100}, unbounded coords=discard ] {\polarisXhostXbfgsXrecXcondiagXuserXdim};
            \addplot[varseries,rbfgs] table [ x=\xvar, y=\yvar, restrict expr to domain={\thisrow{num_variables}}{1000:1000}, unbounded coords=discard ] {\polarisXhostXbfgsXrecXcondiagXuserXdim};
            \addplot[varseries,rbfgs] table [ x=\xvar, y=\yvar, restrict expr to domain={\thisrow{num_variables}}{10000:10000}, unbounded coords=discard ] {\polarisXhostXbfgsXrecXcondiagXuserXdim};
            \addplot[varseries,rbfgs] table [ x=\xvar, y=\yvar, restrict expr to domain={\thisrow{num_variables}}{100000:100000}, unbounded coords=discard ] {\polarisXhostXbfgsXrecXcondiagXuserXdim};
            \addplot[varseries,rbfgs] table [ x=\xvar, y=\yvar, restrict expr to domain={\thisrow{num_variables}}{1000000:1000000}, unbounded coords=discard ] {\polarisXhostXbfgsXrecXcondiagXuserXdim};
            \addplot[varseries,rbfgs] table [ x=\xvar, y=\yvar, restrict expr to domain={\thisrow{num_variables}}{10000000:10000000}, unbounded coords=discard ] {\polarisXhostXbfgsXrecXcondiagXuserXdim};

            \addplot[varseries,ddense] table [ x=\xvar, y=\yvar, restrict expr to domain={\thisrow{num_variables}}{100:100}, unbounded coords=discard ] {\polarisXhostXdbfgsXinplaceXcondiagXuserXdim};
            \addplot[varseries,ddense] table [ x=\xvar, y=\yvar, restrict expr to domain={\thisrow{num_variables}}{1000:1000}, unbounded coords=discard ] {\polarisXhostXdbfgsXinplaceXcondiagXuserXdim};
            \addplot[varseries,ddense] table [ x=\xvar, y=\yvar, restrict expr to domain={\thisrow{num_variables}}{10000:10000}, unbounded coords=discard ] {\polarisXhostXdbfgsXinplaceXcondiagXuserXdim};
            \addplot[varseries,ddense] table [ x=\xvar, y=\yvar, restrict expr to domain={\thisrow{num_variables}}{100000:100000}, unbounded coords=discard ] {\polarisXhostXdbfgsXinplaceXcondiagXuserXdim};
            \addplot[varseries,ddense] table [ x=\xvar, y=\yvar, restrict expr to domain={\thisrow{num_variables}}{1000000:1000000}, unbounded coords=discard ] {\polarisXhostXdbfgsXinplaceXcondiagXuserXdim};
            \addplot[varseries,ddense] table [ x=\xvar, y=\yvar, restrict expr to domain={\thisrow{num_variables}}{10000000:10000000}, unbounded coords=discard ] {\polarisXhostXdbfgsXinplaceXcondiagXuserXdim};

        \addplot[histseries,ddense,host,hist5] table [ x=\xvar, y=\yvar, restrict expr to domain={\thisrow{history size}}{5:5}, unbounded coords=discard ] {\polarisXhostXdbfgsXinplaceXcondiagXuserXdim};
        \addplot[histseries,ddense,host,hist10] table [ x=\xvar, y=\yvar, restrict expr to domain={\thisrow{history size}}{10:10}, unbounded coords=discard ] {\polarisXhostXdbfgsXinplaceXcondiagXuserXdim};
        \addplot[histseries,ddense,host,hist20] table [ x=\xvar, y=\yvar, restrict expr to domain={\thisrow{history size}}{20:20}, unbounded coords=discard ] {\polarisXhostXdbfgsXinplaceXcondiagXuserXdim};
        \addplot[histseries,ddense,host,hist50] table [ x=\xvar, y=\yvar, restrict expr to domain={\thisrow{history size}}{50:50}, unbounded coords=discard ] {\polarisXhostXdbfgsXinplaceXcondiagXuserXdim};

        \addplot[histseries,rbfgs,host,hist5] table [ x=\xvar, y=\yvar, restrict expr to domain={\thisrow{history size}}{5:5}, unbounded coords=discard ] {\polarisXhostXbfgsXrecXcondiagXuserXdim};
        \addplot[histseries,rbfgs,host,hist10] table [ x=\xvar, y=\yvar, restrict expr to domain={\thisrow{history size}}{10:10}, unbounded coords=discard ] {\polarisXhostXbfgsXrecXcondiagXuserXdim};
        \addplot[histseries,rbfgs,host,hist20] table [ x=\xvar, y=\yvar, restrict expr to domain={\thisrow{history size}}{20:20}, unbounded coords=discard ] {\polarisXhostXbfgsXrecXcondiagXuserXdim};
        \addplot[histseries,rbfgs,host,hist50] table [ x=\xvar, y=\yvar, restrict expr to domain={\thisrow{history size}}{50:50}, unbounded coords=discard ] {\polarisXhostXbfgsXrecXcondiagXuserXdim};

        \addplot[histseries,cd,host,hist5] table [ x=\xvar, y=\yvar, restrict expr to domain={\thisrow{history size}}{5:5}, unbounded coords=discard ] {\polarisXhostXbfgsXcdXcondiagXuserXdim};
        \addplot[histseries,cd,host,hist10] table [ x=\xvar, y=\yvar, restrict expr to domain={\thisrow{history size}}{10:10}, unbounded coords=discard ] {\polarisXhostXbfgsXcdXcondiagXuserXdim};
        \addplot[histseries,cd,host,hist20] table [ x=\xvar, y=\yvar, restrict expr to domain={\thisrow{history size}}{20:20}, unbounded coords=discard ] {\polarisXhostXbfgsXcdXcondiagXuserXdim};
        \addplot[histseries,cd,host,hist50] table [ x=\xvar, y=\yvar, restrict expr to domain={\thisrow{history size}}{50:50}, unbounded coords=discard ] {\polarisXhostXbfgsXcdXcondiagXuserXdim};

        \path (axis cs:0.5e-4,4) node [anchor=west] {\small$n=10^2$};
        \path (axis cs:1e-4,11) node [anchor=center] {\small$10^3$};
        \path (axis cs:2.5e-4,23) node [anchor=center] {\small$10^4$};
        \path (axis cs:2e-3,19) node [anchor=center] {\small$10^5$};
        \path (axis cs:4e-2,10) node [anchor=center] {\small$10^6$};
        \path (axis cs:5e-1,8) node [anchor=north] {\small$10^7$};

        \node (halfBsy) at (axis cs:1,21.5) {};
        \node (quarterBsy) at (axis cs:1,10.75) {};
        \node (rightaxis) at (rel axis cs:1,0) {};
        \path (halfBsy -| rightaxis) node [anchor=south east,inner sep=1pt] {\small$\tfrac{1}{2}B_{\text{S}}$};
        \path (quarterBsy -| rightaxis) node [anchor=south east,inner sep=1pt] {\small$\tfrac{1}{4}B_{\text{S}}$};
        
        
    \end{semilogxaxis}
\end{tikzpicture}%

%% file: figure/polaris-solve-host-bfgs-mpifour-condiag-user.tex
\begin{tikzpicture}
    \begin{semilogxaxis}[
        ymin=0,
        ymax=80,
        legend pos=outer north east,
        extra y ticks={45,90},
        extra y tick labels={{},{}},
        extra y tick style={grid=major,major tick length=0pt},
        xlabel={\strut$T_{\text{step}}$ (s)},
        ylabel={$B_{\text{e}}$ (GB/s)},
        ylabel style={anchor=east,rotate=-90,at={(0,0.9)}},
        xlabel style={anchor=north west,at={(1,0)}},
        ]

        \addlegendimage{empty legend}
        \addlegendentry{\textbf{method}}
        \addlegendimage{histseries,rbfgs,host,hist50}
        \addlegendentry{recursive}
        
        \addlegendimage{histseries,cd,host,hist50}
        \addlegendentry{comp.\ dense}

        \addlegendimage{histseries,ddense,host,hist50}
        \addlegendentry{dense}

        \addlegendimage{empty legend}
        \addlegendentry{\textbf{history size $m$}}
        
        \addlegendimage{histseries,host,hist5}
        \addlegendentry{5}
        
        \addlegendimage{histseries,host,hist10}
        \addlegendentry{10}
        
        \addlegendimage{histseries,host,hist20}
        \addlegendentry{20}
        
        \addlegendimage{histseries,host,hist50}
        \addlegendentry{50}
        
            \addplot[varseries,cd] table [ x=\xvar, y=\yvar, restrict expr to domain={\thisrow{num_variables}}{1000:1000}, unbounded coords=discard ] {\polarisXmpifourXhostXbfgsXcdXcondiagXuserXdim};
            \addplot[varseries,cd] table [ x=\xvar, y=\yvar, restrict expr to domain={\thisrow{num_variables}}{10000:10000}, unbounded coords=discard ] {\polarisXmpifourXhostXbfgsXcdXcondiagXuserXdim};
            \addplot[varseries,cd] table [ x=\xvar, y=\yvar, restrict expr to domain={\thisrow{num_variables}}{100000:100000}, unbounded coords=discard ] {\polarisXmpifourXhostXbfgsXcdXcondiagXuserXdim};
            \addplot[varseries,cd] table [ x=\xvar, y=\yvar, restrict expr to domain={\thisrow{num_variables}}{1000000:1000000}, unbounded coords=discard ] {\polarisXmpifourXhostXbfgsXcdXcondiagXuserXdim};
            \addplot[varseries,cd] table [ x=\xvar, y=\yvar, restrict expr to domain={\thisrow{num_variables}}{10000000:10000000}, unbounded coords=discard ] {\polarisXmpifourXhostXbfgsXcdXcondiagXuserXdim};

            \addplot[varseries,rbfgs] table [ x=\xvar, y=\yvar, restrict expr to domain={\thisrow{num_variables}}{1000:1000}, unbounded coords=discard ] {\polarisXmpifourXhostXbfgsXrecXcondiagXuserXdim};
            \addplot[varseries,rbfgs] table [ x=\xvar, y=\yvar, restrict expr to domain={\thisrow{num_variables}}{10000:10000}, unbounded coords=discard ] {\polarisXmpifourXhostXbfgsXrecXcondiagXuserXdim};
            \addplot[varseries,rbfgs] table [ x=\xvar, y=\yvar, restrict expr to domain={\thisrow{num_variables}}{100000:100000}, unbounded coords=discard ] {\polarisXmpifourXhostXbfgsXrecXcondiagXuserXdim};
            \addplot[varseries,rbfgs] table [ x=\xvar, y=\yvar, restrict expr to domain={\thisrow{num_variables}}{1000000:1000000}, unbounded coords=discard ] {\polarisXmpifourXhostXbfgsXrecXcondiagXuserXdim};
            \addplot[varseries,rbfgs] table [ x=\xvar, y=\yvar, restrict expr to domain={\thisrow{num_variables}}{10000000:10000000}, unbounded coords=discard ] {\polarisXmpifourXhostXbfgsXrecXcondiagXuserXdim};

            \addplot[varseries,ddense] table [ x=\xvar, y=\yvar, restrict expr to domain={\thisrow{num_variables}}{1000:1000}, unbounded coords=discard ] {\polarisXmpifourXhostXdbfgsXinplaceXcondiagXuserXdim};
            \addplot[varseries,ddense] table [ x=\xvar, y=\yvar, restrict expr to domain={\thisrow{num_variables}}{10000:10000}, unbounded coords=discard ] {\polarisXmpifourXhostXdbfgsXinplaceXcondiagXuserXdim};
            \addplot[varseries,ddense] table [ x=\xvar, y=\yvar, restrict expr to domain={\thisrow{num_variables}}{100000:100000}, unbounded coords=discard ] {\polarisXmpifourXhostXdbfgsXinplaceXcondiagXuserXdim};
            \addplot[varseries,ddense] table [ x=\xvar, y=\yvar, restrict expr to domain={\thisrow{num_variables}}{1000000:1000000}, unbounded coords=discard ] {\polarisXmpifourXhostXdbfgsXinplaceXcondiagXuserXdim};
            \addplot[varseries,ddense] table [ x=\xvar, y=\yvar, restrict expr to domain={\thisrow{num_variables}}{10000000:10000000}, unbounded coords=discard ] {\polarisXmpifourXhostXdbfgsXinplaceXcondiagXuserXdim};

        \addplot[histseries,rbfgs,host,hist5] table [ x=\xvar, y=\yvar, restrict expr to domain={\thisrow{history size}}{5:5}, restrict expr to domain={\thisrow{num_variables}}{1000:10000000}, unbounded coords=discard ] {\polarisXmpifourXhostXbfgsXrecXcondiagXuserXdim};
        \addplot[histseries,rbfgs,host,hist10] table [ x=\xvar, y=\yvar, restrict expr to domain={\thisrow{history size}}{10:10}, restrict expr to domain={\thisrow{num_variables}}{1000:10000000},  unbounded coords=discard ] {\polarisXmpifourXhostXbfgsXrecXcondiagXuserXdim};
        \addplot[histseries,rbfgs,host,hist20] table [ x=\xvar, y=\yvar, restrict expr to domain={\thisrow{history size}}{20:20}, restrict expr to domain={\thisrow{num_variables}}{1000:10000000},  unbounded coords=discard ] {\polarisXmpifourXhostXbfgsXrecXcondiagXuserXdim};
        \addplot[histseries,rbfgs,host,hist50] table [ x=\xvar, y=\yvar, restrict expr to domain={\thisrow{history size}}{50:50}, restrict expr to domain={\thisrow{num_variables}}{1000:10000000},  unbounded coords=discard ] {\polarisXmpifourXhostXbfgsXrecXcondiagXuserXdim};

        \addplot[histseries,ddense,host,hist5] table [ x=\xvar, y=\yvar, restrict expr to domain={\thisrow{history size}}{5:5}, restrict expr to domain={\thisrow{num_variables}}{1000:10000000},  unbounded coords=discard ] {\polarisXmpifourXhostXdbfgsXinplaceXcondiagXuserXdim};
        \addplot[histseries,ddense,host,hist10] table [ x=\xvar, y=\yvar, restrict expr to domain={\thisrow{history size}}{10:10}, restrict expr to domain={\thisrow{num_variables}}{1000:10000000},  unbounded coords=discard ] {\polarisXmpifourXhostXdbfgsXinplaceXcondiagXuserXdim};
        \addplot[histseries,ddense,host,hist20] table [ x=\xvar, y=\yvar, restrict expr to domain={\thisrow{history size}}{20:20}, restrict expr to domain={\thisrow{num_variables}}{1000:10000000},  unbounded coords=discard ] {\polarisXmpifourXhostXdbfgsXinplaceXcondiagXuserXdim};
        \addplot[histseries,ddense,host,hist50] table [ x=\xvar, y=\yvar, restrict expr to domain={\thisrow{history size}}{50:50}, restrict expr to domain={\thisrow{num_variables}}{1000:10000000},  unbounded coords=discard ] {\polarisXmpifourXhostXdbfgsXinplaceXcondiagXuserXdim};

        \addplot[histseries,cd,host,hist5] table [ x=\xvar, y=\yvar, restrict expr to domain={\thisrow{history size}}{5:5}, restrict expr to domain={\thisrow{num_variables}}{1000:10000000},  unbounded coords=discard ] {\polarisXmpifourXhostXbfgsXcdXcondiagXuserXdim};
        \addplot[histseries,cd,host,hist10] table [ x=\xvar, y=\yvar, restrict expr to domain={\thisrow{history size}}{10:10}, restrict expr to domain={\thisrow{num_variables}}{1000:10000000},  unbounded coords=discard ] {\polarisXmpifourXhostXbfgsXcdXcondiagXuserXdim};
        \addplot[histseries,cd,host,hist20] table [ x=\xvar, y=\yvar, restrict expr to domain={\thisrow{history size}}{20:20}, restrict expr to domain={\thisrow{num_variables}}{1000:10000000},  unbounded coords=discard ] {\polarisXmpifourXhostXbfgsXcdXcondiagXuserXdim};
        \addplot[histseries,cd,host,hist50] table [ x=\xvar, y=\yvar, restrict expr to domain={\thisrow{history size}}{50:50}, restrict expr to domain={\thisrow{num_variables}}{1000:10000000},  unbounded coords=discard ] {\polarisXmpifourXhostXbfgsXcdXcondiagXuserXdim};

        \path (axis cs:2.2e-5,14) node [anchor=center] {\small$10^3$};
        \path (axis cs:2.1e-4,30) node [anchor=center] {\small$10^4$};
        \path (axis cs:1.5e-3,50) node [anchor=center] {\small$10^5$};
        \path (axis cs:3.5e-2,25) node [anchor=center] {\small$n=10^6$};
        \path (axis cs:5e-1,8) node [anchor=north] {\small$10^7$};

        \node (halfBsy) at (axis cs:1,90) {};
        \node (quarterBsy) at (axis cs:1,45) {};
        \node (rightaxis) at (rel axis cs:1,0) {};
        \path (halfBsy -| rightaxis) node [anchor=south east,inner sep=1pt] {\small$\tfrac{1}{2}B_{\text{S}}$};
        \path (quarterBsy -| rightaxis) node [anchor=south east,inner sep=1pt] {\small$\tfrac{1}{4}B_{\text{S}}$};
    \end{semilogxaxis}
\end{tikzpicture}%

%% file: figure/polaris-solve-host-bfgs-diag-diag.tex
\begin{tikzpicture}
    \begin{semilogxaxis}[
        ymin=0,
        xlabel={\strut$T_{\text{step}}$ (s)},
        ylabel={$B_{\text{e}}$ (GB/s)},
        legend pos=outer north east,
        extra y ticks={10.75,21.5},
        extra y tick labels={{},{}},
        extra y tick style={grid=major,major tick length=0pt},
        ylabel style={anchor=east,rotate=-90,at={(0,1)}},
        xlabel style={anchor=north west,at={(1,0)}},
        ]

        \addlegendimage{empty legend}
        \addlegendentry{\textbf{method}}
        \addlegendimage{histseries,rbfgs,host,hist50}
        \addlegendentry{recursive}
        
        \addlegendimage{histseries,cd,host,hist50}
        \addlegendentry{comp.\ dense}

        \addlegendimage{histseries,ddense,host,hist50}
        \addlegendentry{dense}

        \addlegendimage{empty legend}
        \addlegendentry{\textbf{history size $m$}}
        
        \addlegendimage{histseries,host,hist5}
        \addlegendentry{5}
        
        \addlegendimage{histseries,host,hist10}
        \addlegendentry{10}
        
        \addlegendimage{histseries,host,hist20}
        \addlegendentry{20}
        
        \addlegendimage{histseries,host,hist50}
        \addlegendentry{50}
        
            \addplot[varseries,cd] table [ x=\xvar, y=\yvar, restrict expr to domain={\thisrow{num_variables}}{100:100}, unbounded coords=discard ] {\polarisXhostXbfgsXcdXdiagXdiagXdim};
            \addplot[varseries,cd] table [ x=\xvar, y=\yvar, restrict expr to domain={\thisrow{num_variables}}{1000:1000}, unbounded coords=discard ] {\polarisXhostXbfgsXcdXdiagXdiagXdim};
            \addplot[varseries,cd] table [ x=\xvar, y=\yvar, restrict expr to domain={\thisrow{num_variables}}{10000:10000}, unbounded coords=discard ] {\polarisXhostXbfgsXcdXdiagXdiagXdim};
            \addplot[varseries,cd] table [ x=\xvar, y=\yvar, restrict expr to domain={\thisrow{num_variables}}{100000:100000}, unbounded coords=discard ] {\polarisXhostXbfgsXcdXdiagXdiagXdim};
            \addplot[varseries,cd] table [ x=\xvar, y=\yvar, restrict expr to domain={\thisrow{num_variables}}{1000000:1000000}, unbounded coords=discard ] {\polarisXhostXbfgsXcdXdiagXdiagXdim};
            \addplot[varseries,cd] table [ x=\xvar, y=\yvar, restrict expr to domain={\thisrow{num_variables}}{10000000:10000000}, unbounded coords=discard ] {\polarisXhostXbfgsXcdXdiagXdiagXdim};

            \addplot[varseries,rbfgs] table [ x=\xvar, y=\yvar, restrict expr to domain={\thisrow{num_variables}}{100:100}, unbounded coords=discard ] {\polarisXhostXbfgsXrecXdiagXdiagXdim};
            \addplot[varseries,rbfgs] table [ x=\xvar, y=\yvar, restrict expr to domain={\thisrow{num_variables}}{1000:1000}, unbounded coords=discard ] {\polarisXhostXbfgsXrecXdiagXdiagXdim};
            \addplot[varseries,rbfgs] table [ x=\xvar, y=\yvar, restrict expr to domain={\thisrow{num_variables}}{10000:10000}, unbounded coords=discard ] {\polarisXhostXbfgsXrecXdiagXdiagXdim};
            \addplot[varseries,rbfgs] table [ x=\xvar, y=\yvar, restrict expr to domain={\thisrow{num_variables}}{100000:100000}, unbounded coords=discard ] {\polarisXhostXbfgsXrecXdiagXdiagXdim};
            \addplot[varseries,rbfgs] table [ x=\xvar, y=\yvar, restrict expr to domain={\thisrow{num_variables}}{1000000:1000000}, unbounded coords=discard ] {\polarisXhostXbfgsXrecXdiagXdiagXdim};
            \addplot[varseries,rbfgs] table [ x=\xvar, y=\yvar, restrict expr to domain={\thisrow{num_variables}}{10000000:10000000}, unbounded coords=discard ] {\polarisXhostXbfgsXrecXdiagXdiagXdim};

            \addplot[varseries,ddense] table [ x=\xvar, y=\yvar, restrict expr to domain={\thisrow{num_variables}}{100:100}, unbounded coords=discard ] {\polarisXhostXdbfgsXinplaceXdiagXdiagXdim};
            \addplot[varseries,ddense] table [ x=\xvar, y=\yvar, restrict expr to domain={\thisrow{num_variables}}{1000:1000}, unbounded coords=discard ] {\polarisXhostXdbfgsXinplaceXdiagXdiagXdim};
            \addplot[varseries,ddense] table [ x=\xvar, y=\yvar, restrict expr to domain={\thisrow{num_variables}}{10000:10000}, unbounded coords=discard ] {\polarisXhostXdbfgsXinplaceXdiagXdiagXdim};
            \addplot[varseries,ddense] table [ x=\xvar, y=\yvar, restrict expr to domain={\thisrow{num_variables}}{100000:100000}, unbounded coords=discard ] {\polarisXhostXdbfgsXinplaceXdiagXdiagXdim};
            \addplot[varseries,ddense] table [ x=\xvar, y=\yvar, restrict expr to domain={\thisrow{num_variables}}{1000000:1000000}, unbounded coords=discard ] {\polarisXhostXdbfgsXinplaceXdiagXdiagXdim};
            \addplot[varseries,ddense] table [ x=\xvar, y=\yvar, restrict expr to domain={\thisrow{num_variables}}{10000000:10000000}, unbounded coords=discard ] {\polarisXhostXdbfgsXinplaceXdiagXdiagXdim};

        \addplot[histseries,ddense,host,hist5] table [ x=\xvar, y=\yvar, restrict expr to domain={\thisrow{history size}}{5:5}, unbounded coords=discard ] {\polarisXhostXdbfgsXinplaceXdiagXdiagXdim};
        \addplot[histseries,ddense,host,hist10] table [ x=\xvar, y=\yvar, restrict expr to domain={\thisrow{history size}}{10:10}, unbounded coords=discard ] {\polarisXhostXdbfgsXinplaceXdiagXdiagXdim};
        \addplot[histseries,ddense,host,hist20] table [ x=\xvar, y=\yvar, restrict expr to domain={\thisrow{history size}}{20:20}, unbounded coords=discard ] {\polarisXhostXdbfgsXinplaceXdiagXdiagXdim};
        \addplot[histseries,ddense,host,hist50] table [ x=\xvar, y=\yvar, restrict expr to domain={\thisrow{history size}}{50:50}, unbounded coords=discard ] {\polarisXhostXdbfgsXinplaceXdiagXdiagXdim};

        \addplot[histseries,rbfgs,host,hist5] table [ x=\xvar, y=\yvar, restrict expr to domain={\thisrow{history size}}{5:5}, unbounded coords=discard ] {\polarisXhostXbfgsXrecXdiagXdiagXdim};
        \addplot[histseries,rbfgs,host,hist10] table [ x=\xvar, y=\yvar, restrict expr to domain={\thisrow{history size}}{10:10}, unbounded coords=discard ] {\polarisXhostXbfgsXrecXdiagXdiagXdim};
        \addplot[histseries,rbfgs,host,hist20] table [ x=\xvar, y=\yvar, restrict expr to domain={\thisrow{history size}}{20:20}, unbounded coords=discard ] {\polarisXhostXbfgsXrecXdiagXdiagXdim};
        \addplot[histseries,rbfgs,host,hist50] table [ x=\xvar, y=\yvar, restrict expr to domain={\thisrow{history size}}{50:50}, unbounded coords=discard ] {\polarisXhostXbfgsXrecXdiagXdiagXdim};

        \addplot[histseries,cd,host,hist5] table [ x=\xvar, y=\yvar, restrict expr to domain={\thisrow{history size}}{5:5}, unbounded coords=discard ] {\polarisXhostXbfgsXcdXdiagXdiagXdim};
        \addplot[histseries,cd,host,hist10] table [ x=\xvar, y=\yvar, restrict expr to domain={\thisrow{history size}}{10:10}, unbounded coords=discard ] {\polarisXhostXbfgsXcdXdiagXdiagXdim};
        \addplot[histseries,cd,host,hist20] table [ x=\xvar, y=\yvar, restrict expr to domain={\thisrow{history size}}{20:20}, unbounded coords=discard ] {\polarisXhostXbfgsXcdXdiagXdiagXdim};
        \addplot[histseries,cd,host,hist50] table [ x=\xvar, y=\yvar, restrict expr to domain={\thisrow{history size}}{50:50}, unbounded coords=discard ] {\polarisXhostXbfgsXcdXdiagXdiagXdim};

        \path (axis cs:0.2e-4,2.1) node [anchor=west] {\small$10^2$};
        \path (axis cs:2.2e-5,13) node [anchor=center] {\small$10^3$};
        \path (axis cs:5e-5,20) node [anchor=center] {\small$n=10^4$};
        \path (axis cs:3e-3,17) node [anchor=center] {\small$10^5$};
        \path (axis cs:4.7e-2,15) node [anchor=center] {\small$10^6$};
        \path (axis cs:5.3e-1,16) node [anchor=north] {\small$10^7$};

        \node (halfBsy) at (axis cs:1,21.5) {};
        \node (quarterBsy) at (axis cs:1,10.75) {};
        \node (rightaxis) at (rel axis cs:1,0) {};
        \path (halfBsy -| rightaxis) node [anchor=south east,inner sep=1pt] {\small$\tfrac{1}{2}B_{\text{S}}$};
        \path (quarterBsy -| rightaxis) node [anchor=south east,inner sep=1pt] {\small$\tfrac{1}{4}B_{\text{S}}$};
        
        
    \end{semilogxaxis}
\end{tikzpicture}%

%% file: figure/polaris-solve-host-bfgs-mpifour-diag-diag.tex
\begin{tikzpicture}
    \begin{semilogxaxis}[
        ymin=0,
        ymax=60,
        legend pos=outer north east,
        extra y ticks={45,90},
        extra y tick labels={{},{}},
        extra y tick style={grid=major,major tick length=0pt},
        xlabel={\strut$T_{\text{step}}$ (s)},
        ylabel={$B_{\text{e}}$ (GB/s)},
        ylabel style={anchor=east,rotate=-90,at={(0,0.9)}},
        xlabel style={anchor=north west,at={(1,0)}},
        ]

        \addlegendimage{empty legend}
        \addlegendentry{\textbf{method}}
        \addlegendimage{histseries,rbfgs,host,hist50}
        \addlegendentry{recursive}
        
        \addlegendimage{histseries,cd,host,hist50}
        \addlegendentry{comp.\ dense}

        \addlegendimage{histseries,ddense,host,hist50}
        \addlegendentry{dense}

        \addlegendimage{empty legend}
        \addlegendentry{\textbf{history size $m$}}
        
        \addlegendimage{histseries,host,hist5}
        \addlegendentry{5}
        
        \addlegendimage{histseries,host,hist10}
        \addlegendentry{10}
        
        \addlegendimage{histseries,host,hist20}
        \addlegendentry{20}
        
        \addlegendimage{histseries,host,hist50}
        \addlegendentry{50}

            \addplot[varseries,cd] table [ x=\xvar, y=\yvar, restrict expr to domain={\thisrow{num_variables}}{1000:1000}, unbounded coords=discard ] {\polarisXmpifourXhostXbfgsXcdXdiagXdiagXdim};
            \addplot[varseries,cd] table [ x=\xvar, y=\yvar, restrict expr to domain={\thisrow{num_variables}}{10000:10000}, unbounded coords=discard ] {\polarisXmpifourXhostXbfgsXcdXdiagXdiagXdim};
            \addplot[varseries,cd] table [ x=\xvar, y=\yvar, restrict expr to domain={\thisrow{num_variables}}{100000:100000}, unbounded coords=discard ] {\polarisXmpifourXhostXbfgsXcdXdiagXdiagXdim};
            \addplot[varseries,cd] table [ x=\xvar, y=\yvar, restrict expr to domain={\thisrow{num_variables}}{1000000:1000000}, unbounded coords=discard ] {\polarisXmpifourXhostXbfgsXcdXdiagXdiagXdim};
            \addplot[varseries,cd] table [ x=\xvar, y=\yvar, restrict expr to domain={\thisrow{num_variables}}{10000000:10000000}, unbounded coords=discard ] {\polarisXmpifourXhostXbfgsXcdXdiagXdiagXdim};

            \addplot[varseries,rbfgs] table [ x=\xvar, y=\yvar, restrict expr to domain={\thisrow{num_variables}}{1000:1000}, unbounded coords=discard ] {\polarisXmpifourXhostXbfgsXrecXdiagXdiagXdim};
            \addplot[varseries,rbfgs] table [ x=\xvar, y=\yvar, restrict expr to domain={\thisrow{num_variables}}{10000:10000}, unbounded coords=discard ] {\polarisXmpifourXhostXbfgsXrecXdiagXdiagXdim};
            \addplot[varseries,rbfgs] table [ x=\xvar, y=\yvar, restrict expr to domain={\thisrow{num_variables}}{100000:100000}, unbounded coords=discard ] {\polarisXmpifourXhostXbfgsXrecXdiagXdiagXdim};
            \addplot[varseries,rbfgs] table [ x=\xvar, y=\yvar, restrict expr to domain={\thisrow{num_variables}}{1000000:1000000}, unbounded coords=discard ] {\polarisXmpifourXhostXbfgsXrecXdiagXdiagXdim};
            \addplot[varseries,rbfgs] table [ x=\xvar, y=\yvar, restrict expr to domain={\thisrow{num_variables}}{10000000:10000000}, unbounded coords=discard ] {\polarisXmpifourXhostXbfgsXrecXdiagXdiagXdim};

            \addplot[varseries,ddense] table [ x=\xvar, y=\yvar, restrict expr to domain={\thisrow{num_variables}}{1000:1000}, unbounded coords=discard ] {\polarisXmpifourXhostXdbfgsXinplaceXdiagXdiagXdim};
            \addplot[varseries,ddense] table [ x=\xvar, y=\yvar, restrict expr to domain={\thisrow{num_variables}}{10000:10000}, unbounded coords=discard ] {\polarisXmpifourXhostXdbfgsXinplaceXdiagXdiagXdim};
            \addplot[varseries,ddense] table [ x=\xvar, y=\yvar, restrict expr to domain={\thisrow{num_variables}}{100000:100000}, unbounded coords=discard ] {\polarisXmpifourXhostXdbfgsXinplaceXdiagXdiagXdim};
            \addplot[varseries,ddense] table [ x=\xvar, y=\yvar, restrict expr to domain={\thisrow{num_variables}}{1000000:1000000}, unbounded coords=discard ] {\polarisXmpifourXhostXdbfgsXinplaceXdiagXdiagXdim};
            \addplot[varseries,ddense] table [ x=\xvar, y=\yvar, restrict expr to domain={\thisrow{num_variables}}{10000000:10000000}, unbounded coords=discard ] {\polarisXmpifourXhostXdbfgsXinplaceXdiagXdiagXdim};

        \addplot[histseries,rbfgs,host,hist5] table [ x=\xvar, y=\yvar, restrict expr to domain={\thisrow{history size}}{5:5}, restrict expr to domain={\thisrow{num_variables}}{1000:10000000}, unbounded coords=discard ] {\polarisXmpifourXhostXbfgsXrecXdiagXdiagXdim};
        \addplot[histseries,rbfgs,host,hist10] table [ x=\xvar, y=\yvar, restrict expr to domain={\thisrow{history size}}{10:10}, restrict expr to domain={\thisrow{num_variables}}{1000:10000000},  unbounded coords=discard ] {\polarisXmpifourXhostXbfgsXrecXdiagXdiagXdim};
        \addplot[histseries,rbfgs,host,hist20] table [ x=\xvar, y=\yvar, restrict expr to domain={\thisrow{history size}}{20:20}, restrict expr to domain={\thisrow{num_variables}}{1000:10000000},  unbounded coords=discard ] {\polarisXmpifourXhostXbfgsXrecXdiagXdiagXdim};
        \addplot[histseries,rbfgs,host,hist50] table [ x=\xvar, y=\yvar, restrict expr to domain={\thisrow{history size}}{50:50}, restrict expr to domain={\thisrow{num_variables}}{1000:10000000},  unbounded coords=discard ] {\polarisXmpifourXhostXbfgsXrecXdiagXdiagXdim};

        \addplot[histseries,ddense,host,hist5] table [ x=\xvar, y=\yvar, restrict expr to domain={\thisrow{history size}}{5:5}, restrict expr to domain={\thisrow{num_variables}}{1000:10000000},  unbounded coords=discard ] {\polarisXmpifourXhostXdbfgsXinplaceXdiagXdiagXdim};
        \addplot[histseries,ddense,host,hist10] table [ x=\xvar, y=\yvar, restrict expr to domain={\thisrow{history size}}{10:10}, restrict expr to domain={\thisrow{num_variables}}{1000:10000000},  unbounded coords=discard ] {\polarisXmpifourXhostXdbfgsXinplaceXdiagXdiagXdim};
        \addplot[histseries,ddense,host,hist20] table [ x=\xvar, y=\yvar, restrict expr to domain={\thisrow{history size}}{20:20}, restrict expr to domain={\thisrow{num_variables}}{1000:10000000},  unbounded coords=discard ] {\polarisXmpifourXhostXdbfgsXinplaceXdiagXdiagXdim};
        \addplot[histseries,ddense,host,hist50] table [ x=\xvar, y=\yvar, restrict expr to domain={\thisrow{history size}}{50:50}, restrict expr to domain={\thisrow{num_variables}}{1000:10000000},  unbounded coords=discard ] {\polarisXmpifourXhostXdbfgsXinplaceXdiagXdiagXdim};

        \addplot[histseries,cd,host,hist5] table [ x=\xvar, y=\yvar, restrict expr to domain={\thisrow{history size}}{5:5}, restrict expr to domain={\thisrow{num_variables}}{1000:10000000},  unbounded coords=discard ] {\polarisXmpifourXhostXbfgsXcdXdiagXdiagXdim};
        \addplot[histseries,cd,host,hist10] table [ x=\xvar, y=\yvar, restrict expr to domain={\thisrow{history size}}{10:10}, restrict expr to domain={\thisrow{num_variables}}{1000:10000000},  unbounded coords=discard ] {\polarisXmpifourXhostXbfgsXcdXdiagXdiagXdim};
        \addplot[histseries,cd,host,hist20] table [ x=\xvar, y=\yvar, restrict expr to domain={\thisrow{history size}}{20:20}, restrict expr to domain={\thisrow{num_variables}}{1000:10000000},  unbounded coords=discard ] {\polarisXmpifourXhostXbfgsXcdXdiagXdiagXdim};
        \addplot[histseries,cd,host,hist50] table [ x=\xvar, y=\yvar, restrict expr to domain={\thisrow{history size}}{50:50}, restrict expr to domain={\thisrow{num_variables}}{1000:10000000},  unbounded coords=discard ] {\polarisXmpifourXhostXbfgsXcdXdiagXdiagXdim};

        \path (axis cs:3e-5,10) node [anchor=center] {\small$10^3$};
        \path (axis cs:0.6e-4,30) node [anchor=center] {\small$10^4$};
        \path (axis cs:1.7e-3,33) node [anchor=center] {\small$10^5$};
        \path (axis cs:4e-2,23) node [anchor=center] {\small$n=10^6$};
        \path (axis cs:5e-1,22) node [anchor=north] {\small$10^7$};

        \node (halfBsy) at (axis cs:1,90) {};
        \node (quarterBsy) at (axis cs:1,45) {};
        \node (rightaxis) at (rel axis cs:1,0) {};
        \path (halfBsy -| rightaxis) node [anchor=south east,inner sep=1pt] {\small$\tfrac{1}{2}B_{\text{S}}$};
        \path (quarterBsy -| rightaxis) node [anchor=south east,inner sep=1pt] {\small$\tfrac{1}{4}B_{\text{S}}$};
    \end{semilogxaxis}
\end{tikzpicture}%

%% file: figure/polaris-mult-host-mpifour-dbfgs-bfgs.tex
\begin{tikzpicture}
    \begin{semilogxaxis}[
        ymin=0,
        xlabel={\strut$T_{\text{step}}$ (s)},
        ylabel={$B_{\text{e}}$ (GB/s)},
        legend pos=outer north east,
        extra y ticks={45,90},
        extra y tick labels={{},{}},
        extra y tick style={grid=major,major tick length=0pt},
        ylabel style={anchor=east,rotate=-90,at={(0,1)}},
        xlabel style={anchor=north west,at={(1,0)}},
        ]

        \addlegendimage{empty legend}
        \addlegendentry{\textbf{method}}
        \addlegendimage{histseries,rbfgs,host,hist50}
        \addlegendentry{Recursive}

        \addlegendimage{histseries,dense,host,hist50}
        \addlegendentry{Dense}



        \addlegendimage{empty legend}
        \addlegendentry{\textbf{history size $m$}}
        
        \addlegendimage{histseries,host,hist5}
        \addlegendentry{5}
        
        \addlegendimage{histseries,host,hist10}
        \addlegendentry{10}
        
        \addlegendimage{histseries,host,hist20}
        \addlegendentry{20}
        
        \addlegendimage{histseries,host,hist50}
        \addlegendentry{50}
        
            \addplot[varseries,rbfgs] table [ x=\xvar, y=\yvar, restrict expr to domain={\thisrow{num_variables}}{1000:1000}, unbounded coords=discard ] {\polarisXmpifourXmultXhostXbfgsXrecXdiagXuserXdim};
            \addplot[varseries,rbfgs] table [ x=\xvar, y=\yvar, restrict expr to domain={\thisrow{num_variables}}{10000:10000}, unbounded coords=discard ] {\polarisXmpifourXmultXhostXbfgsXrecXdiagXuserXdim};
            \addplot[varseries,rbfgs] table [ x=\xvar, y=\yvar, restrict expr to domain={\thisrow{num_variables}}{100000:100000}, unbounded coords=discard ] {\polarisXmpifourXmultXhostXbfgsXrecXdiagXuserXdim};
            \addplot[varseries,rbfgs] table [ x=\xvar, y=\yvar, restrict expr to domain={\thisrow{num_variables}}{1000000:1000000}, unbounded coords=discard ] {\polarisXmpifourXmultXhostXbfgsXrecXdiagXuserXdim};
            \addplot[varseries,rbfgs] table [ x=\xvar, y=\yvar, restrict expr to domain={\thisrow{num_variables}}{10000000:10000000}, unbounded coords=discard ] {\polarisXmpifourXmultXhostXbfgsXrecXdiagXuserXdim};

            \addplot[varseries,dense] table [ x=\xvar, y=\yvar, restrict expr to domain={\thisrow{num_variables}}{1000:1000}, unbounded coords=discard ] {\polarisXmpifourXmultXhostXbfgsXdenseXdiagXuserXdim};
            \addplot[varseries,dense] table [ x=\xvar, y=\yvar, restrict expr to domain={\thisrow{num_variables}}{10000:10000}, unbounded coords=discard ] {\polarisXmpifourXmultXhostXbfgsXdenseXdiagXuserXdim};
            \addplot[varseries,dense] table [ x=\xvar, y=\yvar, restrict expr to domain={\thisrow{num_variables}}{100000:100000}, unbounded coords=discard ] {\polarisXmpifourXmultXhostXbfgsXdenseXdiagXuserXdim};
            \addplot[varseries,dense] table [ x=\xvar, y=\yvar, restrict expr to domain={\thisrow{num_variables}}{1000000:1000000}, unbounded coords=discard ] {\polarisXmpifourXmultXhostXbfgsXdenseXdiagXuserXdim};
            \addplot[varseries,dense] table [ x=\xvar, y=\yvar, restrict expr to domain={\thisrow{num_variables}}{10000000:10000000}, unbounded coords=discard ] {\polarisXmpifourXmultXhostXbfgsXdenseXdiagXuserXdim};

        \addplot[histseries,rbfgs,host,hist5] table [ x=\xvar, y=\yvar, restrict expr to domain={\thisrow{history size}}{5:5},restrict expr to domain={\thisrow{num_variables}}{1000:10000000}, unbounded coords=discard ] {\polarisXmpifourXmultXhostXbfgsXrecXdiagXuserXdim};
        \addplot[histseries,rbfgs,host,hist10] table [ x=\xvar, y=\yvar, restrict expr to domain={\thisrow{history size}}{10:10},restrict expr to domain={\thisrow{num_variables}}{1000:10000000},  unbounded coords=discard ] {\polarisXmpifourXmultXhostXbfgsXrecXdiagXuserXdim};
        \addplot[histseries,rbfgs,host,hist20] table [ x=\xvar, y=\yvar, restrict expr to domain={\thisrow{history size}}{20:20},restrict expr to domain={\thisrow{num_variables}}{1000:10000000},  unbounded coords=discard ] {\polarisXmpifourXmultXhostXbfgsXrecXdiagXuserXdim};
        \addplot[histseries,rbfgs,host,hist50] table [ x=\xvar, y=\yvar, restrict expr to domain={\thisrow{history size}}{50:50},restrict expr to domain={\thisrow{num_variables}}{1000:10000000},  unbounded coords=discard ] {\polarisXmpifourXmultXhostXbfgsXrecXdiagXuserXdim};

        \addplot[histseries,dense,host,hist5] table [ x=\xvar, y=\yvar, restrict expr to domain={\thisrow{history size}}{5:5},restrict expr to domain={\thisrow{num_variables}}{1000:10000000},  unbounded coords=discard ] {\polarisXmpifourXmultXhostXbfgsXdenseXdiagXuserXdim};
        \addplot[histseries,dense,host,hist10] table [ x=\xvar, y=\yvar, restrict expr to domain={\thisrow{history size}}{10:10},restrict expr to domain={\thisrow{num_variables}}{1000:10000000},  unbounded coords=discard ] {\polarisXmpifourXmultXhostXbfgsXdenseXdiagXuserXdim};
        \addplot[histseries,dense,host,hist20] table [ x=\xvar, y=\yvar, restrict expr to domain={\thisrow{history size}}{20:20},restrict expr to domain={\thisrow{num_variables}}{1000:10000000},  unbounded coords=discard ] {\polarisXmpifourXmultXhostXbfgsXdenseXdiagXuserXdim};
        \addplot[histseries,dense,host,hist50] table [ x=\xvar, y=\yvar, restrict expr to domain={\thisrow{history size}}{50:50},restrict expr to domain={\thisrow{num_variables}}{1000:10000000},  unbounded coords=discard ] {\polarisXmpifourXmultXhostXbfgsXdenseXdiagXuserXdim};

        \path (axis cs:0.5e-4,5) node [anchor=center] {\small$10^3$};
        \path (axis cs:0.8e-4,26) node [anchor=center] {\small$10^4$};
        \path (axis cs:4e-3,25) node [anchor=center] {\small$10^5$};
        \path (axis cs:4e-2,13) node [anchor=center] {\small$10^6$};
        \path (axis cs:2,8) node [anchor=north] {\small$n=10^7$};

        \node (halfBsy) at (axis cs:1,90) {};
        \node (quarterBsy) at (axis cs:1,45) {};
        \node (rightaxis) at (rel axis cs:1,0) {};
        \path (halfBsy -| rightaxis) node [anchor=south east,inner sep=1pt] {\small$\tfrac{1}{2}B_{\text{S}}$};
        \path (quarterBsy -| rightaxis) node [anchor=south east,inner sep=1pt] {\small$\tfrac{1}{4}B_{\text{S}}$};
        
        
    \end{semilogxaxis}
\end{tikzpicture}%

%% file: figure/polaris-solve-device-bfgs-condiag-user.tex
\begin{tikzpicture}
    \begin{semilogxaxis}[
        ymin=0,
        xlabel={\strut$T_{\text{step}}$ (s)},
        ylabel={$B_{\text{e}}$ (GB/s)},
        legend pos=outer north east,
        extra y ticks={350,700},
        extra y tick labels={{},{}},
        extra y tick style={grid=major,major tick length=0pt},
        ylabel style={anchor=east,rotate=-90,at={(0,0.8)}},
        xlabel style={anchor=north west,at={(1,0)}},
        ]

        \addlegendimage{empty legend}
        \addlegendentry{\textbf{method}}
        \addlegendimage{histseries,rbfgs,device,hist50}
        \addlegendentry{recursive}
        
        \addlegendimage{histseries,cd,device,hist50}
        \addlegendentry{comp.\ dense}

        \addlegendimage{histseries,ddense,device,hist50}
        \addlegendentry{dense}

        \addlegendimage{empty legend}
        \addlegendentry{\textbf{history size $m$}}
        
        \addlegendimage{histseries,device,hist5}
        \addlegendentry{5}
        
        \addlegendimage{histseries,device,hist10}
        \addlegendentry{10}
        
        \addlegendimage{histseries,device,hist20}
        \addlegendentry{20}
        
        \addlegendimage{histseries,device,hist50}
        \addlegendentry{50}
        
            \addplot[varseries,cd] table [ x=\xvar, y=\yvar, restrict expr to domain={\thisrow{num_variables}}{100000:100000}, unbounded coords=discard ] {\polarisXdeviceXbfgsXcdXcondiagXuserXdim};
            \addplot[varseries,cd] table [ x=\xvar, y=\yvar, restrict expr to domain={\thisrow{num_variables}}{1000000:1000000}, unbounded coords=discard ] {\polarisXdeviceXbfgsXcdXcondiagXuserXdim};
            \addplot[varseries,cd] table [ x=\xvar, y=\yvar, restrict expr to domain={\thisrow{num_variables}}{10000000:10000000}, unbounded coords=discard ] {\polarisXdeviceXbfgsXcdXcondiagXuserXdim};

            \addplot[varseries,rbfgs] table [ x=\xvar, y=\yvar, restrict expr to domain={\thisrow{num_variables}}{100000:100000}, unbounded coords=discard ] {\polarisXdeviceXbfgsXrecXcondiagXuserXdim};
            \addplot[varseries,rbfgs] table [ x=\xvar, y=\yvar, restrict expr to domain={\thisrow{num_variables}}{1000000:1000000}, unbounded coords=discard ] {\polarisXdeviceXbfgsXrecXcondiagXuserXdim};
            \addplot[varseries,rbfgs] table [ x=\xvar, y=\yvar, restrict expr to domain={\thisrow{num_variables}}{10000000:10000000}, unbounded coords=discard ] {\polarisXdeviceXbfgsXrecXcondiagXuserXdim};

            \addplot[varseries,ddense] table [ x=\xvar, y=\yvar, restrict expr to domain={\thisrow{num_variables}}{100000:100000}, unbounded coords=discard ] {\polarisXdeviceXdbfgsXinplaceXcondiagXuserXdim};
            \addplot[varseries,ddense] table [ x=\xvar, y=\yvar, restrict expr to domain={\thisrow{num_variables}}{1000000:1000000}, unbounded coords=discard ] {\polarisXdeviceXdbfgsXinplaceXcondiagXuserXdim};
            \addplot[varseries,ddense] table [ x=\xvar, y=\yvar, restrict expr to domain={\thisrow{num_variables}}{10000000:10000000}, unbounded coords=discard ] {\polarisXdeviceXdbfgsXinplaceXcondiagXuserXdim};

        \addplot[histseries,rbfgs,device,hist5] table [ x=\xvar, y=\yvar, restrict expr to domain={\thisrow{history size}}{5:5}, restrict expr to domain={\thisrow{num_variables}}{100000:10000000}, unbounded coords=discard ] {\polarisXdeviceXbfgsXrecXcondiagXuserXdim};
        \addplot[histseries,rbfgs,device,hist10] table [ x=\xvar, y=\yvar, restrict expr to domain={\thisrow{history size}}{10:10}, restrict expr to domain={\thisrow{num_variables}}{100000:10000000},  unbounded coords=discard ] {\polarisXdeviceXbfgsXrecXcondiagXuserXdim};
        \addplot[histseries,rbfgs,device,hist20] table [ x=\xvar, y=\yvar, restrict expr to domain={\thisrow{history size}}{20:20}, restrict expr to domain={\thisrow{num_variables}}{100000:10000000},  unbounded coords=discard ] {\polarisXdeviceXbfgsXrecXcondiagXuserXdim};
        \addplot[histseries,rbfgs,device,hist50] table [ x=\xvar, y=\yvar, restrict expr to domain={\thisrow{history size}}{50:50}, restrict expr to domain={\thisrow{num_variables}}{100000:10000000},  unbounded coords=discard ] {\polarisXdeviceXbfgsXrecXcondiagXuserXdim};

        \addplot[histseries,ddense,device,hist5] table [ x=\xvar, y=\yvar, restrict expr to domain={\thisrow{history size}}{5:5}, restrict expr to domain={\thisrow{num_variables}}{100000:10000000},  unbounded coords=discard ] {\polarisXdeviceXdbfgsXinplaceXcondiagXuserXdim};
        \addplot[histseries,ddense,device,hist10] table [ x=\xvar, y=\yvar, restrict expr to domain={\thisrow{history size}}{10:10}, restrict expr to domain={\thisrow{num_variables}}{100000:10000000},  unbounded coords=discard ] {\polarisXdeviceXdbfgsXinplaceXcondiagXuserXdim};
        \addplot[histseries,ddense,device,hist20] table [ x=\xvar, y=\yvar, restrict expr to domain={\thisrow{history size}}{20:20}, restrict expr to domain={\thisrow{num_variables}}{100000:10000000},  unbounded coords=discard ] {\polarisXdeviceXdbfgsXinplaceXcondiagXuserXdim};
        \addplot[histseries,ddense,device,hist50] table [ x=\xvar, y=\yvar, restrict expr to domain={\thisrow{history size}}{50:50}, restrict expr to domain={\thisrow{num_variables}}{100000:10000000},  unbounded coords=discard ] {\polarisXdeviceXdbfgsXinplaceXcondiagXuserXdim};

        \addplot[histseries,cd,device,hist5] table [ x=\xvar, y=\yvar, restrict expr to domain={\thisrow{history size}}{5:5}, restrict expr to domain={\thisrow{num_variables}}{100000:10000000},  unbounded coords=discard ] {\polarisXdeviceXbfgsXcdXcondiagXuserXdim};
        \addplot[histseries,cd,device,hist10] table [ x=\xvar, y=\yvar, restrict expr to domain={\thisrow{history size}}{10:10}, restrict expr to domain={\thisrow{num_variables}}{100000:10000000},  unbounded coords=discard ] {\polarisXdeviceXbfgsXcdXcondiagXuserXdim};
        \addplot[histseries,cd,device,hist20] table [ x=\xvar, y=\yvar, restrict expr to domain={\thisrow{history size}}{20:20}, restrict expr to domain={\thisrow{num_variables}}{100000:10000000},  unbounded coords=discard ] {\polarisXdeviceXbfgsXcdXcondiagXuserXdim};
        \addplot[histseries,cd,device,hist50] table [ x=\xvar, y=\yvar, restrict expr to domain={\thisrow{history size}}{50:50}, restrict expr to domain={\thisrow{num_variables}}{100000:10000000},  unbounded coords=discard ] {\polarisXdeviceXbfgsXcdXcondiagXuserXdim};

        \path (axis cs:5.5e-4,80) node {\small$10^5$};
        \path (axis cs:1.5e-3,250) node {\small$10^6$};
        \path (axis cs:8e-3,370) node [anchor=north west] {\small$n=10^7$};
        
        
        \node (halfBsy) at (axis cs:1,700) {};
        \node (quarterBsy) at (axis cs:1,350) {};
        \node (rightaxis) at (rel axis cs:1,0) {};
        \path (halfBsy -| rightaxis) node [anchor=north east,inner sep=1pt] {\small$\tfrac{1}{2}B_{\text{S}}$};
        \path (quarterBsy -| rightaxis) node [anchor=south east,inner sep=1pt] {\small$\tfrac{1}{4}B_{\text{S}}$};


        \node (halfBsy) at (axis cs:1,700) {};
        \node (quarterBsy) at (axis cs:1,350) {};
        \node (rightaxis) at (rel axis cs:1,0) {};
        \path (halfBsy -| rightaxis) node [anchor=south east,inner sep=1pt] {\small$\tfrac{1}{2}B_{\text{S}}$};
        \path (quarterBsy -| rightaxis) node [anchor=south east,inner sep=1pt] {\small$\tfrac{1}{4}B_{\text{S}}$};
        
        
    \end{semilogxaxis}
\end{tikzpicture}%

%% file: figure/frontier-solve-device-dbfgs-bfgs-cd-user.tex
\begin{tikzpicture}
    \begin{semilogxaxis}[
        ymin=0,
        xlabel={\strut$T_{\text{step}}$ (s)},
        ylabel={$B_{\text{e}}$ (GB/s)},
        legend pos=outer north east,
        extra y ticks={350,700},
        extra y tick labels={{},{}},
        extra y tick style={grid=major,major tick length=0pt},
        ylabel style={anchor=east,rotate=-90,at={(0,0.9)}},
        xlabel style={anchor=north west,at={(1,0)}},
        ]

        \addlegendimage{empty legend}
        \addlegendentry{\textbf{method}}
        \addlegendimage{histseries,rbfgs,device,hist50}
        \addlegendentry{recursive}

        \addlegendimage{histseries,cd,device,hist50}
        \addlegendentry{comp.\ dense}

        \addlegendimage{histseries,ddense,device,hist50}
        \addlegendentry{dense}

        \addlegendimage{empty legend}
        \addlegendentry{\textbf{history size $m$}}
        
        \addlegendimage{histseries,device,hist5}
        \addlegendentry{5}
        
        \addlegendimage{histseries,device,hist10}
        \addlegendentry{10}
        
        \addlegendimage{histseries,device,hist20}
        \addlegendentry{20}
        
        \addlegendimage{histseries,device,hist50}
        \addlegendentry{50}
        
            \addplot[varseries,rbfgs] table [ x=\xvar, y=\yvar, restrict expr to domain={\thisrow{num_variables}}{100000:100000}, unbounded coords=discard ] {\frontierXdeviceXbfgsXrecXcondiagXuserXdim};
            \addplot[varseries,rbfgs] table [ x=\xvar, y=\yvar, restrict expr to domain={\thisrow{num_variables}}{1000000:1000000}, unbounded coords=discard ] {\frontierXdeviceXbfgsXrecXcondiagXuserXdim};
            \addplot[varseries,rbfgs] table [ x=\xvar, y=\yvar, restrict expr to domain={\thisrow{num_variables}}{10000000:10000000}, unbounded coords=discard ] {\frontierXdeviceXbfgsXrecXcondiagXuserXdim};
            \addplot[varseries,rbfgs] table [ x=\xvar, y=\yvar, restrict expr to domain={\thisrow{num_variables}}{100000000:100000000}, unbounded coords=discard ] {\frontierXdeviceXbfgsXrecXcondiagXuserXdim};

            \addplot[varseries,cd] table [ x=\xvar, y=\yvar, restrict expr to domain={\thisrow{num_variables}}{100000:100000}, unbounded coords=discard ] {\frontierXdeviceXbfgsXcdXcondiagXuserXdim};
            \addplot[varseries,cd] table [ x=\xvar, y=\yvar, restrict expr to domain={\thisrow{num_variables}}{1000000:1000000}, unbounded coords=discard ] {\frontierXdeviceXbfgsXcdXcondiagXuserXdim};
            \addplot[varseries,cd] table [ x=\xvar, y=\yvar, restrict expr to domain={\thisrow{num_variables}}{10000000:10000000}, unbounded coords=discard ] {\frontierXdeviceXbfgsXcdXcondiagXuserXdim};
            \addplot[varseries,cd] table [ x=\xvar, y=\yvar, restrict expr to domain={\thisrow{num_variables}}{100000000:100000000}, unbounded coords=discard ] {\frontierXdeviceXbfgsXcdXcondiagXuserXdim};

            \addplot[varseries,ddense] table [ x=\xvar, y=\yvar, restrict expr to domain={\thisrow{num_variables}}{100000:100000}, unbounded coords=discard ] {\frontierXdeviceXdbfgsXinplaceXcondiagXuserXdim};
            \addplot[varseries,ddense] table [ x=\xvar, y=\yvar, restrict expr to domain={\thisrow{num_variables}}{1000000:1000000}, unbounded coords=discard ] {\frontierXdeviceXdbfgsXinplaceXcondiagXuserXdim};
            \addplot[varseries,ddense] table [ x=\xvar, y=\yvar, restrict expr to domain={\thisrow{num_variables}}{10000000:10000000}, unbounded coords=discard ] {\frontierXdeviceXdbfgsXinplaceXcondiagXuserXdim};
            \addplot[varseries,ddense] table [ x=\xvar, y=\yvar, restrict expr to domain={\thisrow{num_variables}}{100000000:100000000}, unbounded coords=discard ] {\frontierXdeviceXdbfgsXinplaceXcondiagXuserXdim};

        \addplot[histseries,rbfgs,device,hist5] table [ x=\xvar, y=\yvar, restrict expr to domain={\thisrow{history size}}{5:5},restrict expr to domain={\thisrow{num_variables}}{100000:100000000}, unbounded coords=discard ] {\frontierXdeviceXbfgsXrecXcondiagXuserXdim};
        \addplot[histseries,rbfgs,device,hist10] table [ x=\xvar, y=\yvar, restrict expr to domain={\thisrow{history size}}{10:10},restrict expr to domain={\thisrow{num_variables}}{100000:100000000},  unbounded coords=discard ] {\frontierXdeviceXbfgsXrecXcondiagXuserXdim};
        \addplot[histseries,rbfgs,device,hist20] table [ x=\xvar, y=\yvar, restrict expr to domain={\thisrow{history size}}{20:20},restrict expr to domain={\thisrow{num_variables}}{100000:100000000},  unbounded coords=discard ] {\frontierXdeviceXbfgsXrecXcondiagXuserXdim};
        \addplot[histseries,rbfgs,device,hist50] table [ x=\xvar, y=\yvar, restrict expr to domain={\thisrow{history size}}{50:50},restrict expr to domain={\thisrow{num_variables}}{100000:100000000},  unbounded coords=discard ] {\frontierXdeviceXbfgsXrecXcondiagXuserXdim};

        \addplot[histseries,ddense,device,hist5] table [ x=\xvar, y=\yvar, restrict expr to domain={\thisrow{history size}}{5:5},restrict expr to domain={\thisrow{num_variables}}{100000:100000000},  unbounded coords=discard ] {\frontierXdeviceXdbfgsXinplaceXcondiagXuserXdim};
        \addplot[histseries,ddense,device,hist10] table [ x=\xvar, y=\yvar, restrict expr to domain={\thisrow{history size}}{10:10},restrict expr to domain={\thisrow{num_variables}}{100000:100000000},  unbounded coords=discard ] {\frontierXdeviceXdbfgsXinplaceXcondiagXuserXdim};
        \addplot[histseries,ddense,device,hist20] table [ x=\xvar, y=\yvar, restrict expr to domain={\thisrow{history size}}{20:20},restrict expr to domain={\thisrow{num_variables}}{100000:100000000},  unbounded coords=discard ] {\frontierXdeviceXdbfgsXinplaceXcondiagXuserXdim};
        \addplot[histseries,ddense,device,hist50] table [ x=\xvar, y=\yvar, restrict expr to domain={\thisrow{history size}}{50:50},restrict expr to domain={\thisrow{num_variables}}{100000:100000000},  unbounded coords=discard ] {\frontierXdeviceXdbfgsXinplaceXcondiagXuserXdim};

        \addplot[histseries,cd,device,hist5] table [ x=\xvar, y=\yvar, restrict expr to domain={\thisrow{history size}}{5:5},restrict expr to domain={\thisrow{num_variables}}{100000:100000000},  unbounded coords=discard ] {\frontierXdeviceXbfgsXcdXcondiagXuserXdim};
        \addplot[histseries,cd,device,hist10] table [ x=\xvar, y=\yvar, restrict expr to domain={\thisrow{history size}}{10:10},restrict expr to domain={\thisrow{num_variables}}{100000:100000000},  unbounded coords=discard ] {\frontierXdeviceXbfgsXcdXcondiagXuserXdim};
        \addplot[histseries,cd,device,hist20] table [ x=\xvar, y=\yvar, restrict expr to domain={\thisrow{history size}}{20:20},restrict expr to domain={\thisrow{num_variables}}{100000:100000000},  unbounded coords=discard ] {\frontierXdeviceXbfgsXcdXcondiagXuserXdim};
        \addplot[histseries,cd,device,hist50] table [ x=\xvar, y=\yvar, restrict expr to domain={\thisrow{history size}}{50:50},restrict expr to domain={\thisrow{num_variables}}{100000:100000000},  unbounded coords=discard ] {\frontierXdeviceXbfgsXcdXcondiagXuserXdim};

        \path (axis cs:6e-4,60) node [anchor=center] {\small$10^4$};
        \path (axis cs:1.5e-3,200) node [anchor=center] {\small$10^5$};
        \path (axis cs:1.5e-2,300) node [anchor=center] {\small$n=10^6$};
        \path (axis cs:1e-1,300) node [anchor=north] {\small$10^7$};

        \node (halfBsy) at (axis cs:1,700) {};
        \node (quarterBsy) at (axis cs:1,350) {};
        \node (rightaxis) at (rel axis cs:1,0) {};
        \path (halfBsy -| rightaxis) node [anchor=south east,inner sep=1pt] {\small$\tfrac{1}{2}B_{\text{S}}$};
        \path (quarterBsy -| rightaxis) node [anchor=south east,inner sep=1pt] {\small$\tfrac{1}{4}B_{\text{S}}$};
        
        
    \end{semilogxaxis}
\end{tikzpicture}%

%% file: figure/polaris-solve-device-bfgs-diag-diag.tex
\begin{tikzpicture}
    \begin{semilogxaxis}[
        ymin=0,
        xlabel={\strut$T_{\text{step}}$ (s)},
        ylabel={$B_{\text{e}}$ (GB/s)},
        legend pos=outer north east,
        extra y ticks={350,700},
        extra y tick labels={{},{}},
        extra y tick style={grid=major,major tick length=0pt},
        ylabel style={anchor=east,rotate=-90,at={(0,0.85)}},
        xlabel style={anchor=north west,at={(1,0)}},
        ]

        \addlegendimage{empty legend}
        \addlegendentry{\textbf{method}}
        \addlegendimage{histseries,rbfgs,device,hist50}
        \addlegendentry{recursive}
        
        \addlegendimage{histseries,cd,device,hist50}
        \addlegendentry{comp.\ dense}

        \addlegendimage{histseries,ddense,device,hist50}
        \addlegendentry{dense}

        \addlegendimage{empty legend}
        \addlegendentry{\textbf{history size $m$}}
        
        \addlegendimage{histseries,device,hist5}
        \addlegendentry{5}
        
        \addlegendimage{histseries,device,hist10}
        \addlegendentry{10}
        
        \addlegendimage{histseries,device,hist20}
        \addlegendentry{20}
        
        \addlegendimage{histseries,device,hist50}
        \addlegendentry{50}
        
            \addplot[varseries,cd] table [ x=\xvar, y=\yvar, restrict expr to domain={\thisrow{num_variables}}{100000:100000}, unbounded coords=discard ] {\polarisXdeviceXbfgsXcdXdiagXdiagXdim};
            \addplot[varseries,cd] table [ x=\xvar, y=\yvar, restrict expr to domain={\thisrow{num_variables}}{1000000:1000000}, unbounded coords=discard ] {\polarisXdeviceXbfgsXcdXdiagXdiagXdim};
            \addplot[varseries,cd] table [ x=\xvar, y=\yvar, restrict expr to domain={\thisrow{num_variables}}{10000000:10000000}, unbounded coords=discard ] {\polarisXdeviceXbfgsXcdXdiagXdiagXdim};

            \addplot[varseries,rbfgs] table [ x=\xvar, y=\yvar, restrict expr to domain={\thisrow{num_variables}}{100000:100000}, unbounded coords=discard ] {\polarisXdeviceXbfgsXrecXdiagXdiagXdim};
            \addplot[varseries,rbfgs] table [ x=\xvar, y=\yvar, restrict expr to domain={\thisrow{num_variables}}{1000000:1000000}, unbounded coords=discard ] {\polarisXdeviceXbfgsXrecXdiagXdiagXdim};
            \addplot[varseries,rbfgs] table [ x=\xvar, y=\yvar, restrict expr to domain={\thisrow{num_variables}}{10000000:10000000}, unbounded coords=discard ] {\polarisXdeviceXbfgsXrecXdiagXdiagXdim};

            \addplot[varseries,ddense] table [ x=\xvar, y=\yvar, restrict expr to domain={\thisrow{num_variables}}{100000:100000}, unbounded coords=discard ] {\polarisXdeviceXdbfgsXinplaceXdiagXdiagXdim};
            \addplot[varseries,ddense] table [ x=\xvar, y=\yvar, restrict expr to domain={\thisrow{num_variables}}{1000000:1000000}, unbounded coords=discard ] {\polarisXdeviceXdbfgsXinplaceXdiagXdiagXdim};
            \addplot[varseries,ddense] table [ x=\xvar, y=\yvar, restrict expr to domain={\thisrow{num_variables}}{10000000:10000000}, unbounded coords=discard ] {\polarisXdeviceXdbfgsXinplaceXdiagXdiagXdim};

        \addplot[histseries,rbfgs,device,hist5] table [ x=\xvar, y=\yvar, restrict expr to domain={\thisrow{history size}}{5:5}, restrict expr to domain={\thisrow{num_variables}}{100000:10000000}, unbounded coords=discard ] {\polarisXdeviceXbfgsXrecXdiagXdiagXdim};
        \addplot[histseries,rbfgs,device,hist10] table [ x=\xvar, y=\yvar, restrict expr to domain={\thisrow{history size}}{10:10}, restrict expr to domain={\thisrow{num_variables}}{100000:10000000},  unbounded coords=discard ] {\polarisXdeviceXbfgsXrecXdiagXdiagXdim};
        \addplot[histseries,rbfgs,device,hist20] table [ x=\xvar, y=\yvar, restrict expr to domain={\thisrow{history size}}{20:20}, restrict expr to domain={\thisrow{num_variables}}{100000:10000000},  unbounded coords=discard ] {\polarisXdeviceXbfgsXrecXdiagXdiagXdim};
        \addplot[histseries,rbfgs,device,hist50] table [ x=\xvar, y=\yvar, restrict expr to domain={\thisrow{history size}}{50:50}, restrict expr to domain={\thisrow{num_variables}}{100000:10000000},  unbounded coords=discard ] {\polarisXdeviceXbfgsXrecXdiagXdiagXdim};

        \addplot[histseries,ddense,device,hist5] table [ x=\xvar, y=\yvar, restrict expr to domain={\thisrow{history size}}{5:5}, restrict expr to domain={\thisrow{num_variables}}{100000:10000000},  unbounded coords=discard ] {\polarisXdeviceXdbfgsXinplaceXdiagXdiagXdim};
        \addplot[histseries,ddense,device,hist10] table [ x=\xvar, y=\yvar, restrict expr to domain={\thisrow{history size}}{10:10}, restrict expr to domain={\thisrow{num_variables}}{100000:10000000},  unbounded coords=discard ] {\polarisXdeviceXdbfgsXinplaceXdiagXdiagXdim};
        \addplot[histseries,ddense,device,hist20] table [ x=\xvar, y=\yvar, restrict expr to domain={\thisrow{history size}}{20:20}, restrict expr to domain={\thisrow{num_variables}}{100000:10000000},  unbounded coords=discard ] {\polarisXdeviceXdbfgsXinplaceXdiagXdiagXdim};
        \addplot[histseries,ddense,device,hist50] table [ x=\xvar, y=\yvar, restrict expr to domain={\thisrow{history size}}{50:50}, restrict expr to domain={\thisrow{num_variables}}{100000:10000000},  unbounded coords=discard ] {\polarisXdeviceXdbfgsXinplaceXdiagXdiagXdim};

        \addplot[histseries,cd,device,hist5] table [ x=\xvar, y=\yvar, restrict expr to domain={\thisrow{history size}}{5:5}, restrict expr to domain={\thisrow{num_variables}}{100000:10000000},  unbounded coords=discard ] {\polarisXdeviceXbfgsXcdXdiagXdiagXdim};
        \addplot[histseries,cd,device,hist10] table [ x=\xvar, y=\yvar, restrict expr to domain={\thisrow{history size}}{10:10}, restrict expr to domain={\thisrow{num_variables}}{100000:10000000},  unbounded coords=discard ] {\polarisXdeviceXbfgsXcdXdiagXdiagXdim};
        \addplot[histseries,cd,device,hist20] table [ x=\xvar, y=\yvar, restrict expr to domain={\thisrow{history size}}{20:20}, restrict expr to domain={\thisrow{num_variables}}{100000:10000000},  unbounded coords=discard ] {\polarisXdeviceXbfgsXcdXdiagXdiagXdim};
        \addplot[histseries,cd,device,hist50] table [ x=\xvar, y=\yvar, restrict expr to domain={\thisrow{history size}}{50:50}, restrict expr to domain={\thisrow{num_variables}}{100000:10000000},  unbounded coords=discard ] {\polarisXdeviceXbfgsXcdXdiagXdiagXdim};

        \path (axis cs:6.9e-4,100) node {\small$10^5$};
        \path (axis cs:2.2e-3,250) node {\small$10^6$};
        \path (axis cs:9.8e-3,300) node [anchor=north west] {\small$n=10^7$};
        
        
        \node (halfBsy) at (axis cs:1,700) {};
        \node (quarterBsy) at (axis cs:1,350) {};
        \node (rightaxis) at (rel axis cs:1,0) {};
        \path (halfBsy -| rightaxis) node [anchor=north east,inner sep=1pt] {\small$\tfrac{1}{2}B_{\text{S}}$};
        \path (quarterBsy -| rightaxis) node [anchor=south east,inner sep=1pt] {\small$\tfrac{1}{4}B_{\text{S}}$};


        \node (halfBsy) at (axis cs:1,700) {};
        \node (quarterBsy) at (axis cs:1,350) {};
        \node (rightaxis) at (rel axis cs:1,0) {};
        \path (halfBsy -| rightaxis) node [anchor=south east,inner sep=1pt] {\small$\tfrac{1}{2}B_{\text{S}}$};
        \path (quarterBsy -| rightaxis) node [anchor=south east,inner sep=1pt] {\small$\tfrac{1}{4}B_{\text{S}}$};
        
        
    \end{semilogxaxis}
\end{tikzpicture}%

%% file: figure/polaris-solve-device-bfgs-mpifour-condiag-user.tex
\begin{tikzpicture}
    \begin{semilogxaxis}[
        ymin=0,
        ymax=2500,
        legend pos=outer north east,
        extra y ticks={1400,2800},
        extra y tick labels={{},{}},
        extra y tick style={grid=major,major tick length=0pt},
        xlabel={\strut$T_{\text{step}}$ (s)},
        ylabel={$B_{\text{e}}$ (GB/s)},
        ylabel style={anchor=east,rotate=-90,at={(0,0.9)}},
        xlabel style={anchor=north west,at={(1.01,0)}},
        ]

        \addlegendimage{empty legend}
        \addlegendentry{\textbf{method}}
        \addlegendimage{histseries,rbfgs,device,hist50}
        \addlegendentry{recursive}
        
        \addlegendimage{histseries,cd,device,hist50}
        \addlegendentry{comp.\ dense}

        \addlegendimage{histseries,ddense,device,hist50}
        \addlegendentry{dense}

        \addlegendimage{empty legend}
        \addlegendentry{\textbf{history size $m$}}
        
        \addlegendimage{histseries,device,hist5}
        \addlegendentry{5}
        
        \addlegendimage{histseries,device,hist10}
        \addlegendentry{10}
        
        \addlegendimage{histseries,device,hist20}
        \addlegendentry{20}
        
        \addlegendimage{histseries,device,hist50}
        \addlegendentry{50}
        
            \addplot[varseries,cd] table [ x=\xvar, y=\yvar, restrict expr to domain={\thisrow{num_variables}}{1000000:1000000}, unbounded coords=discard ] {\polarisXmpifourXdeviceXbfgsXcdXcondiagXuserXdim};
            \addplot[varseries,cd] table [ x=\xvar, y=\yvar, restrict expr to domain={\thisrow{num_variables}}{10000000:10000000}, unbounded coords=discard ] {\polarisXmpifourXdeviceXbfgsXcdXcondiagXuserXdim};
            \addplot[varseries,cd] table [ x=\xvar, y=\yvar, restrict expr to domain={\thisrow{num_variables}}{100000000:100000000}, unbounded coords=discard ] {\polarisXmpifourXdeviceXbfgsXcdXcondiagXuserXdim};

            \addplot[varseries,rbfgs] table [ x=\xvar, y=\yvar, restrict expr to domain={\thisrow{num_variables}}{1000000:1000000}, unbounded coords=discard ] {\polarisXmpifourXdeviceXbfgsXrecXcondiagXuserXdim};
            \addplot[varseries,rbfgs] table [ x=\xvar, y=\yvar, restrict expr to domain={\thisrow{num_variables}}{10000000:10000000}, unbounded coords=discard ] {\polarisXmpifourXdeviceXbfgsXrecXcondiagXuserXdim};
            \addplot[varseries,rbfgs] table [ x=\xvar, y=\yvar, restrict expr to domain={\thisrow{num_variables}}{100000000:100000000}, unbounded coords=discard ] {\polarisXmpifourXdeviceXbfgsXrecXcondiagXuserXdim};

            \addplot[varseries,ddense] table [ x=\xvar, y=\yvar, restrict expr to domain={\thisrow{num_variables}}{1000000:1000000}, unbounded coords=discard ] {\polarisXmpifourXdeviceXdbfgsXinplaceXcondiagXuserXdim};
            \addplot[varseries,ddense] table [ x=\xvar, y=\yvar, restrict expr to domain={\thisrow{num_variables}}{10000000:10000000}, unbounded coords=discard ] {\polarisXmpifourXdeviceXdbfgsXinplaceXcondiagXuserXdim};
            \addplot[varseries,ddense] table [ x=\xvar, y=\yvar, restrict expr to domain={\thisrow{num_variables}}{100000000:100000000}, unbounded coords=discard ] {\polarisXmpifourXdeviceXdbfgsXinplaceXcondiagXuserXdim};

        \addplot[histseries,ddense,device,hist5] table [ x=\xvar, y=\yvar, restrict expr to domain={\thisrow{history size}}{5:5},restrict expr to domain={\thisrow{num_variables}}{1000000:100000000}, unbounded coords=discard ] {\polarisXmpifourXdeviceXdbfgsXinplaceXcondiagXuserXdim};
        \addplot[histseries,ddense,device,hist10] table [ x=\xvar, y=\yvar, restrict expr to domain={\thisrow{history size}}{10:10},restrict expr to domain={\thisrow{num_variables}}{1000000:100000000},  unbounded coords=discard ] {\polarisXmpifourXdeviceXdbfgsXinplaceXcondiagXuserXdim};
        \addplot[histseries,ddense,device,hist20] table [ x=\xvar, y=\yvar, restrict expr to domain={\thisrow{history size}}{20:20},restrict expr to domain={\thisrow{num_variables}}{1000000:100000000},  unbounded coords=discard ] {\polarisXmpifourXdeviceXdbfgsXinplaceXcondiagXuserXdim};
        \addplot[histseries,ddense,device,hist50] table [ x=\xvar, y=\yvar, restrict expr to domain={\thisrow{history size}}{50:50},restrict expr to domain={\thisrow{num_variables}}{1000000:100000000},  unbounded coords=discard ] {\polarisXmpifourXdeviceXdbfgsXinplaceXcondiagXuserXdim};

        \addplot[histseries,rbfgs,device,hist5] table [ x=\xvar, y=\yvar, restrict expr to domain={\thisrow{history size}}{5:5},restrict expr to domain={\thisrow{num_variables}}{1000000:100000000}, unbounded coords=discard ] {\polarisXmpifourXdeviceXbfgsXrecXcondiagXuserXdim};
        \addplot[histseries,rbfgs,device,hist10] table [ x=\xvar, y=\yvar, restrict expr to domain={\thisrow{history size}}{10:10},restrict expr to domain={\thisrow{num_variables}}{1000000:100000000},  unbounded coords=discard ] {\polarisXmpifourXdeviceXbfgsXrecXcondiagXuserXdim};
        \addplot[histseries,rbfgs,device,hist20] table [ x=\xvar, y=\yvar, restrict expr to domain={\thisrow{history size}}{20:20},restrict expr to domain={\thisrow{num_variables}}{1000000:100000000},  unbounded coords=discard ] {\polarisXmpifourXdeviceXbfgsXrecXcondiagXuserXdim};
        \addplot[histseries,rbfgs,device,hist50] table [ x=\xvar, y=\yvar, restrict expr to domain={\thisrow{history size}}{50:50},restrict expr to domain={\thisrow{num_variables}}{1000000:100000000},  unbounded coords=discard ] {\polarisXmpifourXdeviceXbfgsXrecXcondiagXuserXdim};

        \addplot[histseries,cd,device,hist5] table [ x=\xvar, y=\yvar, restrict expr to domain={\thisrow{history size}}{5:5},restrict expr to domain={\thisrow{num_variables}}{1000000:100000000},  unbounded coords=discard ] {\polarisXmpifourXdeviceXbfgsXcdXcondiagXuserXdim};
        \addplot[histseries,cd,device,hist10] table [ x=\xvar, y=\yvar, restrict expr to domain={\thisrow{history size}}{10:10},restrict expr to domain={\thisrow{num_variables}}{1000000:100000000},  unbounded coords=discard ] {\polarisXmpifourXdeviceXbfgsXcdXcondiagXuserXdim};
        \addplot[histseries,cd,device,hist20] table [ x=\xvar, y=\yvar, restrict expr to domain={\thisrow{history size}}{20:20},restrict expr to domain={\thisrow{num_variables}}{1000000:100000000},  unbounded coords=discard ] {\polarisXmpifourXdeviceXbfgsXcdXcondiagXuserXdim};
        \addplot[histseries,cd,device,hist50] table [ x=\xvar, y=\yvar, restrict expr to domain={\thisrow{history size}}{50:50},restrict expr to domain={\thisrow{num_variables}}{1000000:100000000},  unbounded coords=discard ] {\polarisXmpifourXdeviceXbfgsXcdXcondiagXuserXdim};

        \path (axis cs:1e-3,400) node {\small$10^6$};
        \path (axis cs:0.4e-2,1200) node {\small$n=10^7$};
        \path (axis cs:3.e-2,1500) node {\small$10^8$};
        
        
        \node (halfBsy) at (axis cs:1,2800) {};
        \node (quarterBsy) at (axis cs:1,1400) {};
        \node (rightaxis) at (rel axis cs:1,0) {};
        \path (halfBsy -| rightaxis) node [anchor=north east,inner sep=1pt] {\small$\tfrac{1}{2}B_{\text{S}}$};
        \path (quarterBsy -| rightaxis) node [anchor=south east,inner sep=1pt] {\small$\tfrac{1}{4}B_{\text{S}}$};
        
        
    \end{semilogxaxis}
\end{tikzpicture}%

%% file: figure/polaris-solve-device-bfgs-mpifour-diag-diag.tex
\begin{tikzpicture}
    \begin{semilogxaxis}[
        ymin=0,
        ymax=2200,
        legend pos=outer north east,
        extra y ticks={1400,2800},
        extra y tick labels={{},{}},
        extra y tick style={grid=major,major tick length=0pt},
        xlabel={\strut$T_{\text{step}}$ (s)},
        ylabel={$B_{\text{e}}$ (GB/s)},
        ylabel style={anchor=east,rotate=-90,at={(0,1)}},
        xlabel style={anchor=north west,at={(1,0)}},
        ]

        \addlegendimage{empty legend}
        \addlegendentry{\textbf{method}}
        \addlegendimage{histseries,rbfgs,device,hist50}
        \addlegendentry{recursive}
        
        \addlegendimage{histseries,cd,device,hist50}
        \addlegendentry{comp.\ dense}


        \addlegendimage{histseries,ddense,device,hist50}
        \addlegendentry{dense}

        \addlegendimage{empty legend}
        \addlegendentry{\textbf{history size $m$}}
        
        \addlegendimage{histseries,device,hist5}
        \addlegendentry{5}
        
        \addlegendimage{histseries,device,hist10}
        \addlegendentry{10}
        
        \addlegendimage{histseries,device,hist20}
        \addlegendentry{20}
        
        \addlegendimage{histseries,device,hist50}
        \addlegendentry{50}
        
            \addplot[varseries,cd] table [ x=\xvar, y=\yvar, restrict expr to domain={\thisrow{num_variables}}{1000000:1000000}, unbounded coords=discard ] {\polarisXmpifourXdeviceXbfgsXcdXdiagXdiagXdim};
            \addplot[varseries,cd] table [ x=\xvar, y=\yvar, restrict expr to domain={\thisrow{num_variables}}{10000000:10000000}, unbounded coords=discard ] {\polarisXmpifourXdeviceXbfgsXcdXdiagXdiagXdim};
            \addplot[varseries,cd] table [ x=\xvar, y=\yvar, restrict expr to domain={\thisrow{num_variables}}{100000000:100000000}, unbounded coords=discard ] {\polarisXmpifourXdeviceXbfgsXcdXdiagXdiagXdim};

            \addplot[varseries,rbfgs] table [ x=\xvar, y=\yvar, restrict expr to domain={\thisrow{num_variables}}{1000000:1000000}, unbounded coords=discard ] {\polarisXmpifourXdeviceXbfgsXrecXdiagXdiagXdim};
            \addplot[varseries,rbfgs] table [ x=\xvar, y=\yvar, restrict expr to domain={\thisrow{num_variables}}{10000000:10000000}, unbounded coords=discard ] {\polarisXmpifourXdeviceXbfgsXrecXdiagXdiagXdim};
            \addplot[varseries,rbfgs] table [ x=\xvar, y=\yvar, restrict expr to domain={\thisrow{num_variables}}{100000000:100000000}, unbounded coords=discard ] {\polarisXmpifourXdeviceXbfgsXrecXdiagXdiagXdim};

            \addplot[varseries,ddense] table [ x=\xvar, y=\yvar, restrict expr to domain={\thisrow{num_variables}}{1000000:1000000}, unbounded coords=discard ] {\polarisXmpifourXdeviceXdbfgsXinplaceXdiagXdiagXdim};
            \addplot[varseries,ddense] table [ x=\xvar, y=\yvar, restrict expr to domain={\thisrow{num_variables}}{10000000:10000000}, unbounded coords=discard ] {\polarisXmpifourXdeviceXdbfgsXinplaceXdiagXdiagXdim};
            \addplot[varseries,ddense] table [ x=\xvar, y=\yvar, restrict expr to domain={\thisrow{num_variables}}{100000000:100000000}, unbounded coords=discard ] {\polarisXmpifourXdeviceXdbfgsXinplaceXdiagXdiagXdim};

        \addplot[histseries,ddense,device,hist5] table [ x=\xvar, y=\yvar, restrict expr to domain={\thisrow{history size}}{5:5},restrict expr to domain={\thisrow{num_variables}}{1000000:100000000}, unbounded coords=discard ] {\polarisXmpifourXdeviceXdbfgsXinplaceXdiagXdiagXdim};
        \addplot[histseries,ddense,device,hist10] table [ x=\xvar, y=\yvar, restrict expr to domain={\thisrow{history size}}{10:10},restrict expr to domain={\thisrow{num_variables}}{1000000:100000000},  unbounded coords=discard ] {\polarisXmpifourXdeviceXdbfgsXinplaceXdiagXdiagXdim};
        \addplot[histseries,ddense,device,hist20] table [ x=\xvar, y=\yvar, restrict expr to domain={\thisrow{history size}}{20:20},restrict expr to domain={\thisrow{num_variables}}{1000000:100000000},  unbounded coords=discard ] {\polarisXmpifourXdeviceXdbfgsXinplaceXdiagXdiagXdim};
        \addplot[histseries,ddense,device,hist50] table [ x=\xvar, y=\yvar, restrict expr to domain={\thisrow{history size}}{50:50},restrict expr to domain={\thisrow{num_variables}}{1000000:100000000},  unbounded coords=discard ] {\polarisXmpifourXdeviceXdbfgsXinplaceXdiagXdiagXdim};

        \addplot[histseries,rbfgs,device,hist5] table [ x=\xvar, y=\yvar, restrict expr to domain={\thisrow{history size}}{5:5},restrict expr to domain={\thisrow{num_variables}}{1000000:100000000}, unbounded coords=discard ] {\polarisXmpifourXdeviceXbfgsXrecXdiagXdiagXdim};
        \addplot[histseries,rbfgs,device,hist10] table [ x=\xvar, y=\yvar, restrict expr to domain={\thisrow{history size}}{10:10},restrict expr to domain={\thisrow{num_variables}}{1000000:100000000},  unbounded coords=discard ] {\polarisXmpifourXdeviceXbfgsXrecXdiagXdiagXdim};
        \addplot[histseries,rbfgs,device,hist20] table [ x=\xvar, y=\yvar, restrict expr to domain={\thisrow{history size}}{20:20},restrict expr to domain={\thisrow{num_variables}}{1000000:100000000},  unbounded coords=discard ] {\polarisXmpifourXdeviceXbfgsXrecXdiagXdiagXdim};
        \addplot[histseries,rbfgs,device,hist50] table [ x=\xvar, y=\yvar, restrict expr to domain={\thisrow{history size}}{50:50},restrict expr to domain={\thisrow{num_variables}}{1000000:100000000},  unbounded coords=discard ] {\polarisXmpifourXdeviceXbfgsXrecXdiagXdiagXdim};

        \addplot[histseries,cd,device,hist5] table [ x=\xvar, y=\yvar, restrict expr to domain={\thisrow{history size}}{5:5},restrict expr to domain={\thisrow{num_variables}}{1000000:100000000},  unbounded coords=discard ] {\polarisXmpifourXdeviceXbfgsXcdXdiagXdiagXdim};
        \addplot[histseries,cd,device,hist10] table [ x=\xvar, y=\yvar, restrict expr to domain={\thisrow{history size}}{10:10},restrict expr to domain={\thisrow{num_variables}}{1000000:100000000},  unbounded coords=discard ] {\polarisXmpifourXdeviceXbfgsXcdXdiagXdiagXdim};
        \addplot[histseries,cd,device,hist20] table [ x=\xvar, y=\yvar, restrict expr to domain={\thisrow{history size}}{20:20},restrict expr to domain={\thisrow{num_variables}}{1000000:100000000},  unbounded coords=discard ] {\polarisXmpifourXdeviceXbfgsXcdXdiagXdiagXdim};
        \addplot[histseries,cd,device,hist50] table [ x=\xvar, y=\yvar, restrict expr to domain={\thisrow{history size}}{50:50},restrict expr to domain={\thisrow{num_variables}}{1000000:100000000},  unbounded coords=discard ] {\polarisXmpifourXdeviceXbfgsXcdXdiagXdiagXdim};

        \path (axis cs:1.3e-3,330) node {\small$10^6$};
        \path (axis cs:0.6e-2,1000) node {\small$n=10^7$};
        \path (axis cs:0.5e-1,1300) node {\small$10^8$};
        
        
        \node (halfBsy) at (axis cs:1,2800) {};
        \node (quarterBsy) at (axis cs:1,1400) {};
        \node (rightaxis) at (rel axis cs:1,0) {};
        \path (halfBsy -| rightaxis) node [anchor=north east,inner sep=1pt] {\small$\tfrac{1}{2}B_{\text{S}}$};
        \path (quarterBsy -| rightaxis) node [anchor=south east,inner sep=1pt] {\small$\tfrac{1}{4}B_{\text{S}}$};
        
        
    \end{semilogxaxis}
\end{tikzpicture}%

%% file: figure/polaris-mult-device-mpifour-dbfgs-bfgs.tex
\begin{tikzpicture}
    \begin{semilogxaxis}[
        ymin=0,
        xlabel={\strut$T_{\text{step}}$ (s)},
        ylabel={$B_{\text{e}}$ (GB/s)},
        legend pos=outer north east,
        extra y ticks={1400,2800},
        extra y tick labels={{},{}},
        extra y tick style={grid=major,major tick length=0pt},
        ylabel style={anchor=east,rotate=-90,at={(0,0.9)}},
        xlabel style={anchor=north west,at={(1,0)}},
        ]

        \addlegendimage{empty legend}
        \addlegendentry{\textbf{method}}
        \addlegendimage{histseries,rbfgs,device,hist50}
        \addlegendentry{Recursive}

        \addlegendimage{histseries,dense,device,hist50}
        \addlegendentry{Dense}



        \addlegendimage{empty legend}
        \addlegendentry{\textbf{history size $m$}}
        
        \addlegendimage{histseries,device,hist5}
        \addlegendentry{5}
        
        \addlegendimage{histseries,device,hist10}
        \addlegendentry{10}
        
        \addlegendimage{histseries,device,hist20}
        \addlegendentry{20}
        
        \addlegendimage{histseries,device,hist50}
        \addlegendentry{50}
        
            \addplot[varseries,rbfgs] table [ x=\xvar, y=\yvar, restrict expr to domain={\thisrow{num_variables}}{1000000:1000000}, unbounded coords=discard ] {\polarisXmpifourXmultXdeviceXbfgsXrecXdiagXuserXdim};
            \addplot[varseries,rbfgs] table [ x=\xvar, y=\yvar, restrict expr to domain={\thisrow{num_variables}}{10000000:10000000}, unbounded coords=discard ] {\polarisXmpifourXmultXdeviceXbfgsXrecXdiagXuserXdim};

            \addplot[varseries,dense] table [ x=\xvar, y=\yvar, restrict expr to domain={\thisrow{num_variables}}{1000000:1000000}, unbounded coords=discard ] {\polarisXmpifourXmultXdeviceXbfgsXdenseXdiagXuserXdim};
            \addplot[varseries,dense] table [ x=\xvar, y=\yvar, restrict expr to domain={\thisrow{num_variables}}{10000000:10000000}, unbounded coords=discard ] {\polarisXmpifourXmultXdeviceXbfgsXdenseXdiagXuserXdim};

        \addplot[histseries,rbfgs,device,hist5] table [ x=\xvar, y=\yvar, restrict expr to domain={\thisrow{history size}}{5:5},restrict expr to domain={\thisrow{num_variables}}{1000000:10000000}, unbounded coords=discard ] {\polarisXmpifourXmultXdeviceXbfgsXrecXdiagXuserXdim};
        \addplot[histseries,rbfgs,device,hist10] table [ x=\xvar, y=\yvar, restrict expr to domain={\thisrow{history size}}{10:10},restrict expr to domain={\thisrow{num_variables}}{1000000:10000000},  unbounded coords=discard ] {\polarisXmpifourXmultXdeviceXbfgsXrecXdiagXuserXdim};
        \addplot[histseries,rbfgs,device,hist20] table [ x=\xvar, y=\yvar, restrict expr to domain={\thisrow{history size}}{20:20},restrict expr to domain={\thisrow{num_variables}}{1000000:10000000},  unbounded coords=discard ] {\polarisXmpifourXmultXdeviceXbfgsXrecXdiagXuserXdim};
        \addplot[histseries,rbfgs,device,hist50] table [ x=\xvar, y=\yvar, restrict expr to domain={\thisrow{history size}}{50:50},restrict expr to domain={\thisrow{num_variables}}{1000000:10000000},  unbounded coords=discard ] {\polarisXmpifourXmultXdeviceXbfgsXrecXdiagXuserXdim};

        \addplot[histseries,dense,device,hist5] table [ x=\xvar, y=\yvar, restrict expr to domain={\thisrow{history size}}{5:5},restrict expr to domain={\thisrow{num_variables}}{1000000:10000000},  unbounded coords=discard ] {\polarisXmpifourXmultXdeviceXbfgsXdenseXdiagXuserXdim};
        \addplot[histseries,dense,device,hist10] table [ x=\xvar, y=\yvar, restrict expr to domain={\thisrow{history size}}{10:10},restrict expr to domain={\thisrow{num_variables}}{1000000:10000000},  unbounded coords=discard ] {\polarisXmpifourXmultXdeviceXbfgsXdenseXdiagXuserXdim};
        \addplot[histseries,dense,device,hist20] table [ x=\xvar, y=\yvar, restrict expr to domain={\thisrow{history size}}{20:20},restrict expr to domain={\thisrow{num_variables}}{1000000:10000000},  unbounded coords=discard ] {\polarisXmpifourXmultXdeviceXbfgsXdenseXdiagXuserXdim};
        \addplot[histseries,dense,device,hist50] table [ x=\xvar, y=\yvar, restrict expr to domain={\thisrow{history size}}{50:50},restrict expr to domain={\thisrow{num_variables}}{1000000:10000000},  unbounded coords=discard ] {\polarisXmpifourXmultXdeviceXbfgsXdenseXdiagXuserXdim};

        \path (axis cs:5e-3,100) node [anchor=center] {\small$10^6$};
        \path (axis cs:1.5e-2,400) node [anchor=north] {\small$n=10^7$};

        \node (halfBsy) at (axis cs:1,2800) {};
        \node (quarterBsy) at (axis cs:1,1400) {};
        \node (rightaxis) at (rel axis cs:1,0) {};
        \path (halfBsy -| rightaxis) node [anchor=south east,inner sep=1pt] {\small$\tfrac{1}{2}B_{\text{S}}$};
        \path (quarterBsy -| rightaxis) node [anchor=south east,inner sep=1pt] {\small$\tfrac{1}{4}B_{\text{S}}$};
        
        
    \end{semilogxaxis}
\end{tikzpicture}%